\documentclass[preprint,12pt]{elsarticle}
\usepackage{graphicx}

\usepackage{amssymb}
 \usepackage{amsthm}
\usepackage{amsmath}
\usepackage{lineno}

\usepackage{enumerate}
\usepackage{here}
\usepackage{indentfirst}
\usepackage{bm}
\usepackage{url}
\usepackage{color}
\usepackage{ulem}
\usepackage{here}
\usepackage{tikz}
\usetikzlibrary{shapes.geometric}
\usetikzlibrary {shapes.misc}
\usetikzlibrary{positioning}
\usepackage[hang,small,bf]{caption}
\usepackage[subrefformat=parens]{subcaption}
\def\XXint#1#2#3{{\setbox0=\hbox{$#1{#2#3}{\int}$ }
		\vcenter{\hbox{$#2#3$ }}\kern-.6\wd0}}

\newcommand{\dbar}[1]{ \hat{#1}   }

\journal{Elsevier}

\begin{document}

\begin{frontmatter}



\title{
	Topology optimization for acoustic structures
	\\considering viscous and thermal boundary layers
	\\using a sequential linearized Navier--Stokes model
}

\author{Yuki~Noguchi\corref{cor1}} 
\ead{noguchi@mech.t.u-tokyo.ac.jp}
\cortext[cor1]{Corresponding author.
	Tel.: +81-3-5841-0294;
	Fax: +81-3-5841-0294.}
\author{Takayuki~Yamada \corref{}}

\address{Department of Strategic Studies, Institute of Engineering Innovation, School of Engineering, The University of Tokyo, Yayoi 2-11-16, Bunkyo--ku, Tokyo 113-8656, Japan.}

\begin{abstract}
This \textcolor {black}{study} proposes a level set-based topology optimization method for designing acoustic structures with viscous and thermal boundary layers in perspective. \textcolor {black}{Acoustic} waves propagating in a narrow channel are damped by viscous and thermal boundary layers. To estimate these viscothermal effects, we first introduce a sequential linearized Navier--Stokes model based on three weakly coupled Helmholtz equations for viscous, thermal, and acoustic pressure fields. Then, the optimization problem is formulated, where a sound-absorbing structure comprising air and an isothermal rigid medium is targeted, and its sound absorption coefficient is set as an objective function. The adjoint variable method and the concept of the topological derivative are used to \textcolor {black}{approximately} obtain design sensitivity. A level set-based topology optimization method is used to solve the optimization problem. Two-dimensional numerical examples are provided to support the validity of the proposed method. \textcolor {black}{Moreover}, the mechanisms that lead to the high absorption coefficient of the optimized design are discussed.	
\end{abstract}


\end{frontmatter}


\section{Introduction}
\label{intro}
Controlling the propagation behavior of acoustic waves is important to achieve a comfortable environment, and various types of acoustic devices are used to obtain desired acoustic performance.
For example, unwanted sounds can be reduced using sound barrier walls or sound-absorbing materials.
Acoustic metamaterials \cite{liu2000locally}, \textcolor {black}{artificial composite structures exhibiting extraordinary properties that cannot be attained in homogeneous acoustic media that naturally exist,
} have expanded the possibilities in acoustic wave control. In particular, negative refraction is achieved by a metamaterial that simultaneously possesses negative mass density and negative bulk modulus \cite{li2004double}. Novel acoustic devices with acoustic cloaking structure \cite{zigoneanu2014three} and acoustic hyperlenses \cite{li2009experimental} have been realized owing to the unusual properties of metamaterials.

\textcolor {black}{As represented by the abovementioned example, acoustic structures, i.e., composite structures that efficiently control acoustic wave propagation behavior, have been widely used.} The performance of acoustic structures, such as transmission coefficients, reflection coefficients, and effective material parameters, is heavily influenced by structural design. Therefore, topology optimization is a powerful tool for designing acoustic structures with exceptional features. Its design offers the highest degree of freedom of any structural optimization method and has been used to solve acoustic problems \textcolor {black}{after} Bends{\o}e and Kikuchi \cite{bendsoe1988generating} developed it for linear elasticity problems. \textcolor {black}{Many} optimization examples for \textcolor {black}{designing} acoustic devices are available. Wadbro and Berggren optimized an acoustic horn \cite{wadbro2006topology}. Du and Olhoff \cite{du2007minimization} optimized a bi-material structure design to \textcolor {black}{minimize} sound radiation. Topology optimization has also been used to optimize acoustic metamaterials. Lu et al. \cite{lu2013topology} \textcolor {black}{reported} an optimized unit cell configuration with negative bulk modulus. Dong et al. \cite{dong2019systematic} used genetic algorithm-based topology optimization to develop a double-negative acoustic metamaterial that induced negative refraction. A topology optimization method for a planar type of acoustic metamaterials was proposed based on the estimation using a two-scale homogenization method \cite{noguchi2021topology}. This method has been extended for metamaterials composed of multiple unit cells \cite{noguchi2021level}. \textcolor {black}{Moreover}, topology optimization can address acoustic-elastic coupled problems. A comparative review of topology optimization for acoustic-elastic coupled systems has been previously reported \cite{dilgen2019topology}.

Acoustic devices that produce audible sound frequently have narrow channels via which acoustic waves propagate. A typical hearing aid device has compact microphones and speakers that can be worn in the ear. Wind musical instruments have pipes that generate standing waves, and the sound is produced by blowing into them. Almost all of the acoustic metamaterials discussed above, including topology-optimized ones, are composed of a periodic array of microstructures, with tiny channels and cavities included in each unit cell to induce local resonances.

The behavior of acoustic waves propagating in these narrow channels differs from that in bulk, and thermal and viscous boundary layers influence acoustic waves. Airborne sound is impacted by viscous forces that are proportional to the velocity gradient owing to the viscosity of air. When the nonslip boundary condition is imposed on the channel's surface, the velocity is driven to zero. Then, using a thin boundary layer, a significant velocity gradient is formed. Moreover, acoustic pressure perturbation generates temperature perturbation in the bulk region. Owing to the difference in thermal conductivities between air and solid, this perturbation becomes zero at the channel's surface, and the transition in the temperature field from the surface to the bulk yields a substantial temperature gradient, \textcolor {black}{thus} resulting in the thermal boundary layer. These boundary layers dampen the acoustic pressure, and the damping increases as the surface-to-volume ratio increases.

Acoustic metamaterials are no exception when it comes to viscous and thermal boundary layer phenomena. Moleron et al. \cite{moleron2016visco} theoretically and experimentally investigated these effects on rigid slab-based metamaterials' transmission and reflection characteristics. They \textcolor {black}{reported} that losses induced by boundary layers affect resonances in metamaterials and they are substantially dependent on geometrical parameters. Christiansen and Sigmund \cite{christiansen2016experimental} established \textcolor {black}{the} experimental validation of a topology-optimized acoustic metamaterial. The sample used in the experiment was \textcolor {black}{3D} printed based on an optimized design developed using a density-based topology optimization algorithm \cite{christiansen2016designing} without considering the viscothermal losses \textcolor {black}{with the aim of} achieving a negative refractive wave propagation behavior. In addition to the desired negative refraction behavior, they detected unexpected sound attenuation, which was considered to be produced by the boundary layer effects. Henr\'{i}quez et al. \cite{henriquez2017viscothermal} observed that an acoustic metamaterial intended to demonstrate double-negative behavior did not perform as anticipated owing to sound absorption and the existence of viscous and thermal boundary layers. Furthermore, Huang et al. \cite{huang2020compact} proposed a broadband sound-absorbing metamaterial composed of multiple numbers of neck-embedded Helmholtz resonators, which are designed to generate coherent coupling of resonances while accounting for viscous and thermal losses.

These examples show that considering the viscous and thermal boundary layer effects is critical when designing acoustic structures. Although the abovementioned studies of topology optimization for acoustics did not consider these effects, an optimization approach with viscothermal losses could explore more realistic designs that demonstrate desirable functions. An adequate numerical analysis model is necessary to conduct such an optimization. There are numerous acoustic analysis models available that account for boundary layer effects. A set of linearized Navier--Stokes equations is used in a full linearized Navier--Stokes (FLNS) model. This linearization is based on \textcolor {black}{small} fluctuations in the still air equilibrium condition. The finite element method (FEM) can be used to generate numerical solutions; however, the computational cost is high because the equations are strongly coupled with state fields. Kirchhoff \cite{Kirchhoff} studied analytical solutions to the linearized Navier--Stokes equations defined in an infinite cylinder with a rigid isothermal wall as its surface. Tijdeman \cite{tijdeman1975propagation} validates Zwikker and Kosten's \cite{zwikker1949sound} simplified Kirchhoff solution, and the low reduced frequency (LRF) model was proposed. This model can efficiently obtain the acoustic pressure by deriving the pressure field from \textcolor {black}{1D or 2D} problems and \textcolor {black}{analytically} accounting for viscothermal effects. However, it is only valid if acoustic waves propagate in layers or tubes with cross-sections smaller than the acoustic wavelength. Bossart et al. \cite{bossart2003hybrid} proposed the boundary layer impedance model based on the isentropic Helmholtz equation, which is \textcolor {black}{extensively} used to characterize undamped acoustic waves in bulk with a boundary condition equivalent to the viscothermal effects. This boundary condition depends on the estimation of the tangential wavenumber defined on the channel surface, which necessitates iterative acoustic pressure computation. Furthermore, the boundary layer impedance model is only valid if the thickness of the boundary layer is substantially thinner than all of the geometry's characteristic lengths. \textcolor {black}{Recently,} Berggren et al. \cite{berggren2018acoustic} proposed a similar method that requires only one solution of the isentropic Helmholtz equation with a Wentzell boundary condition; however, their model does not apply to surfaces with large curvatures because the boundary condition is derived based on the assumption that the surfaces are flat.

A sequential linearized Navier--Stokes (SLNS) model suggested by Kampinga et al. \cite{kampinga2010viscothermal} is another model for viscothermal acoustics. The governing equation of this model is derived from the set of equations used in the FLNS model. They considered a case in which the acoustic wavelength is much larger than the viscous and thermal boundary layer thicknesses for airborne sound in the audible frequency range. This assumption simplifies the equations in the FLNS model and weakly coupled Helmholtz equations are derived corresponding to the viscous and thermal fields, in addition to the acoustic pressure. This model does not impose any assumptions about the geometry as opposed to the previously discussed models, except for the FLNS model. Because three types of weakly coupled systems must be solved without dimension reduction, it requires \textcolor {black}{additional} computational effort compared to the LRF model and models proposed by Bossart et al. \cite{bossart2003hybrid} and Berggren et al. \cite{berggren2018acoustic}. However, its computational cost is lower than that of the FLNS model owing to the simplification of the FLNS equations. Furthermore, the acoustic-elastic coupled analysis can be conducted with the SLNS model in which the deformation of the solid medium is considered, as found in \cite{kampinga2010viscothermal}.

Structural optimization methods have recently been combined with these analysis models for viscothermal acoustics. Christensen \cite{christensen2017topology} conducted topology optimization for a hearing aid application example using the LRF model. Tissot et al. \cite{tissot2020optimal} proposed a shape optimization method in which the viscothermal acoustic model proposed by Berggren et al. \cite{berggren2018acoustic} was used, and XFEM cut elements were used to impose the abovementioned boundary conditions. However, these optimization methods have restrictions on the structure's geometry because of their analysis models, as previously mentioned. Andersen et al. \cite{andersen2019shape} proposed a shape optimization method for acoustic structures by considering viscothermal losses estimated using a dissipative boundary element method (BEM). Although this method has no geometry limitation, it was emphasized in \cite{andersen2019shape} that the adjoint system cannot be solved in the BEM's framework. Rather than using the sensitivity based on the adjoint variable method, the finite difference method was introduced in \cite{andersen2019shape}; however, it costs \textcolor {black}{considerable} computational time when a large number of design variables is considered. Generally, the design variables in topology optimization are larger than that of shape optimization; the direct use of this approach to topology optimization seems difficult.

As explained, the SLNS model can be applied to any geometry for audible sounds. A topology optimization method incorporated with the SLNS model and associated adjoint variable-based sensitivity analysis enables optimal design of acoustic structures, including metamaterials, considering the viscous and thermal boundary layer effects without restricting the optimized structure design. 
In this \textcolor {black}{study}, we propose a level set-based topology optimization method, \textcolor {black}{which was} proposed by Yamada et al. \cite{yamada2010topology}, combined with the SLNS model for a viscothermal acoustics system.
The rest of this paper is organized as follows. Section \ref{sec: SLNS model} provides a brief overview of the SLNS model and its derivation from the FLNS model. In Section \ref{sec: Formulation of Topology optimization}, we explain the design settings and formulate the optimization problem. As the viscous and thermal boundary layers play an important role in optimization, we target a sound-absorbing structure comprising air and an isothermal rigid medium, and its sound absorption coefficients over a specific frequency range are set as an objective function. In addition, the design sensitivity is \textcolor {black}{approximately} derived based on the adjoint variable method and the topological derivative concept \cite{Sokolowski19991251}, and its explicit formula is obtained. Subsequently, a level set-based topology optimization method is introduced in Section \ref{sec: Level set-based topology optimization}. Numerical implementation, including the FEM-based discretization for state and adjoint variables and the optimization algorithm, is discussed in Section \ref{sec:Numerical implementation}. Two \textcolor {black}{2D} numerical examples are provided in Section \ref{sec: numerical examples}, corresponding to the maximization of the absorption coefficients in closed and open tubes, respectively. Optimized designs are demonstrated, and their performances are examined using the SLNS and FLNS models. \textcolor {black}{Furthermore}, we discuss the cause of the high absorption coefficients of these designs using the FLNS model. 
\textcolor {black}{We demonstrate the importance of considering boundary layers in the optimization in Appendices A and B by offering optimized designs with and without boundary layers.}
Finally, conclusions are summarized in Section \ref{sec: Conclusion}.

\section{SLNS model}\label{sec: SLNS model}
\begin{figure}[H]
	\centering
	\includegraphics[scale=0.3]{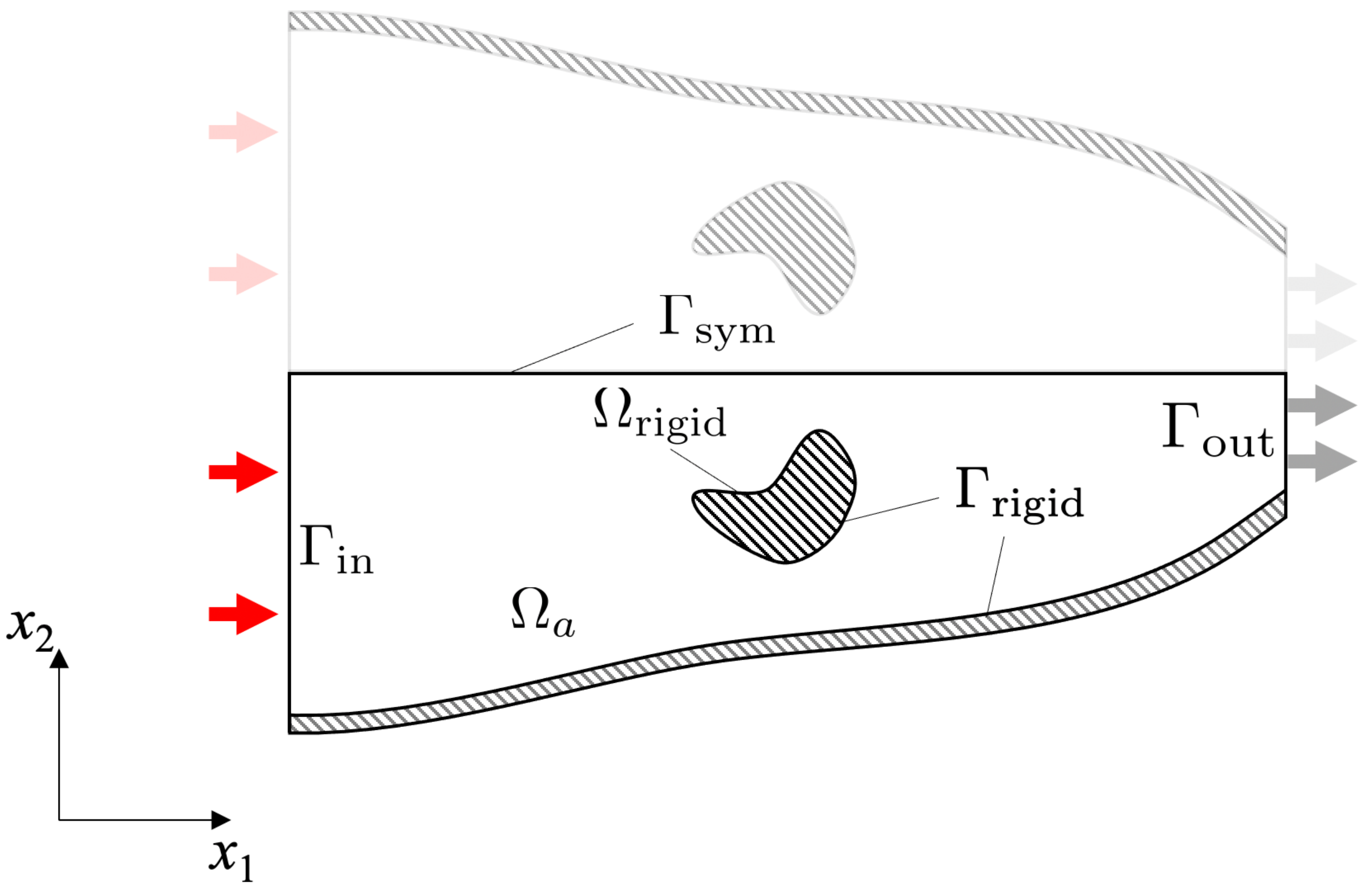}
	\caption{Geometrical settings and boundary conditions. }
\label{fig:Geom}       
\end{figure}

In this section, we briefly introduce an approximate method for the viscothermal acoustic model, \textcolor {black}{known as} the SLNS model. This model is derived from the FLNS model, which is based on the linear acoustic assumption and Newton--Fourier ideal gas assumptions \cite{pierce2019acoustics}. The details of this derivation are provided in \cite{kampinga2010viscothermal}. Several analysis models for viscothermal acoustics, including the FLNS and SLNS models, are compared in \cite{thibault2020viscothermal}.

Figure \ref{fig:Geom} illustrates the geometrical settings. We consider the situation that a rigid medium, which is non-deformable and isothermal, denoted by $\Omega_\mathrm{rigid}$, is surrounded by an air-filled region, $\Omega_a$. The incident acoustic wave $P_\mathrm{in}$ impinges on a boundary, $\Gamma_\mathrm{in}$, and it is transmitted through an outlet, $\Gamma_\mathrm{out}$. The boundary of $\Omega_\mathrm{rigid}$ and a side wall are considered as rigid surfaces on which acoustic waves are reflected, and viscous and thermal boundary layer effects appear. A mirror-symmetric geometry with respect to a boundary, $\Gamma_\mathrm{sym}$, is considered, and half of the entire geometry is targeted for numerical analysis.

We first introduce the FLNS model. 
Let $(\check{\rho}, \check{p},\check{T},\check{\bm{v}},\check{H},\check{s})$ denote the time-dependent density, pressure, temperature, velocity, specific enthalpy, and specific entropy of air.
Assuming the harmonic oscillation with the angular frequency $\omega$, these time-dependent fields can be decomposed into their quiescent and oscillating parts, as follows:
\begin{align}
	\check{\psi} &= \psi_0 + \mathrm{Re}\left\{  \psi\exp(i\omega t)\right\},
\end{align}
where $\check{\psi}$ represents a total field corresponding to $(\check{\rho}, \check{p},\check{T},\check{\bm{v}},\check{H},\check{s})$, $\psi_0$ is its quiescent value, and $\psi$ is the complex amplitude of the oscillating part of $\check{\psi}$.
\textcolor {black}{The following set of equations for the oscillating part of variables is obtained based on the linear acoustic assumption and the constitutive laws for a Newton-Fourier ideal gas:
}
\begin{align}
	i\omega \rho_0 \bm{v} - \nabla \cdot \sigma &= \bm{f}~~~\mathrm{in~}\Omega_a, \label{eq: FLNS 1}\\
	i\omega \rho_0 C_p T + \nabla \cdot \bm{q} -i\omega p &=Q~~~\mathrm{in~}\Omega_a,\label{eq: FLNS 2}\\
	\nabla \cdot \bm{v} -i\omega \frac{T}{T_0}+i\omega \frac{p}{p_0}&=0~~~\mathrm{in~}\Omega_a,\label{eq: FLNS 3}
\end{align}
which correspond to the \textcolor {black}{momentum, entropy, and continuity equations}, respectively.
$\bm{f}$ and $Q$ are the body force and heat source, respectively, $\sigma$ is the stress tensor, and $\bm{q}$ is the heat flow vector. For a Newton-Fourier ideal gas, the following constitutive laws hold:
\begin{align}
	&\sigma = \lambda(\nabla \cdot \bm{v})\bm{I} + 2\mu \varepsilon(\bm{v}) -p\bm{I},\nonumber\\
	&\bm{q}= -\kappa \nabla T,\nonumber\\
	&\frac{\rho}{\rho_0}=\frac{p}{p_0}-\frac{T}{T_0},\nonumber\\
	&H=C_p T,\label{eq: constitutive laws}
\end{align}
where $\bm{I}$ represents the identity tensor of the second order, and $\varepsilon(\bm{v})=\frac{1}{2}\left\{\nabla \bm{v}	 + (\nabla \bm{v})^T \right\}$ extracts the symmetric part of the gradient of $\bm{v}$.
$\mu$ and $\lambda$ are the dynamic viscosity and the second viscosity coefficients, respectively, 
$\kappa$ represents the heat conduction coefficient, and
$C_p$ is the specific heat at constant pressure.
\textcolor {black}{Equations (\ref{eq: FLNS 1})--(\ref{eq: FLNS 3}) are} used as the governing equations for the FLNS model.
Notably, they are fully coupled with the variables $(\bm{v},T,p)$.
Hereafter, the body source $\bm{f}$ and heat source $Q$ are set to zero for simplicity.

The basic boundary conditions used in the viscothermal acoustic analysis are summarized in \cite{kampinga2010viscothermal}.
In this paper, the following boundary conditions are introduced:
\begin{align}
	&\bm{v}\cdot \bm{n} =0 ~~~\mathrm{on~}\Gamma_\mathrm{rigid},\label{eq: BC_FLNS_first}\\
	&\bm{v}\cdot \bm{t} =0 ~~~\mathrm{on~}\Gamma_\mathrm{rigid},\label{eq: BC_FLNS_v2}\\
	&T=0 ~~~\mathrm{on~}\Gamma_\mathrm{rigid},\label{eq: BC_FLNS_T0}\\
	&\bm{v}\cdot \bm{n} =0 ~~~\mathrm{on~}\Gamma_\mathrm{sym},\\
	&(\sigma \cdot \bm{n})\cdot \bm{t}=0 ~~~\mathrm{on~}\Gamma_\mathrm{sym},\\
	&\bm{q}\cdot \bm{n}=0 ~~~\mathrm{on~}\Gamma_\mathrm{sym},\\
	&(\sigma \cdot \bm{n})\cdot \bm{n} =-\rho_0 c_0(\bm{v}\cdot \bm{n}) - 2P_\mathrm{in} ~~~\mathrm{on~}\Gamma_\mathrm{in},\label{eq: BC_FLNS_incident1}\\
	&(\sigma \cdot \bm{n})\cdot \bm{t}=0 ~~~\mathrm{on~}\Gamma_\mathrm{in},\label{eq: BC_FLNS_incident2}\\
	&\bm{q}\cdot \bm{n}=0 ~~~\mathrm{on~}\Gamma_\mathrm{in},\label{eq: BC_FLNS_incident3}\\
	&(\sigma \cdot \bm{n})\cdot \bm{n} =-\rho_0 c_0(\bm{v}\cdot \bm{n}) ~~~\mathrm{on~}\Gamma_\mathrm{out},\label{eq: BC_FLNS_non1}\\
	&(\sigma \cdot \bm{n})\cdot \bm{t}=0 ~~~\mathrm{on~}\Gamma_\mathrm{out},\label{eq: BC_FLNS_non2}\\
	&\bm{q}\cdot \bm{n}=0 ~~~\mathrm{on~}\Gamma_\mathrm{out},\label{eq: BC_FLNS_last}
\end{align}
where $\bm{n}$ is an outward unit normal vector defined in $\Omega_a$, and $\bm{t}$ is the unit tangential vector. The non-slip boundary condition for $\bm{v}$ and the isothermal boundary condition for $T$ are applied to the rigid surfaces, $\Gamma_\mathrm{rigid}$, as expressed by Eqs.~(\ref{eq: BC_FLNS_first})--(\ref{eq: BC_FLNS_T0}). The incident wave condition is applied to $\Gamma_\mathrm{in}$, \textcolor {black}{where $P_\mathrm{in}$ represents the incident wave expression.}
\textcolor {black}{The} non-reflecting boundary condition is applied to $\Gamma_\mathrm{out}$ to reduce the reflected waves on that boundary. Based on the mirror-symmetric geometry, the symmetric boundary condition is applied to $\Gamma_\mathrm{sym}$.

Next, the SLNS model is provided by imposing further assumptions to the FLNS model.
It is well known that the wavelength of an acoustic wave is much longer than the thickness of the viscous and thermal boundary layers for audible sounds.
This property can be expressed in terms of three types of wavenumbers: acoustic, thermal, and viscous wavenumbers ($k_0$, $k_h$, and $k_v$), which are defined as
\begin{align}
	k_0 = \frac{\omega}{c_0},~~~
	k_v^2 = -\frac{i\omega\rho_0}{\mu},~~~
	k_h^2 = -\frac{i\omega\rho_0 C_p}{\kappa}.
\end{align}
The acoustic wavenumber is given as a real number, whereas thermal and viscous wavenumbers are complex numbers.
Using these wavenumbers, the abovementioned property of the viscous and thermal boundary layers are described as
\begin{align}
	|\dbar{k_v}|\gg 1,~~~|\dbar{k_h}|\gg 1, \label{eq: property of layers}
\end{align}
where $\dbar{k_v}=\frac{k_v}{k_0}$ and $\dbar{k_h}=\frac{k_h}{k_0}$, respectively.
The set of equations used in the SLNS model are derived by ignoring small terms considering the viscous and thermal layer's property \textcolor {black}{(Eq.~(\ref{eq: property of layers})).}
\begin{align}
	u_\varphi + k_{\varphi}^{-2}\nabla^2 u_\varphi &= 1~~~\mathrm{in~}\Omega_a,\label{eq: SLNS_uphi}\\
	\nabla \cdot \left(\frac{u_v}{\rho_0}\nabla p\right) + \frac{\omega^2}{K_0}\left\{\gamma - (\gamma -1)u_h\right\}p &= 0~~~\mathrm{in~}\Omega_a,\label{eq: SLNS_p}
\end{align}
with $\varphi=v,h$. \textcolor {black}{$\gamma$ represents the ratio of specific heats, and $K_0$ represents the bulk modulus of the medium.}
Equations~(\ref{eq: SLNS_uphi}) and (\ref{eq: SLNS_p}) are utilized as the governing equations in the SLNS model.
Different from those in the FLNS model, these equations are weakly coupled for the variables $(u_v,u_h,p)$. That is, once the systems for $u_v$ and $u_h$, which are independent from each other, are solved, the acoustic pressure $p$ can be sequentially obtained. Importantly, the isentropic Helmholtz equation for acoustic waves without considering the boundary layer effects is covered if $u_v=1$ and $u_h=1$ are substituted into Eq.~(\ref{eq: SLNS_p}).

Corresponding to the boundary conditions in the FLNS model expressed in \textcolor {black}{Eqs.}~(\ref{eq: BC_FLNS_first})--(\ref{eq: BC_FLNS_last}), 
the boundary conditions for $(u_v,u_h,p)$ are summarized as follows:
\begin{align}
	&u_\varphi = 0~~~\mathrm{on~}\Gamma_\mathrm{rigid},\\
	&\bm{n}\cdot \nabla p = 0~~~\mathrm{on~}\Gamma_\mathrm{rigid},\\
	&\bm{n}\cdot \nabla u_\varphi = 0~~~\mathrm{on~}{\Gamma_\mathrm{sym}},\\
	&\bm{n}\cdot \nabla p = 0~~~\mathrm{on~}\Gamma_\mathrm{sym},\\
	&\bm{n}\cdot \nabla u_\varphi = 0~~~\mathrm{on~}\Gamma_\mathrm{in},\\
	&\bm{n}\cdot \nabla p + ik_0 p  = 2ik_0 P_\mathrm{in}~~~\mathrm{on~}\Gamma_\mathrm{in},\\
	&\bm{n}\cdot \nabla u_\varphi = 0~~~\mathrm{on~}\Gamma_\mathrm{out},\\
	&\bm{n}\cdot \nabla p + ik_0 p  = 0~~~\mathrm{on~}\Gamma_\mathrm{out},
\end{align}
with $\varphi = v,h$.

Finally, the weak forms of the set of equations used in FLNS and SLNS models are established. To distinguish the pressure $p$ in the FLNS model from that in the SLNS model, the notation $p_f$ is used for the FLNS model. Considering the boundary conditions in Eqs.~(\ref{eq: BC_FLNS_first})--(\ref{eq: BC_FLNS_last}), the weak forms in the FLNS model can be expressed as follows:
\begin{align}
	&i\omega \rho_0\int_{\Omega_a}\bm{v}\cdot \tilde{\bm{v}} d\Omega + 2\mu \int_{\Omega_a}\varepsilon(\bm{v}): \varepsilon(\tilde{\bm{v}})d\Omega +\lambda \int_{\Omega_a}(\nabla \cdot {\bm{v}}) (\nabla \cdot \tilde{\bm{v}}) d\Omega\nonumber\\
	&+\rho_0 c_0\int_{\Gamma_\mathrm{in}\cup\Gamma_\mathrm{out}}(\bm{v}\cdot \bm{n}) (\tilde{\bm{v}}\cdot \bm{n}) d\Gamma
	-\int_{\Omega_a}p_f(\nabla \cdot \tilde{\bm{v}}) d\Omega \nonumber\\
	&= -\int_{\Gamma_\mathrm{in}} 2 P_\mathrm{in}\bm{n}\cdot \tilde{\bm{v}}  d\Gamma
	~~~\forall \tilde{\bm{v}} \in W_u,\label{eq: weak_FLNS_disp}\\
	&-\frac{i\omega \rho_0 C_p}{T_0}\int_{\Omega_a}T\tilde{T}d\Omega -\frac{\kappa}{T_0}\int_{\Omega_a}\nabla T \cdot \nabla \tilde{T}d\Omega + \frac{i\omega}{T_0}\int_{\Omega_a}p_f\tilde{T}d\Omega = 0
	~~~\forall \tilde{T} \in W_T,\label{eq: weak_FLNS_T}\\
	&-\frac{i\omega}{p_0}\int_{\Omega_a}p_f\tilde{p_f}d\Omega - \int_{\Omega_a} (\nabla \cdot \bm{v})  \tilde{p_f}d\Omega + \frac{i \omega}{T_0}\int_{\Omega_a}T\tilde{p_f}d\Omega = 0
	~~~\forall \tilde{p_f} \in L^2(\Omega_a),\label{eq: weak_FLNS_p}
\end{align}
where $\tilde{\bm{v}}\in W_u$, $\tilde{T}\in W_T$, and $\tilde{p_f}\in L^2(\Omega_a)$ are the test functions corresponding to $(\bm{v},T,p_f)$. 
$W_u = \left\{ \tilde{\bm{v}}\in \{H^1(\Omega_a)\}^d |~ \tilde{\bm{v}}=0~\mathrm{on~}\Gamma_\mathrm{rigid},~\tilde{\bm{v}}\cdot \bm{n}=0~\mathrm{on~}\Gamma_\mathrm{sym}\right\}$\\
and $W_T = \left\{ \tilde{T}\in H^1(\Omega_a) |~ \tilde{T}=0~\mathrm{on~}\Gamma_\mathrm{rigid}\right\}$ are the subspaces of the Sobolev space satisfying the Dirichlet boundary conditions, where $d=2$ or $3$ represents the number of spatial dimensions. 

Likewise, the weak forms in the SLNS model can be expressed as
\begin{align}
	&\int_{\Omega_a}k_\varphi^{-2}\nabla u_\varphi \cdot \nabla \tilde{u_\varphi} d\Omega -\int_{\Omega_a}u_\varphi \tilde{u_\varphi} d\Omega = -\int_{\Omega_a}\tilde{u_\varphi} d\Omega ~~~\forall \tilde{u_\varphi} \in V_u,\\
	&\int_{\Omega_a}\left(\frac{u_v}{\rho_0}\right)\nabla p \cdot \nabla \tilde{p}d \Omega - \int_{\Omega_a}\frac{\omega^2}{K_0}\{\gamma-(\gamma-1)u_h
	\} p\tilde{p}d\Omega \nonumber\\
	&+ \int_{\Gamma_\mathrm{in}\cup\Gamma_\mathrm{out}}u_v\frac{i k_0}{\rho_0}p\tilde{p}d\Gamma
	= \int_{\Gamma_\mathrm{in}}u_v\frac{2ik_0}{\rho_0}P_\mathrm{in}	\tilde{p}d\Gamma
	~~~\forall \tilde{p} \in H^1(\Omega),
\end{align}
with $\varphi = v,h$.
$\tilde{u_\varphi}\in V_u$ and $\tilde{p}$ are test functions corresponding to $(u_\varphi,p)$.
Similar to $W_T$, $V_u$ is defined as $V_u = \left\{ \tilde{u_\varphi}\in H^1(\Omega_a) |~ \tilde{u_\varphi}=0~\mathrm{on~}\Gamma_\mathrm{rigid}\right\}$.

Although the FLNS model requires fewer assumptions and thus is more accurate than the SLNS model, it is based on the set of equations strongly coupled with each other, whereas the SLNS model uses weakly coupled equations. 
Furthermore, the number of unknown variables in these models is different.
Depending on the number of spatial dimensions ($d = 2$ or $3$), 
the FLNS model requires $(d+2)$ variables, depending on the component of the velocity field, while the SLNS model requires only three variables since all of their equations are the scalar Helmholtz equations. 
Therefore, in this research, we propose a topology optimization method based on the SLNS model and use it for optimizing the acoustic structure.

\textcolor {black}{\ref{sec: Detail SLNS} presents more details on the derivation of the SLNS model from the FLNS model.}

\section{Topology optimization for the viscothermal acoustics problem}\label{sec: Formulation of Topology optimization}

\subsection{Design settings}
\begin{figure}[H]
	\centering
	\includegraphics[scale=0.6]{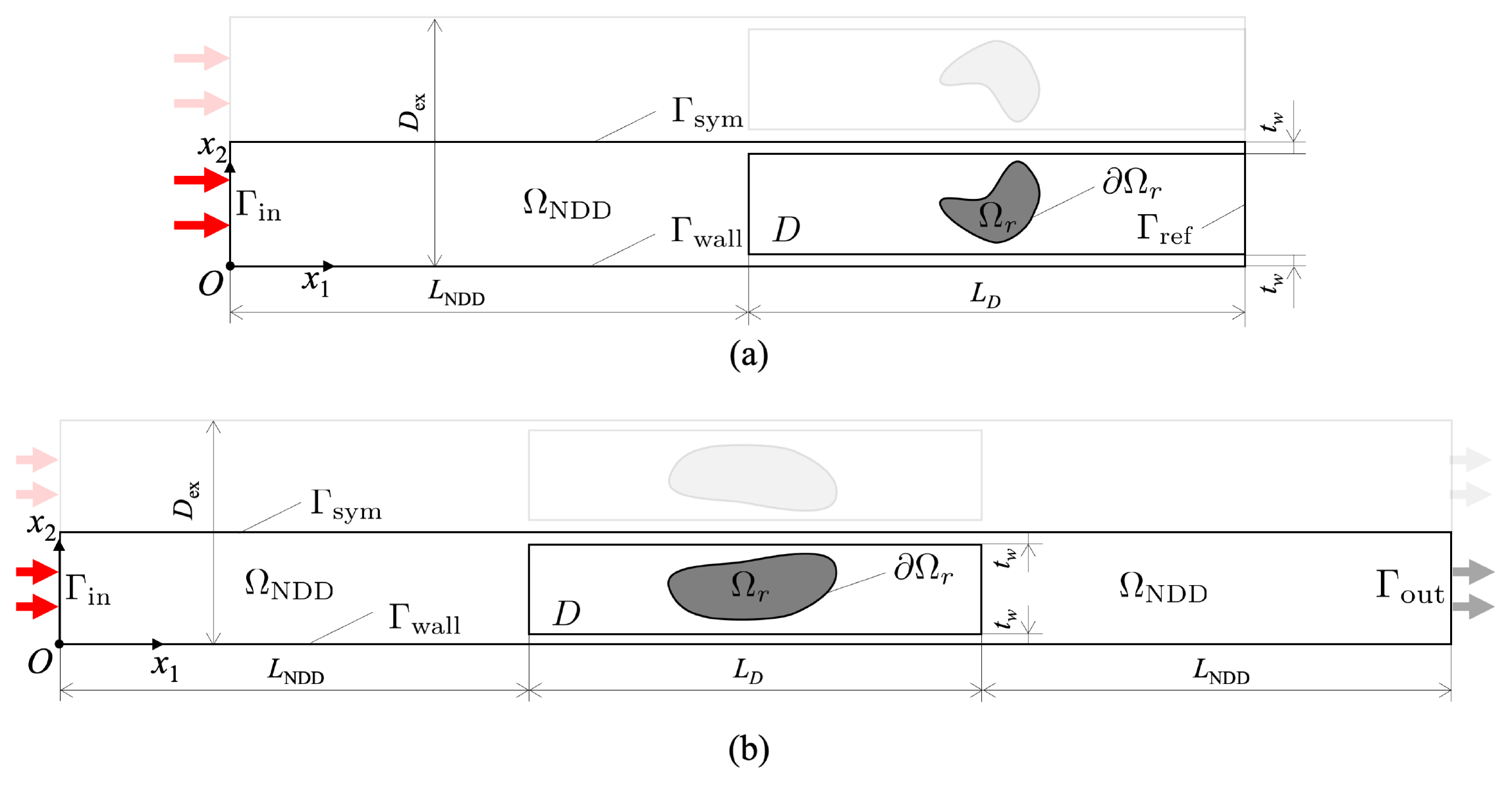}
	\caption{Geometrical settings and boundary conditions for the system of viscothermal acoustics. 
	(a) Case 1. (b) Case 2.}
	\label{fig:Design settings}       
\end{figure}
In this section, the design settings in topology optimization for the viscothermal acoustics problem are explained. Figure~\ref{fig:Design settings} shows the settings of the design domain and boundary conditions. 
We consider two types of two-dimensional design problems, Case 1 and Case 2. 
We set a design domain $D$, which is composed of the rigid domain $\Omega_r$ and air-filled region $D\setminus \overline{\Omega_r}$, and a non-design domain $\Omega_\mathrm{NDD}$ filled with air.
Symmetry structures are considered in both cases, and their half domain is set as a computational domain. The dimensions of geometries are denoted by $D_\mathrm{ex}$, $L_\mathrm{NDD}$, $L_D$, and $t_w$, respectively, which are detailed later in Section~\ref{sec: numerical examples}. 
Figure~\ref{fig:Design settings}(a) shows the settings for Case 1, which corresponds to a closed tube. An incident plane wave, $P_\mathrm{in}=\exp(-i k_0 x_1)$, impinges on $\Gamma_\mathrm{in}$, and it reflects on the right end, $\Gamma_\mathrm{ref}$. The bottom boundary $\Gamma_\mathrm{wall}$ is set as a rigid wall. Then, the Dirichlet boundary conditions for $u_v$ and $u_h$ are applied to $\Gamma_\mathrm{wall}$ and the surface of the rigid domain in $D$, denoted by $\partial \Omega_r$. The symmetry condition is imposed on the top boundary $\Gamma_\mathrm{sym}$. It is also applied to $\Gamma_\mathrm{ref}$ to create reflected waves.

Figure~\ref{fig:Design settings}(b) shows the settings for Case 2, which corresponds to an open tube. An incident plane wave impinges on $\Gamma_\mathrm{in}$ similarly, but it is transmitted through an outlet located at the right end, $\Gamma_\mathrm{out}$. The other settings of boundary conditions are the same as those in Case 1.

Based on the abovementioned boundary conditions and the SLNS model introduced in Section~\ref{sec: SLNS model}, the weak forms in Case 1 are summarized as follows:
\begin{align}
	&\int_{\Omega_a}k_\varphi^{-2}\nabla u_\varphi \cdot \nabla \tilde{u_\varphi} d\Omega -\int_{\Omega_a}u_\varphi \tilde{u_\varphi} d\Omega = -\int_{\Omega_a}\tilde{u_\varphi} d\Omega ~~~\forall \tilde{u_\varphi} \in V_u,\nonumber\\
	&\int_{\Omega_a}\left(\frac{u_v}{\rho_0}\right)\nabla p \cdot \nabla \tilde{p}d \Omega - \int_{\Omega_a}\frac{\omega^2}{K_0}\{
	\gamma - (\gamma-1)u_h
	\} p\tilde{p}d\Omega \nonumber\\
	&+ \int_{\Gamma_\mathrm{in}}u_v\frac{i k_0}{\rho_0}p\tilde{p}d\Gamma
	= \int_{\Gamma_\mathrm{in}}u_v\frac{2ik_0}{\rho_0}P_\mathrm{in}	\tilde{p}d\Gamma
	~~~\forall \tilde{p} \in H^1(\Omega_a),\nonumber
\end{align}
with $\varphi = v,h$.
$\Omega_a=\Omega_\mathrm{NDD}\cup (D\setminus \overline{\Omega_r})$ represents air-filled regions, and $V_u$ is defined as 
$V_u = \left\{ \tilde{u_\varphi}\in H^1(\Omega_a) |~ \tilde{u_\varphi}=0~\mathrm{on~}\Gamma_\mathrm{wall}\cup \partial \Omega_r\right\}$.
Similarly, the weak forms in Case 2 are summarized as follows:
\begin{align}
	&\int_{\Omega_a}k_\varphi^{-2}\nabla u_\varphi \cdot \nabla \tilde{u_\varphi} d\Omega -\int_{\Omega_a}u_\varphi \tilde{u_\varphi} d\Omega = -\int_{\Omega_a}\tilde{u_\varphi} d\Omega ~~~\forall \tilde{u_\varphi} \in V_u,\nonumber\\
	&\int_{\Omega_a}\left(\frac{u_v}{\rho_0}\right)\nabla p \cdot \nabla \tilde{p}d \Omega - \int_{\Omega_a}\frac{\omega^2}{K_0}\{
	\gamma-(\gamma-1)u_h
	\}
	 p\tilde{p}d\Omega \nonumber\\
	&+ \int_{\Gamma_\mathrm{in} \cup \Gamma_\mathrm{out} }u_v\frac{i k_0}{\rho_0}p\tilde{p}d\Gamma
	= \int_{\Gamma_\mathrm{in}}u_v\frac{2ik_0}{\rho_0}P_\mathrm{in}	\tilde{p}d\Gamma
	~~~\forall \tilde{p} \in H^1(\Omega_a),\nonumber
\end{align}
with $\varphi = v,h$.

\subsection{Formulation of the optimization problem}\label{sec: formulation of the optimization problem}

As the viscous and thermal boundary layers play an important role, the sound absorption coefficient is maximized based on the above design settings. To estimate the absorption coefficient, we introduce S-parameters, defined as
\begin{align}
	S_{11} &=\frac{\int_{\Gamma_\mathrm{in}}(p-P^\mathrm{in})P^{\mathrm{in}\ast}  d\Gamma } {\int_{\Gamma_\mathrm{in}}P^\mathrm{in}P^{\mathrm{in}\ast}  d\Gamma},\\
	S_{21} &=\frac{\int_{\Gamma_\mathrm{out}} p P^{\mathrm{in}\ast}  d\Gamma } {\int_{\Gamma_\mathrm{in}}P^\mathrm{in}P^{\mathrm{in}\ast}  d\Gamma},
\end{align}
where $\ast$ represents the complex conjugate of the variable.
We note that $S_{11}$ is used in both Case 1 and 2, but $S_{21}$ is used only in Case 2.
Then, the absorption coefficient $\alpha$ can be estimated as follows:
\begin{equation}  \label{eq: cases alpha}
	\alpha=
	\left\{
	\begin{alignedat}{2}
		& 1-|S_{11}|^2    &\quad&    \text{in Case 1}  \\
		& 1-\left(|S_{11}|^2 + |S_{21}|^2    \right)         &     &    \text{in Case 2}
	\end{alignedat}
	\right.
\end{equation}
Since $\alpha$ depends on the angular frequency $\omega$, we maximize $\alpha$ over a certain frequency band, $\omega_{init} \le \omega \le \omega_{fin}$ with $0< \omega_i\le \omega_f$.
To do so, $\omega$ is discretized by a frequency step, $\Delta \omega = {(\omega_{fin}-\omega_{init})}/{n}$, and we set discretized angular frequency as $\omega_k=k\Delta \omega + \omega_{init}$ with $0\le k \le n$.

The objective of topology optimization in this research is to obtain the optimized $\Omega_r$ that maximizes the sound absorption coefficient over a certain frequency range. 
Therefore, we set the averaged value of $\alpha$ as an objective function $J$, and the optimization problem is formulated as follows:
\begin{align}
	\min_{\Omega_r}~&J = -\frac{1}{n+1}\sum_{k=0}^{n}\alpha(\omega_k) \nonumber\\
	\mathrm{subject~to~}&\mathrm{Governing~equations~in~}\Omega_a\text{~for~}\omega_k~(0\le k \le n).
	\label{eq: optimization problem}
\end{align}
\textcolor {black}{
This optimization problem is unstable if there is no constraint on the minimum thickness for the air-filled region in the optimized design because air-filled cavities are likely to be created, and their entrances tend to shrink to form the structure like the Helmholtz resonator during the optimization. They are sometimes closed when updating the design variables, worsening the value of the sound absorption coefficient. Thus, the optimization calculation will be unstable, and the result with a high value of the sound absorption coefficient could not be obtained. For simplicity, we placed the non-design domain $\Omega_\mathrm{NDD}$ as explained in the previous subsection. Due to the non-design domain and the symmetry condition applied on $\Gamma_\mathrm{sym}$, the optimized design contains at least three air-filled channels considering the mirror symmetry.
}

\textcolor {black}{
In their previous research, Andersen et al. \cite{andersen2019shape} conducted shape optimization to maximize sound absorption with a similar setting to Case 1.  We conducted topology optimization with the setting of \cite{andersen2019shape} in \ref{sec: Benchmark} to show the validity of the proposed method.
}

\subsection{Sensitivity analysis}
The design sensitivity is required to minimize the objective function. We conduct the sensitivity analysis based on the concept of the topological derivative \cite{Sokolowski19991251}. 
Generally, the topological derivative measures the change in the rate of an objective function when a circular infinitesimal inclusion domain, $\Omega_{\varepsilon}$, appears in the target system. In this research, the topological derivative is considered when a circular-shaped rigid domain, $\Omega_{\varepsilon}$, with the radius $\varepsilon$ is inserted into the viscothermal acoustic system. Let $(J+\delta J)$ denote the objective function when $\Omega_{\varepsilon}$ appears, whereas $J$ denotes the objective function without $\Omega_{\varepsilon}$. The topological derivative $D_T J$ is then defined as follows:
\begin{align}
	(J + \delta J) - J = g(\varepsilon)D_T J + o(g(\varepsilon)), \label{eq: Def of DtJ}
\end{align}
where $g(\varepsilon)$ is a function depending on the radius of $\Omega_\varepsilon$.
As discussed in~\ref{sec: Appendix: Derivation of DtJ}, $D_T J$ with $g(\varepsilon)=\pi \varepsilon^2$ is approximately obtained as follows:
\begin{align}
	D_T J &= \sum_{k=0}^n 2\mathrm{Re}\left[\frac{2}{\rho_0}\nabla p(\omega_k)\cdot \nabla q(\omega_k) - \frac{\omega^2}{K_0}p(\omega_k) q(\omega_k)
	\right]
	\label{eq: DtJ_mainpart}
\end{align}
where $q$ represents an adjoint acoustic pressure, and its definition is explained in \ref{sec: Appendix: Derivation of DtJ}.
\textcolor {black}{This formula is only an approximated topological derivative, and it is} the same as the topological derivative obtained without considering the boundary layer effects; however, we numerically confirmed its usability to estimate the variation in the objective function when a small rigid domain with a finite radius appeared (\ref{sec: numerical dJ}).
\textcolor {black}{According to the discussion in \ref{sec: numerical dJ}, a gap exists between the values of the approximated topological derivative and the finite difference when their magnitudes are large. However, their trends in the magnitude relationship are the same; thus, there seems to be no problem in their practical use for minimizing the objective function.
}

\section{Level set-based topology optimization}\label{sec: Level set-based topology optimization}
To solve the optimization problem in Eq.~(\ref{eq: optimization problem}), a level set-based topology optimization method proposed by Yamada et al. \cite{yamada2010topology} is introduced.
This method expresses the shape and topology of the structure by a level set function, and it is updated using a reaction--diffusion equation with the use of the topological derivative. 

We first define the level set function $\phi$. 
As explained, the design domain $D$ contains the rigid domain $\Omega_r$ and the air-filled region $D\setminus\overline{\Omega_r}$.
These domains and rigid surfaces, $\partial \Omega_r$, are represented by $\phi$ as follows:
\begin{eqnarray}
\left\{
\begin{array}{ll}
0<\phi(\bm{x})\le 1 &\mathrm{if}~~\bm{x}\in \Omega_r\\
\phi(\bm{x})= 0 &\mathrm{if}~~\bm{x}\in \partial \Omega_r\\
-1\le \phi(\bm{x})< 0 &\mathrm{if}~~\bm{x}\in D\setminus\overline{\Omega_r}.\label{eq:profile of LSF}
\end{array}
\right. 
\end{eqnarray}
Notably, this level set function is different from a signed distance function, which is often used in level set-based shape optimization \cite{allaire2004structural}.

The optimization problem for minimizing the objective function $J$ by optimizing the material distribution in the design domain $D$ can be formulated as follows:
\textcolor {black}{
\begin{align}
\inf_{\phi}~~~&J \label{eq: optimization problem_chiphi}
\end{align}
}
\textcolor {black}{We define the characteristic function $\chi_\phi$ represented by $\phi$ as follows:}
\begin{eqnarray}
\chi_\phi=
\left\{
\begin{array}{ll}
1 &\mathrm{if}~~\phi\ge 0 \\
0 &\mathrm{if}~~\phi< 0
\end{array}
\right. 
\end{eqnarray}
To find the distribution of $\phi$ that minimizes the objective function $J$, a fictitious time is introduced, and the optimization problem is replaced with a time-evolution problem. 
Let $t_s$ denote a fictitious time in the optimization.
If we assume that a partial derivative of the level set function with respect to time $\frac{\partial \phi}{\partial t_s}$ is proportional to  the design sensitivity $J'$, the time-evolution equation for $\phi$ is expressed as
\begin{align}
\frac{\partial \phi}{\partial t_s}=-K_\phi {J}',
\end{align}
where $K_\phi>0$ is a positive constant. 
To regularize the optimization problem in Eq.~(\ref{eq: optimization problem_chiphi}), a regularization term is introduced into the time-evolution equation, as follows:
\begin{align}
\frac{\partial \phi}{\partial t_s}=-K_\phi ({J}'-\tau \nabla^2 \phi),\label{eq:reaction-diffusion eq}
\end{align}
where $\tau >0$ is a regularization parameter that controls the regularization effect. 
The larger value of $\tau$ yields a smoother distribution of $\phi$. The value of $\tau$ should be chosen based on the size of the computational domain so that the minimization of $J$ is not disturbed by the regularization effect. 
Eq.~(\ref{eq:reaction-diffusion eq}) is a reaction--diffusion equation with the diffusion and reaction terms. The reaction term corresponds to the design sensitivity $J'$, and the diffusion term corresponds to the regularization term.

Instead of using Eq.~(\ref{eq:reaction-diffusion eq}), a nondimensional form of the reaction--diffusion equation is solved in $D$, equipped with boundary conditions for $\phi$ \textcolor {black}{on a part of the boundary of the design domain denoted by $\Gamma_{\phi n}~(\subset \partial D)$ and the other parts $\partial D \setminus \Gamma_{\phi n}$}. 
By setting an appropriate initial condition corresponding to the initial configuration of the optimization, the system for $\phi$ can be summarized as follows:
\begin{eqnarray}
\left\{
\begin{array}{ll}
\cfrac{\partial \phi}{\partial t_s}=-K_\phi (C_J {J}'-\tau L_\phi^2 \nabla^2 \phi) &\mathrm{in}~~D, \\
\bm{n}\cdot\nabla \phi =0 &\mathrm{on}~~\Gamma_{\phi n},\\
\phi =-1 &\mathrm{on}~~\partial D \setminus \Gamma_{\phi n},\\
\phi(\bm{x},t_s=0)=\phi_0(\bm{x}),
\label{eq:reaction-diffusion equation with B.C}
\end{array}
\right. 
\end{eqnarray}
where $C_J$ is a coefficient to normalize the design sensitivity and is the inverse of the averaged value of the design sensitivity in $D$, $C_J=\frac{|D|}{\int_{D}|J'|d\Omega}$.
$L_\phi$ represents the characteristic length, which normalizes the effect of the regularization term.
For simplicity, we imposed the Neumann boundary condition on $\Gamma_{\phi n}\subset \partial D$ and the Dirchlet boundary condition on the other parts of $\partial D$; however, other boundary conditions can also be applied. The fourth line shows the initial condition at which the initial level-set function, $\phi_0(\bm{x})$, represents the initial configuration.

The design sensitivity $J'$ is defined by the topological derivative, $D_T J$. 
According to the definition of $D_T J$ expressed in Eq.~(\ref{eq: Def of DtJ}) and the form of the reaction--diffusion equation with the profile of $\phi$ in Eq.~(\ref{eq:profile of LSF}), $J'$ can be written as follows:
\begin{align}
	J'=D_T J(1-\chi_\phi).
\end{align}

\section{Numerical implementation}\label{sec:Numerical implementation}

\subsection{FEM-based discretization for the state and adjoint problems}\label{sec:Remesh and rigid domain}
To obtain the state and adjoint variables, the governing and adjoint equations are discretized using FEM. 
The finite element program used in this research is implemented by FreeFEM \cite{MR3043640}, which is an open-source PDE solver.

The weak forms of state and adjoint problems are defined in the air-filled domain, $\Omega_a$; however, the reaction--diffusion equation is solved in the design domain, including the rigid domain, $\Omega_r$. To simplify the numerical implementation, we introduce a fictitious acoustic material in the rigid domain instead of eliminating the rigid region from the computational domain. For example, the weak forms of the state problem in Case 1 are modified as
\begin{align}
	&\int_{\Omega}k_\varphi^{-2}\nabla u_\varphi \cdot \nabla \tilde{u_\varphi} d\Omega -\int_{\Omega}u_\varphi \tilde{u_\varphi} d\Omega = -\int_{\Omega_a}\tilde{u_\varphi} d\Omega ~~~\forall \tilde{u_\varphi} \in \tilde{V_u},\nonumber\\
	&\int_{\Omega_a}\left(\frac{u_v}{\rho_0}\right)\nabla p \cdot \nabla \tilde{p}d \Omega - \int_{\Omega_a}\frac{\omega^2}{K_0}\{\gamma-(\gamma-1)u_h\} p\tilde{p}d\Omega \nonumber\\
	&+\int_{\Omega_r}\left(\frac{1}{\rho_r}\right)\nabla p \cdot \nabla \tilde{p}d \Omega - \int_{\Omega_r}\frac{\omega^2}{K_r} p\tilde{p}d\Omega \nonumber\\
	&+ \int_{\Gamma_\mathrm{in}}u_v\frac{i k_0}{\rho_0}p\tilde{p}d\Gamma
	= \int_{\Gamma_\mathrm{in}}u_v\frac{2ik_0}{\rho_0}P_\mathrm{in}	\tilde{p}d\Gamma
	~~~\forall \tilde{p} \in H^1(\Omega),\nonumber
\end{align}
where $\tilde{V_u} = \left\{ \tilde{u_\varphi}\in H^1(\Omega) |~ \tilde{u_\varphi}=0~\mathrm{on~}\Gamma_\mathrm{wall}\cup \partial \Omega_r\right\}$ is the space $V_u$ extended to the whole region, $\Omega$.
$(\rho_r,K_r)$ represents the mass density and bulk modulus of the fictitious acoustic material. 
According to \cite{yoon2020topology},
they should be sufficiently large compared to $(\rho_0, K_0)$. Thus, two coefficients are introduced: $c_{r1} \gg 1$ and $c_{r2}\gg 1$, defined as $(\rho_r,K_r)=(c_{r1}\rho_0,c_{r2}K_0)$. The settings of these values are explained in Section \ref{sec: numerical examples}. 
Similarly, the adjoint problems are solved. All variables in the state and adjoint systems based on the SLNS model are discretized by the quadratic and continuous finite element.

In the finite element analysis, body-fitted meshes, that is, the elements fitted to the surface of the rigid regions, are required to implement the Dirichlet boundary conditions for $u_h$ and $u_v$ on $\partial \Omega_r$. 
This can be numerically implemented by using a zero-level isosurface of the level set function. Furthermore, since the boundary layer effects appear in the air-filled region near the rigid surface $\partial \Omega_r$, it is necessary to discretize such a region with finer meshes. Thus, a function $u_e$ is introduced, which is an approximated solution of the Eikonal equation \cite{churbanov2019numerical} to obtain the signed distance function, defined as follows:
\begin{align}
	u_e = \alpha_e \log(v_e)(1-2\chi_\phi)
\end{align}
where $v_e$ is the solution of the following PDE:
\begin{align}
-\alpha_e^2\nabla^2 v_e + v_e &= 0~~~\mathrm{in~}\Omega,\nonumber\\
v_e &= 1~~~\mathrm{on~}\Gamma_\mathrm{wall}\cup \partial \Omega_r,\label{eq: PDE of approximiated eikonal}
\end{align}
where $\alpha_e$ is a positive constant. 
$u_e$ approaches to the signed distance function when $\alpha_e \to 0$ \cite{churbanov2019numerical}. 
\textcolor {black}{We set $\alpha_e$ to be almost twice the value of the element size in the design domain to obtain the smooth distribution of $v_e$.}
Equation~(\ref{eq: PDE of approximiated eikonal}) satisfies the maximum principle, and $0 < v_e(\bm{x}) < 1$ holds for $\bm{x}\in \Omega$. Therefore, the function $u_e$ is positive in rigid regions, while it is negative in air-filled regions, similar to the definition of $\phi$.
Based on the distribution of $u_e$, the region where $-\beta_e < u_e < 0$ with a positive constant $\beta_e$ is discretized by finer meshes, while the other regions are discretized by course meshes.
The maximum size of elements in the region where $-\beta_e < u_e < 0$ is set to $1/5$ of the viscous wavelength $\lambda_v$ at $\omega_{fin}$, defined with $\lambda_v=\frac{2\pi}{ |k_v(\omega_{fin})| }$.
We use an open-source mesher, Mmg \cite{dapogny2014three} for the implementation of this remeshing procedure.
\subsection{Fictitious time-directional sensitivity filtering}
The sound absorption coefficient $\alpha$ strongly depends on the configuration of the rigid region $\Omega_r$; therefore, the direct use of the design sensitivity $J'$ makes the optimization procedure unstable. 
To avoid this, we average the $J'$ along the direction of the fictitious time $t_s$, as follows:
\begin{align}
	\overline{J'}(t_s) = \alpha_t J'(t_s) + (1-\alpha_t)\overline{J'}(t_s-\Delta t_s),\label{eq: time-directional filtering}
\end{align}
where $0 < \alpha_t \le 1$ is a parameter to adjust the effect of averaging, and $\Delta  t_s$ represents a fictitious time step.
By using $\overline{J'}$ instead of $J'$ in the reaction--diffusion equation, the level set function is updated based on the current and past distributions of the design sensitivity, depending on the value of $\alpha_t$. 
By using the smaller value of $\alpha_t$, the level set function is updated based on the design sensitivity at earlier iterations, which stabilizes the optimization procedure. We note that setting $\alpha_t=1$ corresponds to the use of the original sensitivity $J'$ because $\overline{J'}=J'$ for all $t_s$.

\subsection{Optimization flowchart}
\begin{figure}[H]
	\centering
	\begin{tikzpicture}[font=\scriptsize]
	\small
	\tikzset{Terminal/.style={rounded rectangle,  draw,  text centered, text width=1.5cm, minimum height=0.8cm}};
	\tikzset{Process/.style={rectangle,  draw,  text centered, text width=5cm, minimum height=0.8cm}};
	\tikzset{ProcessSmall/.style={rectangle,  draw,  text centered, text width=1.5cm, minimum height=0.8cm}};
	\tikzset{Decision/.style={diamond,  draw,  text centered, aspect=3,text width=1.5cm, minimum height=1.0cm}};
	\tikzstyle{arrow} = [thick,->,>=stealth]
	\node[Terminal](a)at (0,0){Start};
	\node[Process, below=0.6 of a.center](b){Initialize variables};
	\node[Process, below=0.6 of b.center](c){Remesh based on $\phi=0$ and solve the approximated Eikonal equation};
	\node[Process, below=0.6 of c.center](d){Remesh based on $u_e$};
	\node[Process, below=0.6 of d.center](e){$k=0$};
	\node[Process, below=0.6 of e.center](f){Solve state problem at $\omega_k$};
	\node[Process, below=0.6 of f.center](g){Solve adjoint problem at $\omega_k$};
	\node[Decision, below=0.6 of g.center](h){$k=n$?};
	\node[ProcessSmall, left=of g,  xshift=-5 ](ha){$k=k+1$};
	\node[Process, below=0.8 of h.center](i){Calculate $J$};
	\node[Decision, below=0.6 of i.center](j){Converged?};
	\node[Process, below=0.8 of j.center](k){Compute $J'$ and update $\overline{J'}$};
	\node[Terminal, right=of j,  xshift=5 ](ja){End};
	\node[Process, below=0.8 of k.center](l){Solve RDE to update $\phi$};
	\draw[-, thick]  (a) --(b);
	\draw[-, thick]  (b) --(c);
	\draw[-, thick]  (c) --(d);
	\draw[-, thick]  (d) --(e);
	\draw[-, thick]  (e) --(f);
	\draw[-, thick]  (f) --(g);
	\draw[-, thick]  (g) --(h);
	\draw[->, thick]  (h)node[below, xshift=-100]{No}  -| (ha);
	\draw[->,thick]  (ha) |-(f);
	\draw[->, thick]  (h)node[below=-0.2 of h, xshift=-20]{Yes}  -- (i);
	\draw[-, thick]  (i) --(j);
	\draw[->, thick]  (j)node[below=-0.2 of j, xshift=-20]{No}  -- (k);
	\draw[->, thick]  (j)node[right=of j, xshift = -25, yshift=5]{Yes}  -- (ja);
	\draw[-, thick]  (k) --(l);
	\draw[->, thick]  (l.west) -- ++(-3.5,0)  |- (c);
	\end{tikzpicture}
	\caption{Optimization flowchart}
	\label{fig:Opt flowchart}       
\end{figure}
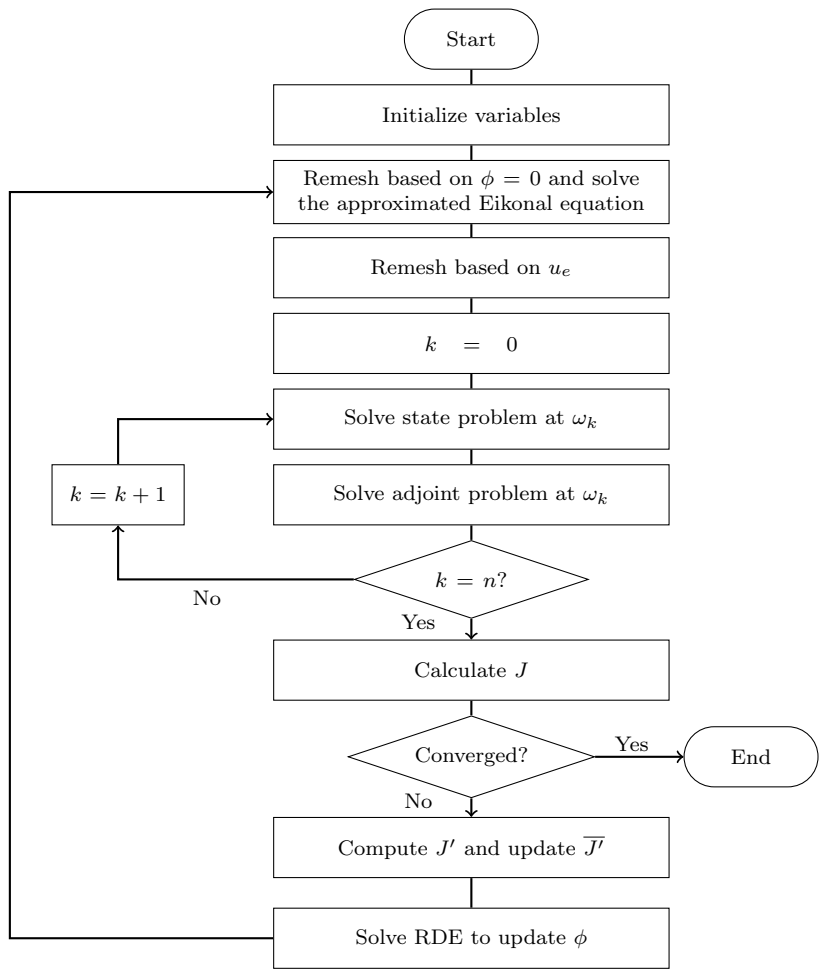

Figure~\ref{fig:Opt flowchart} presents the optimization flowchart. 
First, the computational domain is discretized using finite elements, and variables used in the optimization are initialized, including the level set function \textcolor {black}{$\phi$.} 
\textcolor {black}{We set the initial mesh as a background mesh on which the nodal value of $\phi$ is defined with P1 elements.}
Then, we proceed to the remeshing procedure using Mmg. 
By refining the meshes using the zero-level isosurface of $\phi$, the approximated signed distance function $u_e$ is obtained by solving the system expressed in Eq.~(\ref{eq: PDE of approximiated eikonal}). 
Then, remeshing based on $u_e$ brings us the discretization of finite elements, considering the boundary layers. 
The state and adjoint problems are solved for each $\omega_k$ with $0\le k\le n$, and the objective function $J$ is calculated. 
We introduce the 10-iteration moving average of the relative error between the values of $J$ for two consecutive iterations. 
If this value gets sufficiently small after several optimization iterations, the optimization calculation is terminated. 
Otherwise, the design sensitivity $J'$ is computed using the solution of state and adjoint problems. 
At this process, fictitious time-directional sensitivity filtering is applied
based on Eq.~(\ref{eq: time-directional filtering}), and the modified sensitivity $\overline{J'}$ is computed. 
\textcolor {black}{$\phi$ defined in the background mesh is mapped to the refined mesh and updated by solving the reaction--diffusion equation in Eq.~(\ref{eq:reaction-diffusion equation with B.C}) based on the distribution of the design sensitivity.}
\textcolor {black}{Subsequently}, we return to the remeshing procedure. These processes are repeated until the value of $J$ converges.

\section{Numerical examples}\label{sec: numerical examples}
In this section, two-dimensional numerical examples corresponding to Case 1 and Case 2 are provided to demonstrate the validity of the proposed optimization method.

First, parameters used in both cases are explained.
The dimensions of geometries shown in Figure~\ref{fig:Design settings} are set as $D_\mathrm{ex}=0.03$[m], $L_\mathrm{NDD}=0.06$[m], $L_D=0.06$[m], and $t_w=1.5\times10^{-3}$[m] in both cases.
The angular frequencies of incident waves are set as $\omega_{init}=2\pi \times 3000\mathrm{[rad~s^{-1}]}$ and  $\omega_{fin}=2\pi \times 6000\mathrm{[rad~s^{-1}]}$ with $n=20$ so that the absorption coefficients from 3000[Hz] to 6000[Hz] are targeted.
A set of parameters of air is obtained from \cite{kampinga2010viscothermal}, as follows:
The sound speed and quiescent mass density of air are set to $c_0=341.2\mathrm{[m~s^{-1}]}$ and $\rho_0=1.225\mathrm{[kg~m^{-3}]}$; then, the bulk modulus of air can be estimated as $K_0 = \rho_0 c_0^2$.
The heat conduction coefficient of air is $\kappa=25.18 \times 10^{-3}\mathrm{[W~m^{-1}K^{-1}]}$.
The quiescent temperature and pressure are $T_0=294.3$[K] and $p_0=1.015\times 10^5$[Pa], respectively.
Dynamic viscosity is set as $\mu = 18.29\times 10^{-6}[\mathrm{Pa~s}]$, whereas the second viscosity is set as $\lambda =-1.22\times 10^{-6}[\mathrm{Pa~s}]$.
The specific heat at constant volume and that at constant pressure are given as $C_v=693.8\mathrm{[J~kg^{-1}K^{-1}]}$ and $C_p=975.3\mathrm{[J~kg^{-1}K^{-1}]}$, respectively. Thus, the ratio of specific heats is obtained as $\gamma=1.406$.
The coefficients to define the material parameters corresponding to the rigid medium are set as $c_{r1}=1.0\times 10^{13}$ and $c_{r2}=1.0\times 10^{3}$.
The parameter $\beta_e$ to remesh the boundary layer using the approximated signed distance function $u_e$ is set to $\beta_e=0.0003$[m], which is roughly 6 times the thermal boundary layer thickness $\delta_h$ at 3000[Hz], defined as $\delta_h = \frac{-1}{\mathrm{Im}(k_h)}$.
The regularization parameter in the reaction--diffusion equation is set to $\tau=5\times 10^{-5}$ with the characteristic length $L_\phi = D_\mathrm{ex}$, whereas the parameter used in the fictitious time-directional sensitivity filtering is set to $\alpha_t=0.01$.

\subsection{Optimization result for Case 1}\label{sec: Case1}
\begin{figure}[H]
	\centering
	\includegraphics[scale=0.5]{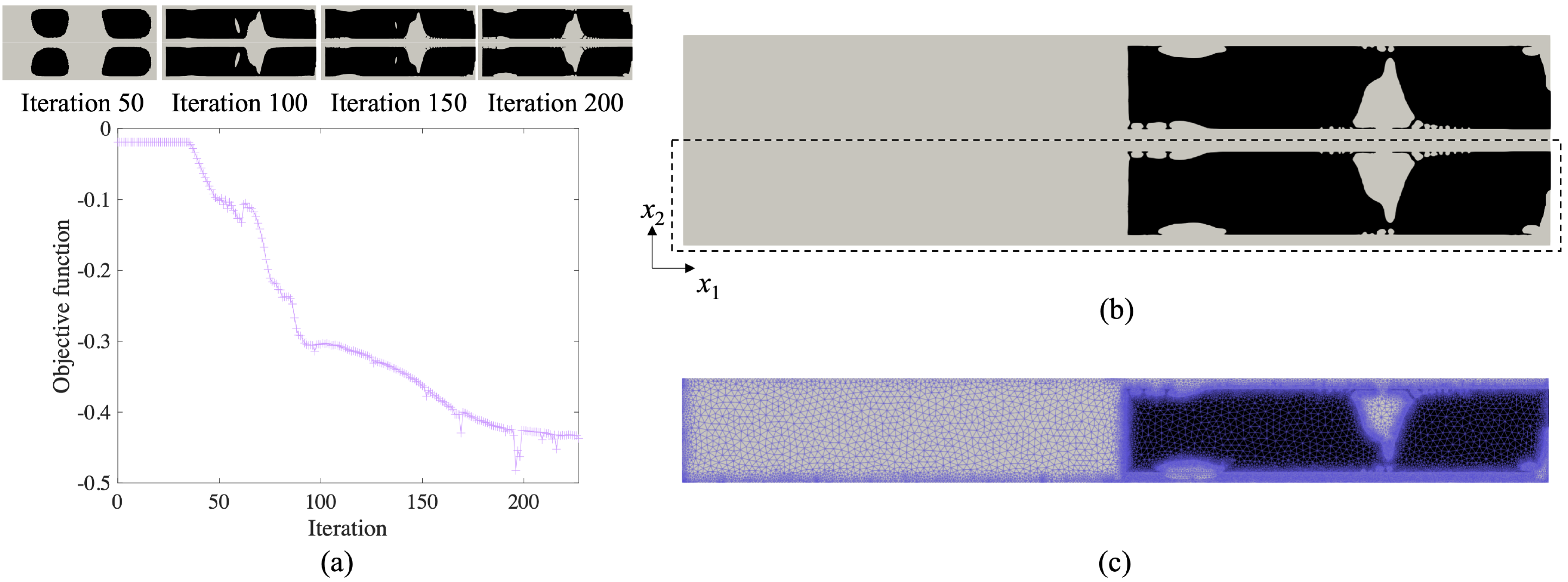}
	\caption{\textcolor {black}{Optimization results for Case 1. 
		(a) History of the objective function and intermediate designs.
		(b) Optimized design. The dotted box represents the computational domain. (c) Finite element discretization around $D$. 
		}	
	}
	\label{fig:Ex1_Optconfig}       
\end{figure}
In Case 1, we started the optimization calculation with $\phi_0(\bm{x})=-1$, that is, the design domain $D$ is entirely filled with air.
We set $\Gamma_{\phi n}$ in Eq.~(\ref{eq:reaction-diffusion equation with B.C}) as the right boundary of the design domain, $\Gamma_{\phi n}=\partial D \cap \Gamma_\mathrm{ref}$, on which the Neumann boundary condition for $\phi$ is applied. 
The value of $J$ at the initial iteration was $-1.90\times 10^{-2}$, whose small absolute value implies that the entire incident wave at the target frequency range is reflected at $\Gamma_\mathrm{ref}$ according to the definition of $J$ in Eq.~(\ref{eq: optimization problem}).

Figure~\ref{fig:Ex1_Optconfig} shows the optimization results for Case 1. 
\textcolor {black}{Figure~\ref{fig:Ex1_Optconfig}(a) represents the history of the objective function and the intermediate designs.}
The black region represents the rigid domain $\Omega_r$, while the gray region represents the air-filled domain, $\Omega_a$. 
\textcolor {black}{The smooth history of the objective function in decreasing trend was observed, indicating the usability of the approximated topological derivative in Eq.~(\ref{eq: DtJ_mainpart}).}
We finally halted the optimization calculation at the \textcolor {black}{227th} 
iteration at which the 10-iteration moving average of the relative error between the values of $J$ for two consecutive iterations was \textcolor {black}{$2.18 \times 10^{-3}$.} 
Figure~\ref{fig:Ex1_Optconfig}\textcolor {black}{(b)} represents the optimized design, where a mirrored image is shown corresponding to the symmetry condition on $\Gamma_\mathrm{sym}$, and the dotted box represents the computational domain.
Figure~\ref{fig:Ex1_Optconfig}\textcolor {black}{(c)} shows the finite element discretization around the design domain. 
The meshes are fitted to the rigid surface, and the boundary layer region is discretized by the finer meshes based on the method explained in Section~\ref{sec:Numerical implementation}. 
The optimized design is characterized by air-filled cavities connected with the non-design domain around $\Gamma_\mathrm{wall}$ and $\Gamma_\mathrm{sym}$ and forms two types of narrow channels. Hereinafter, we call the channel touching on $\Gamma_\mathrm{wall}$ as an external channel and that including $\Gamma_\mathrm{sym}$ as a central channel. In addition, the oscillating boundary of the rigid domain with a high curvature can be confirmed, for example, at the entrances of channels and cavities. 

\begin{figure}[H]
	\centering
	\includegraphics[scale=0.5]{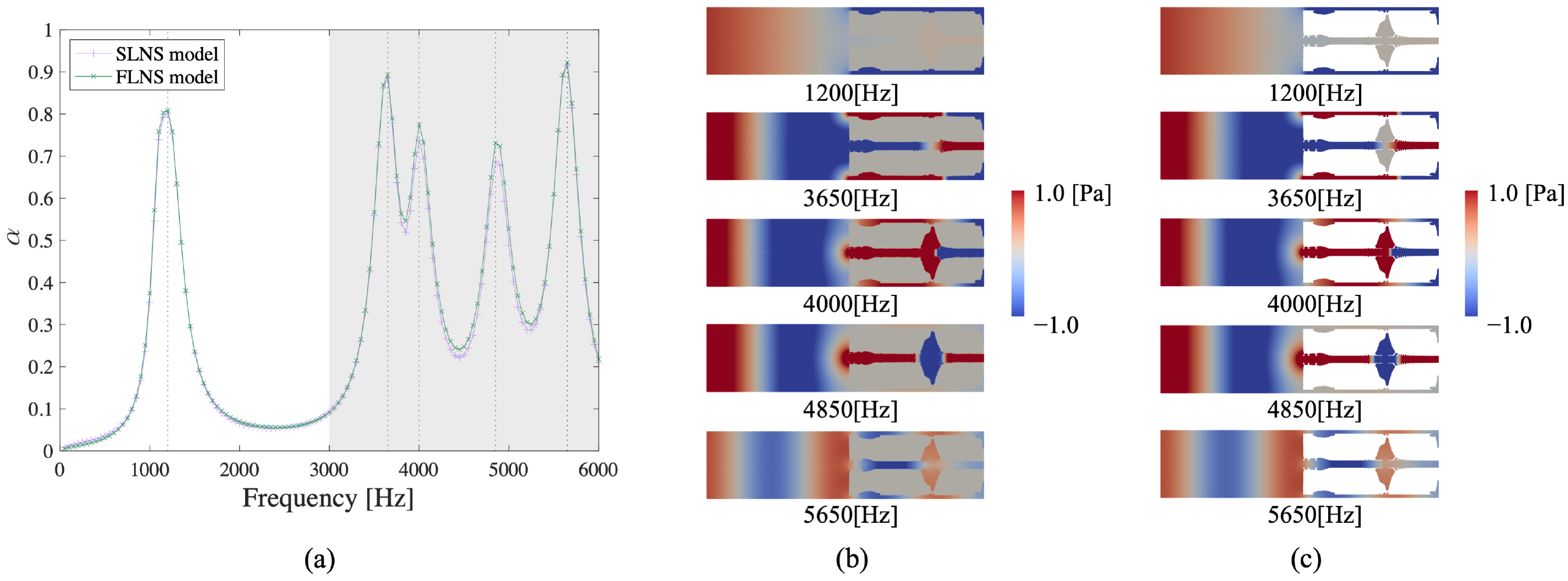}
	\caption{
		\textcolor {black}{
		(a) Frequency response of $\alpha$. (b) Distribution of $\mathrm{Re}(p)$ obtained by the SLNS model. (c) Distribution of $\mathrm{Re}(p_f)$ obtained by the FLNS model. 
		The contour diagrams of (b) and (c) are mirrored by considering the symmetry boundary condition on $\Gamma_\mathrm{sym}$.}	}
	\label{fig:Ex1_freqresponse}       
\end{figure}

The value of $J$ of the optimized design was \textcolor {black}{$-0.437$}, 
that is, the averaged value of the sound absorption coefficient $\alpha$ over $2\pi \times3000 \le \omega \le 2\pi\times 6000$ was
\textcolor {black}{$0.437$.}
Figure~\ref{fig:Ex1_freqresponse}(a) shows the frequency response of $\alpha$ examined by the SLNS and FLNS models for optimized design. This frequency response is examined over $50\mathrm{[Hz]} \le \frac{\omega}{2\pi} \le 6000\mathrm{[Hz]}$, and the value of $\alpha$ is evaluated per 50[Hz]. 
Hereafter, the frequency responses of some quantities are examined in the same manner. The purple and green lines represent results obtained by the SLNS and FLNS models, respectively. A good congruence between them justifies the use of the SLNS model.
\textcolor {black}{The computational time for obtaining the frequency response using the SLNS model was 1154~$\mathrm{[s]}$, whereas that using the FLNS model was 2788~$\mathrm{[s]}$, which were determined by using a single workstation processor (Intel Xeon W CPU 2.5GHz, 28cores, 384GB memory) and showed the efficiency of the SLNS model.}
 In these plots, 5 peaks with high $\alpha$ are obtained, and 4 of them are included in the target frequency range, emphasized by the gray area. 
Figure~\ref{fig:Ex1_freqresponse}(b) shows the distribution of $\mathrm{Re}(p)$ obtained by the SLNS model at these frequencies, whereas Figure~\ref{fig:Ex1_freqresponse}(c) shows that of $\mathrm{Re}(p_f)$ obtained by the FLNS model. 
To observe the absorption performance of the optimized design, the color range in the contour diagrams is limited from $-1$ to $1$[Pa]. 
A similar acoustic pressure distribution observed at each frequency in these figures validates the SLNS model and the representation way for the rigid region explained in Section~\ref{sec:Remesh and rigid domain}. The reflected waves seem to be reduced particularly at 5650 [Hz], where $\alpha$ takes the maximum value (\textcolor {black}{$\alpha=0.92$} 
 estimated by the SLNS model) within the target frequency range.

\begin{figure}[H]
	\centering
	\includegraphics[scale=0.5]{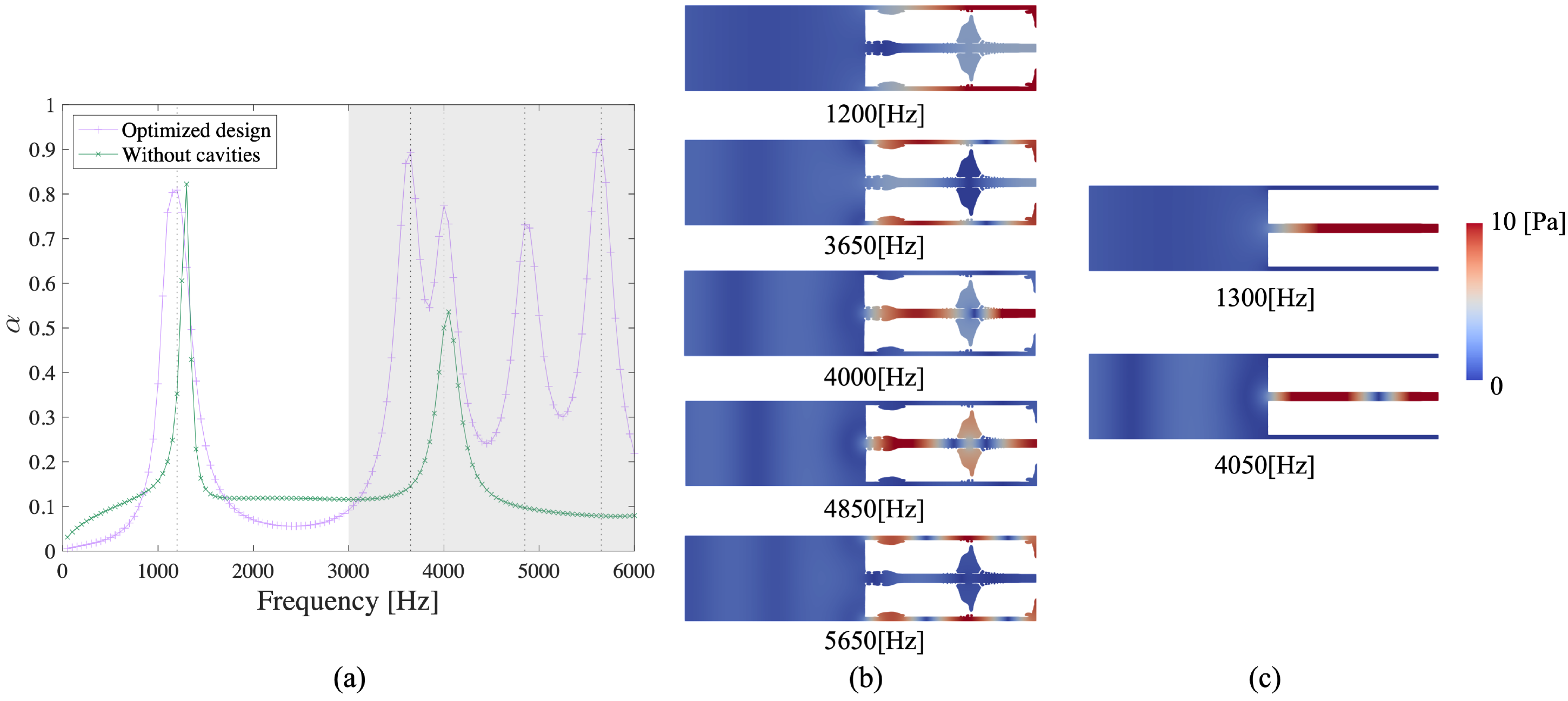}
	\caption{
		\textcolor {black}{
		 (a) Frequency responses of $\alpha$ for the optimized design and the structure without cavities. 
		 (b) Distribution of the absolute value of acoustic pressure $|p_f|$ for the optimized design. 
		 (c) Distribution of the absolute value of acoustic pressure $|p_f|$ for the structure without cavities. 
		 The contour diagrams of (b) and (c) are mirrored by considering the symmetry boundary condition on $\Gamma_\mathrm{sym}$.
		 }
	}
	\label{fig:Ex1_pressure}       
\end{figure}

Next, we discuss the contribution of the optimized structure to the absorption coefficient $\alpha$ by using the results obtained from the FLNS model. To clarify the functions of the optimized design, we compare the frequency response of $\alpha$ and the distribution of acoustic pressure $|p_f|$ in the optimized design with those in the structure without cavities, that is, all of the design domain is filled with the rigid medium.

Figure~\ref{fig:Ex1_pressure}(a) shows the comparison of frequency responses of $\alpha$ for them, and (b) and (c) show the corresponding distribution of $|p_f|$ at frequencies where $\alpha$ is large. 
As shown in the green plot in Figure~\ref{fig:Ex1_pressure}(a), only 2 peaks appear in the structure without cavities, whereas 5 peaks appear in the optimized structure, indicated by the purple plot. 
These 2 peaks are due to the resonances of the closed tube, which are induced in the central channel, as shown in Figure~\ref{fig:Ex1_pressure}(c). 
In the optimized design, the resonance modes similar to them appeared at 1200 [Hz] and 
\textcolor {black}{4000} [Hz]. 
Similar to the case without cavities, the large value of $|p_f|$ is confirmed at the central channel at 
\textcolor {black}{4000} [Hz]; 
however, that is confirmed at the external channel at 1200 [Hz], which brings a slight difference in the position of the first peak of $\alpha$ between these structures. 
The other 3 peaks in the optimized design cannot be seen in the case without cavities. 
As Figure~\ref{fig:Ex1_pressure}(b) shows, these are due to the modes in which one of the channels and the connected cavities resonate together. 
Therefore, the role of cavities in the optimized structure is to induce the resonances in either of two channels depending on the frequency, which resulted in multiple peaks in the frequency response of $\alpha$.

\subsection{Optimization result for Case 2}\label{sec: Case2}
\begin{figure}[H]
	\centering
	\includegraphics[scale=0.5]{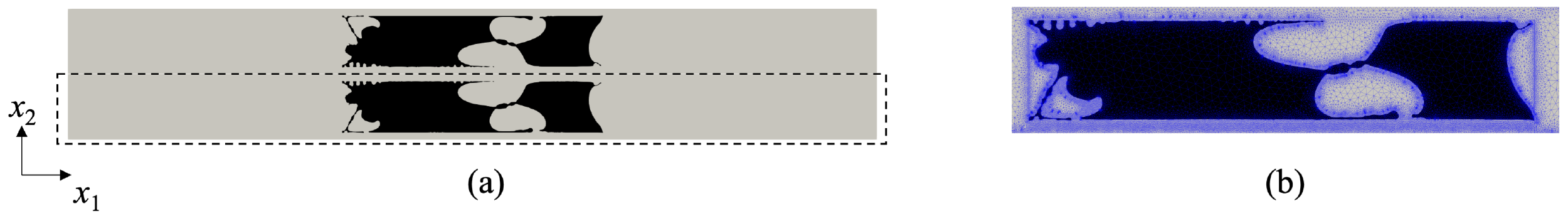}
	\caption{Optimization results for Case 2. 
		(a) Optimized design. The dotted box represents the computational domain. (b) Finite element discretization around $D$. }
	\label{fig:Ex2_Optconfig}       
\end{figure}
In Case 2, we start the optimization calculation with $\phi_0(\bm{x})=1$, that is, the design domain $D$ is filled with the rigid medium.
The boundary $\Gamma_{\phi n}$ is not used in this example, that is, the Dirchlet boundary condition in Eq.~(\ref{eq:reaction-diffusion equation with B.C}) was applied to $\partial D$. 
The value of $J$ at the initial iteration was $-8.61\times 10^{-2}$. 
Figure~\ref{fig:Ex2_Optconfig} shows the optimization results for Case 2. 
The optimization calculation was stopped at 360-th iteration at which the 10-iteration moving average of the relative error between the values of $J$ for two consecutive iterations was $1.81 \times 10^{-2}$. 
Figure~\ref{fig:Ex2_Optconfig}(a) represents the optimized design, where the dotted box represents the computational domain. 
Figure~\ref{fig:Ex2_Optconfig}(b) shows the finite element discretization around the design domain. As shown in Case 1, the optimized design is characterized by the central and external channels connected with air-filled cavities. The oscillating boundary of the rigid domain was also observed in this case, especially at the entrances of the central channel.

\begin{figure}[H]
	\centering
	\includegraphics[scale=0.5]{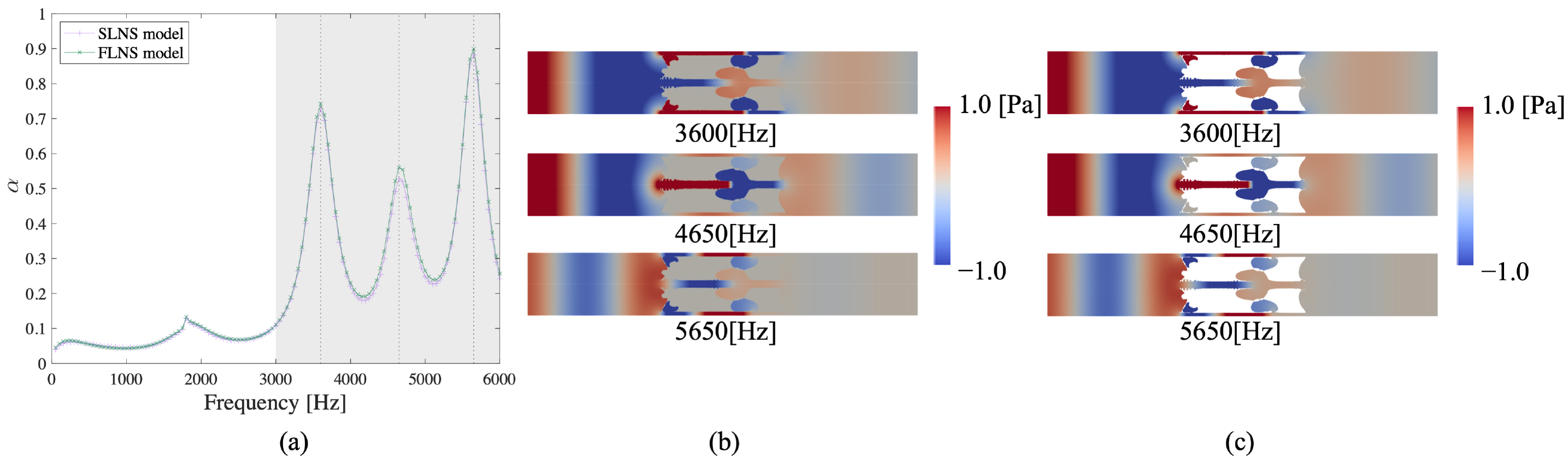}
	\caption{(a) Frequency response of $\alpha$. (b) Distribution of $\mathrm{Re}(p)$ obtained by the SLNS model. (c) Distribution of $\mathrm{Re}(p_f)$ obtained by the FLNS model. 
	The contour diagrams of (b) and (c) are mirrored by considering the symmetry boundary condition on $\Gamma_\mathrm{sym}$.	
}
	\label{fig:Ex2_freqresponse}       
\end{figure}

The value of $J$ of the optimized design was $-0.374$, that is, the average value of the sound absorption coefficient $\alpha$ over the target frequency range was $0.374$. 
Figure~\ref{fig:Ex2_freqresponse}(a) shows the frequency response of $\alpha$ examined by the SLNS and FLNS models for the optimized design. 
The purple and green lines represent results obtained by the SLNS model and FLNS model, respectively. A good congruence between them is confirmed in this case too, which justifies the use of the SLNS model. 
3 peaks can be observed in the target frequency range. 
Figure~\ref{fig:Ex2_freqresponse}(b) shows the distribution of $\mathrm{Re}(p)$ obtained by the SLNS model at these frequencies, whereas Figure~\ref{fig:Ex2_freqresponse}(c) shows that of $\mathrm{Re}(p_f)$ obtained by the FLNS model. 
We can confirm the sound-absorbing behavior from these figures, especially at 5650 [Hz], where $\alpha$ takes the maximum value ($\alpha=0.88$ estimated by the SLNS model) within the target frequency range. 
A similar distribution of acoustic pressure was observed at these frequencies, which also supports the use of the SLNS model.

\begin{figure}[H]
	\centering
	\includegraphics[scale=0.5]{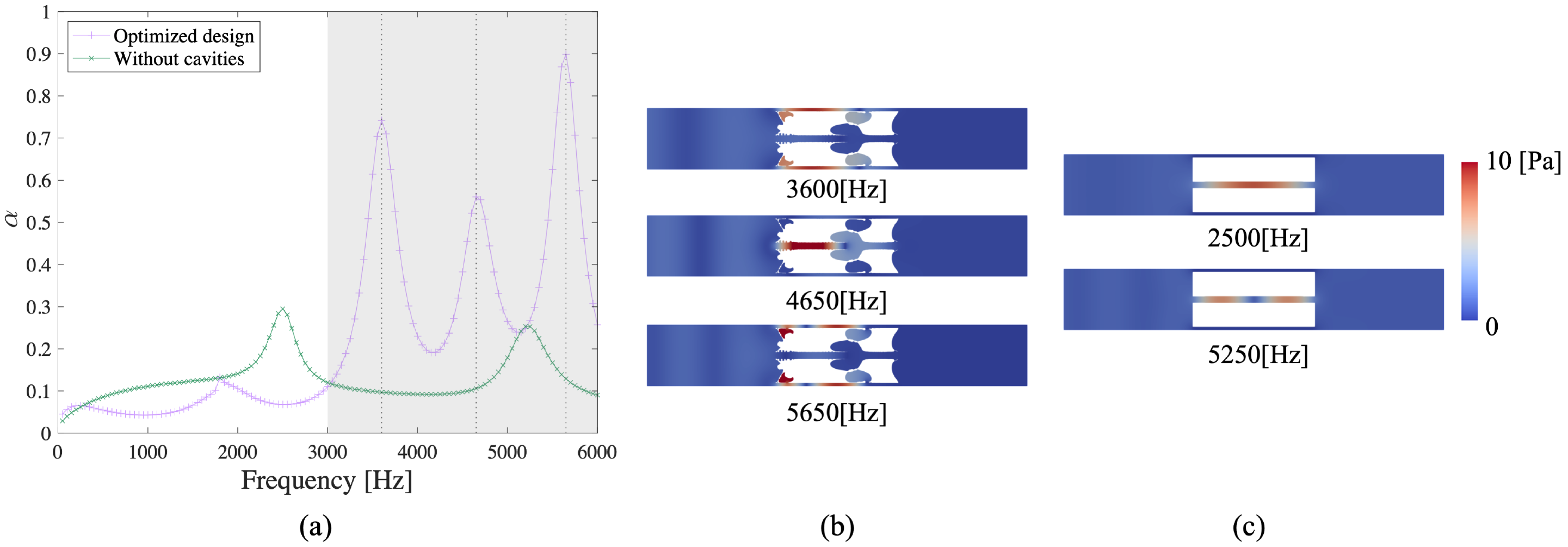}
	\caption{
		 (a) Frequency responses of $\alpha$ for the optimized design and the structure without cavities. 
		 (b) Distribution of the absolute value of acoustic pressure $|p_f|$ for the optimized design. 
		 (c) Distribution of the absolute value of acoustic pressure $|p_f|$ for the structure without cavities. The contour diagrams of (b) and (c) are mirrored by considering the symmetry boundary condition on $\Gamma_\mathrm{sym}$.
	}
	\label{fig:Ex2_pressure}       
\end{figure}
Figure~\ref{fig:Ex2_pressure}(a) shows the frequency responses of $\alpha$ for the optimized design and the structure without cavities that corresponds to the initial configuration, whereas (b) and (c) show the distribution of $|p_f|$ obtained by the FLNS model at frequencies where $\alpha$ takes large values for the structures. 
The frequency curve of the optimized structure is different from that for the initial configuration. 
The two peaks in the plot for the initial configuration are due to the resonances of the open tube, which are induced at the central tube, as shown in Figure~\ref{fig:Ex2_pressure}(c). 
Further, such resonance modes cannot be confirmed in the optimized structure at 3600, 4650, and 5650 [Hz]. In the optimized structure, resonance modes are induced in which one of the two channels (the external or central channels) and its connected cavities form a single resonator, as shown in Figure~\ref{fig:Ex2_pressure}(b). 
The external channel is activated at 3600 [Hz] and 5650 [Hz], where the higher mode of 3600 [Hz] is observed at 5650 [Hz], while the central channel is activated at 4650 [Hz]. These modes confine their acoustic energy in the channels and cavities, leading to a large $\alpha$.

\subsection{Thermal and viscous dissipation in the optimized designs}
\begin{figure}[H]
	\centering
	\includegraphics[scale=0.5]{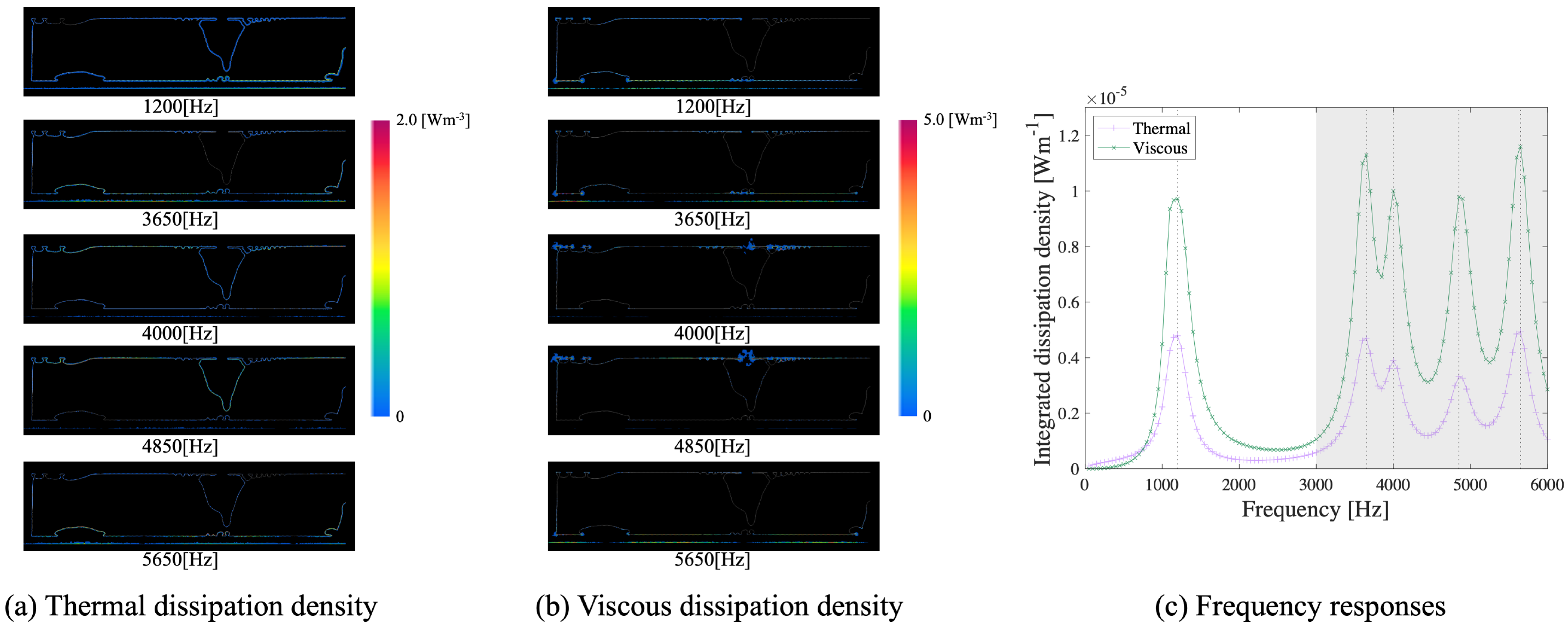}
	\caption{
		\textcolor {black}{
		Thermal and viscous dissipation effects in the optimized design for Case 1.
		(a) Distribution of thermal dissipation density $\Phi_h$ around $D$. (b) Distribution of viscous dissipation density $\Phi_v$ around $D$. (c) Frequency responses of the integrated dissipation densities, $\int_{\Omega_a}\Phi_{h}  d\Omega$ and $\int_{\Omega_a}\Phi_{v}  d\Omega$. 
		The area with a small value of dissipation densities is not colored in (a) and (b).
		}
	}
	\label{fig:Ex1_Loss}       
\end{figure}
\begin{figure}[H]
	\centering
	\includegraphics[scale=0.5]{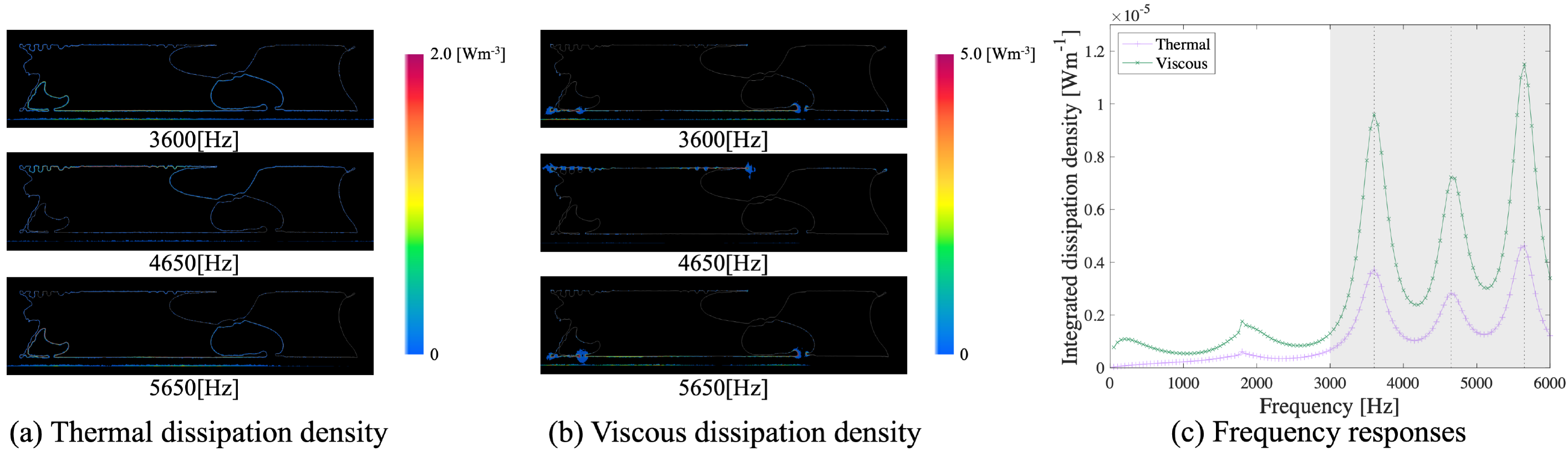}
	\caption{
		\textcolor {black}{
		Thermal and viscous dissipation effects in the optimized design for Case 2.
		(a) Distribution of thermal dissipation density $\Phi_h$ around $D$. (b) Distribution of viscous dissipation density $\Phi_v$ around $D$. (c) Frequency responses of the integrated dissipation densities, $\int_{\Omega_a}\Phi_{h}  d\Omega$ and $\int_{\Omega_a}\Phi_{v}  d\Omega$. 
		The area with a small value of dissipation densities is not colored in (a) and (b).		
	}
	}
	\label{fig:Ex2_Loss}       
\end{figure}
At the resonance frequencies, dissipation effects due to the viscous and thermal boundary layers are present.
To observe them, we introduce the viscous and thermal dissipation densities,  $\Phi_h=\frac{\kappa}{2T_0}|\nabla T|^2$ and $\Phi_v=\frac{1}{2}\mathrm{Re}[ \{\lambda(\nabla\cdot \bm{v})\bm{I}+2\mu\varepsilon(\bm{v}) \}^\ast : \nabla \bm{v}  ]$ \cite{pierce2019acoustics,comsol}.
Note that we took time average to these quantities since the time-harmonic oscillations were targeted.

Figure~\ref{fig:Ex1_Loss} shows the thermal and viscous dissipation effects in the optimized design of Case 1, 
whereas Figure~\ref{fig:Ex2_Loss} shows those of Case 2. 
Figure ~\ref{fig:Ex1_Loss}(a) and (b) represent the spatial distributions of $\Phi_{h}$ and $\Phi_v$ for Case 1, respectively. 
\textcolor {black}{The area with a small value of dissipation densities is not colored.}
Both $\Phi_{h}$ and $\Phi_v$ tend to take larger values in the vicinity of the rigid surfaces; however, their distributions slightly differ. As shown in (a), the thermal dissipation effect strongly appears at a place with large $|p_f|$, for example, on the surface of the cavities. 
Moreover, the viscous dissipation effect mainly appears at a place where the curvature on the rigid surface $\partial \Omega_r$ is large, such as the entrance of cavities. 
Figure~\ref{fig:Ex1_Loss}(c) depicts the frequency response of the integrated dissipation densities in the air-filled domain, $\int_{\Omega_a}\Phi_{h}  d\Omega$ and $\int_{\Omega_a}\Phi_{v}  d\Omega$. 
Within the target frequency range, the viscous dissipation is superior to the thermal one. 
The same trends in the spatial distribution of $\Phi_{h}$ and $\Phi_v$ and the frequency responses of $\int_{\Omega_a}\Phi_{h}  d\Omega$ and $\int_{\Omega_a}\Phi_{v}  d\Omega$ were found in Case 2, as shown in Figure~\ref{fig:Ex2_Loss}(a)--(c).
The optimized designs in both Case 1 and Case 2 contain highly oscillating structural configurations, represented by the structure at the entrance of the central channels. 
They increase the length of boundaries where the viscous and thermal boundary layer effects are present. 
Furthermore, the high curvature surfaces in the optimized design are important to enhance sound absorption via the viscous layer effect. 
The emergence of these structural designs by the proposed optimization method based on the SLNS model indicates that it can perform the optimization calculation by taking into account the influence of the boundary layer.

\textcolor {black}{
In addition to the abovementioned examples, we show the importance of considering the boundary layer effects by conducting topology optimization based on the proposed method and the method ignoring the boundary layers. 
In \ref{sec: Benchmark}, the sound absorption maximization problem is again targeted with a setting based on the previous research \cite{andersen2019shape}.
In \ref{sec: negative refraction}, a metamaterial slab is optimized to realize the negative refraction done in \cite{christiansen2016designing} without including the boundary layer effects.
}

\section{Conclusion}\label{sec: Conclusion}
In this paper, we proposed a topology optimization method for acoustic structures considering the viscous and thermal boundary layers based on the SLNS model. The main results can be summarized as follows:
	
	\begin{enumerate}
		\item We introduced the SLNS model derived from the FLNS model, which was proposed in a previous study \cite{kampinga2010viscothermal}.		
			
		\item An optimization problem is formulated based on the framework of a level set-based topology optimization method. We targeted a sound-absorbing structure, whose absorption coefficient can be estimated by the SLNS model. To achieve a high absorption performance at a certain frequency band, an objective function was formulated as the average of the sound absorption coefficient over that frequency range.
		
		\item A sensitivity analysis was performed based on the adjoint variable method and the concept of the topological derivative. We approximately derived the explicit formula of the topological derivative and discussed its practicality to estimate the change rate of the objective function, as shown in \ref{sec: Appendix: Derivation of DtJ}.
				
		\item We proposed an optimization algorithm for the level set-based topology optimization in which the elements used in finite element analysis were refined based on the distribution of the level set function and the approximated signed distance function for the analysis of viscothermal acoustic systems. Further, we elaborated on the numerical implementations, especially for the representation method of the rigid structure and the filtering method for the design sensitivity.
				
		\item Two-dimensional numerical examples were provided to demonstrate the validity of the proposed optimization method. 
		Two cases of optimization examples, whose design settings corresponded to closed and open tubes, respectively, were examined. 
		In both cases, the optimized structure had \textcolor {black}{narrow channels} and air-filled cavities connected with them. 
		From the acoustic pressure distribution analyzed, we deduced that cavities in the optimized designs, which induced multiple resonance modes within the target frequency range, are important for improving the sound absorption performance. In addition to the abovementioned structural features, a strongly oscillating configuration was seen in the obtained designs. 
		The spatial distribution of the dissipation densities and their frequency responses revealed that such a configuration in the optimized designs can enhance thermal and viscous dissipation in the boundary layer.
	\end{enumerate}

\textcolor {black}{
In summary, the proposed method is characterized to be applicable to acoustic devices with any geometry working for the audible frequency due to the SLNS model. Its computational cost is efficient, thanks to the adjoint variable method shown in \ref{sec: Benchmark}, where the sound absorption maximization problem based on the setting in the previous work \cite{andersen2019shape} is addressed. However, it sometimes reaches the local optimum solution due to the large degree of design freedom. For example, the creation of new channels during sound absorption maximization seems difficult in contrast to the optimization examples based on the LRF model \cite{christensen2017topology,van2000noise}. One of the solutions to avoid this problem is to introduce geometrical constraints in the optimization problem, such as a constraint function for the minimum length scale of air-filled regions \cite{guest2004achieving}.
}

\textcolor {black}{As shown in the example in \ref{sec: negative refraction}, the proposed method} can also be applied to the design of acoustic metamaterials working for audible sounds. 
If one focus on the design of their microstructure, the optimization can be conducted in the unit cell with applying the periodic boundary conditions. 
Although we showed two-dimensional optimization examples, three-dimensional problems can be treated as well since the SLNS model is still applicable. A considerable modification is required when structural vibrations are considered. In this case, the rigid structures that we optimized were replaced with deformable structures, and their vibrations were coupled with acoustic waves propagating in the air on their interfaces. In a previous study \cite{kampinga2010viscothermal}, such an acoustic-elastic coupled system has been analyzed based on the coupling of the equations used in the SLNS model and the elastodynamic equation. Based on the proposed optimization method, we would like to tackle this issue in our future works.

\section*{Acknowledgment}
Funding: This work was supported by Kawai Foundation for Sound Technology \& Music.

\appendix

\section{Sound absorption maximization problem based on the setting in \cite{andersen2019shape}}\label{sec: Benchmark}
\begin{figure}[H]
	\centering
	\includegraphics[scale=0.6]{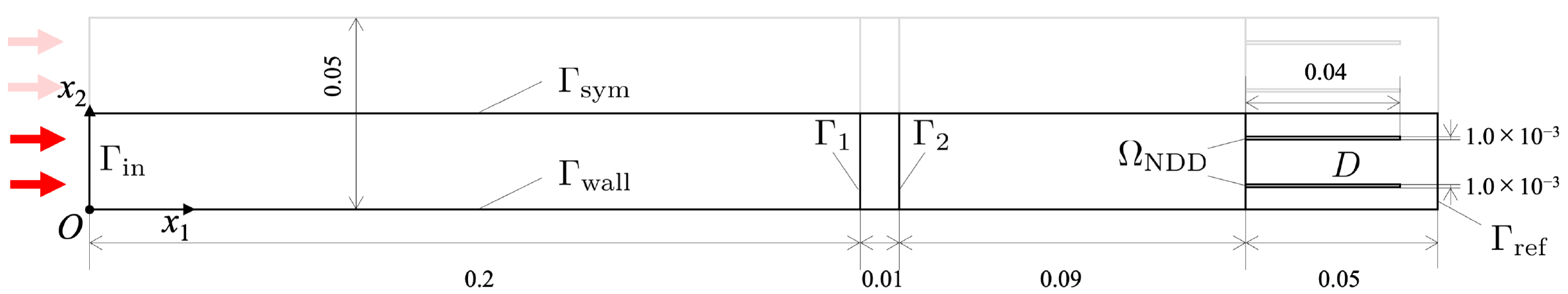}
	\caption{\textcolor {black}{Geometrical settings and boundary conditions for the design of the sound-absorbing structure \cite{andersen2019shape}.}}
	\label{fig:Bench_Geom}       
\end{figure}

\textcolor {black}{We consider a sound absorption maximization problem conducted in \cite{andersen2019shape} and demonstrate the difference between the results with and without boundary layers.
Figure~\ref{fig:Bench_Geom} shows the geometrical settings of the design problem.
As in Case 1, this example corresponds to the closed tube. Moreover, the incident wave $P_\mathrm{in}=\exp (-i k_0 x_1)$ applied on $\Gamma_\mathrm{in}$ is absorbed by the optimized rigid structures $\Omega_r$ in the design domain $D$.
The symmetry condition is imposed on $\Gamma_\mathrm{sym}$ and on the right-end $\Gamma_\mathrm{ref}$ to reflect the waves.
The bottom boundary $\Gamma_\mathrm{wall}$ and the rigid domain boundaries $\partial \Omega_r$ are considered as rigid surfaces.}

\textcolor {black}{
We set air-filled non-design domains $\Omega_\mathrm{NDD}$ to make the optimization calculation stable considering the reason explained in Section~\ref{sec: formulation of the optimization problem}. In \cite{andersen2019shape}, shape optimization problems are conducted under geometrical constraints on the minimum thickness and curvature of the optimized designs by directly calculating the curvature and the thickness of the air-filled channels and cavities. The direct use of this approach requires heavy computational cost during topology optimization. For simplicity, we considered non-design domains herein.
}

\textcolor {black}{In this system, the sound absorption coefficient can be expressed by Eq.~(\ref{eq: cases alpha}). However, we formulate the objective function to be similar to \cite{andersen2019shape} to confirm the validity of the proposed method:
\begin{align}
\alpha &= 1- \left| \frac{ P_1 \exp (-i k_0 d_w) -P_2  }{P_2 - P_1\exp (i k_0 d_w) }
\right|^2, \nonumber\\
P_1 &=\frac{1}{|\Gamma_1|}\int_{\Gamma_1}p d\Gamma, \nonumber\\
P_2 &=\frac{1}{|\Gamma_2|}\int_{\Gamma_2}p d\Gamma, 
\end{align}
where $d_w=0.01$[m] represents the distance between the two boundaries $\Gamma_1$ and $\Gamma_2$ shown in Fig.~\ref{fig:Bench_Geom}.
$P_1$ and $P_2$ are the averaged pressures on $\Gamma_1$ and $\Gamma_2$, respectively.
We focus herein on single-frequency optimization at $\omega_0 = 2 \pi \times 2000~\mathrm{[rad~s^{-1}]}$. 
The objective function to be minimized is set as $J=-\alpha(\omega_0)$.
A sensitivity analysis can be performed similar to the previous cases.}

\textcolor {black}{We then compare the optimization results with and without the boundary layer effects. 
The boundary layer effects can be ignored by setting $u_v = u_h = 1$ in the governing equation for $p$. 
As opposed to the dissipative BEM used in \cite{andersen2019shape}, this setting brings no loss in the system, and $\alpha$ becomes zero for any shape of $\Omega_r$, making the optimization impossible.
Therefore, the bulk loss of air is added to the system without the boundary layers by using the complex sound speed $\frac{c_0}{1- i \tau_\mathrm{loss}}$ instead of $c_0$, where $\tau_\mathrm{loss}$ is a parameter for controlling the loss.}

\textcolor {black}{The parameters that represent air and rigid are the same with the abovementioned settings, except for the case considering the bulk loss of air.
Similarly, the remeshing parameters are set similar to those in cases 1 and 2.
The regularization parameter in the reaction--diffusion equation is set to $\tau=5\times 10^{-4}$ with the characteristic length $L_\phi = 0.05~\mathrm{[m]}$.
The Neumann boundary condition for $\phi$ is imposed on the boundary $\partial \Omega_\mathrm{NDD} \cup \partial D$, whereas the Dirichlet boundary condition for $\phi$ is set on the other boundary in $\partial D$. The boundary $\Gamma_{\phi n}$ appearing in Eq.~(\ref{eq:reaction-diffusion equation with B.C}) is set accordingly.
Time-directional sensitivity filtering is not applied to this example because we focus on single-frequency optimization.
Thus, we set $\alpha_t=1$.
In the case with bulk loss, the loss factor $\tau_\mathrm{loss}$ is set to $5\times 10^{-3}$.}

\begin{figure}[H]
	\centering
	\includegraphics[scale=0.7]{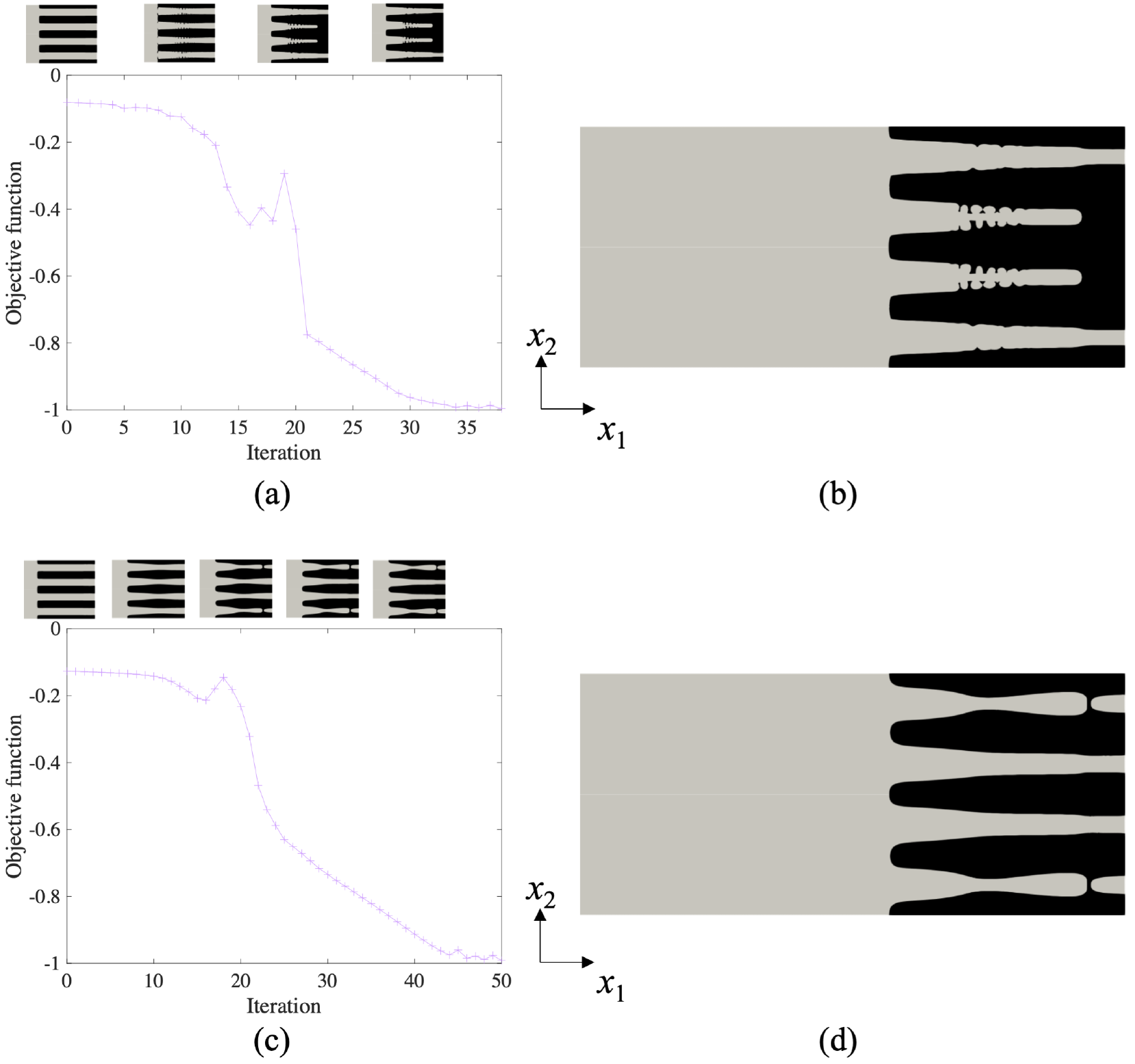}
	\caption{\textcolor {black}{Optimization results: (a) history of the objective function and the intermediate designs (at 0th, 10th, 20th, and 30th iterations) with boundary layers; (b) optimized designs at the 38th iteration with boundary layers; (c) history of the objective function and theh intermediate designs (at 0th, 10th, 20th, 30th, and 40th iterations) without boundary layers; and (d) optimized designs at the 50th iteration without boundary layers. The intermediate and optimized designs above are mirrored by considering the symmetry boundary condition on $\Gamma_\mathrm{sym}$.}
	}
	\label{fig:Bench_optconfig}       
\end{figure}

\textcolor {black}{Figures~\ref{fig:Bench_optconfig}(a) and (b) show the history of the objective function and the optimized design when the boundary layer effects are considered.
We chose an initial configuration with two channels in the computational region (Fig.~\ref{fig:Bench_optconfig}(a)), where the design is mirrored considering the symmetric condition on $\Gamma_\mathrm{sym}$.
The $J$ value at the initial iteration was $-8.10\times 10^{-2}$, denoting a small $\alpha$ value.
We finally halted the optimization calculation at the 38th iteration, at which the 10-iteration moving average of the relative error between the $J$ values for two consecutive iterations was $9.49\times 10^{-3}$.
The computational time for obtaining the optimized result was 15.4 min when using eight workstation processors (Intel Xeon W CPU 2.5GHz, 28cores, 384GB memory).
This is almost 11$\%$ of the CPU time required in \cite{andersen2019shape}, indicating the efficiency of the computational cost in the proposed method. 
However, a simple comparison with the previous method is not possible because the CPU time required for the geometrical constraints is not included in the proposed method. 
Several approaches have been reported to impose the minimum thickness constraint or equivalent constraints restricting channels from collapsing.
For example, there are methods for introducing a filter to the design variable \cite{guest2004achieving} and for using an artificial field satisfying a PDE similar with the thermal conductivity equation to detect internal voids \cite{yamada2021topology}.
Their computational cost is not as heavy as that of the direct approach; thus, the CPU time for obtaining the optimized design by the proposed method will not exceedingly increase, even under the use of these constraints.}

\textcolor {black}{The $J$ value at the optimized design was $-0.996$, which means an almost perfect sound absorption was realized.
Small bump-like structures appeared in the early stage of optimization and remained in the optimized design, similar to those in cases 1 and 2.}

\textcolor {black}{Figures~\ref{fig:Bench_optconfig} (c) and (d) show the history of the objective function and the optimized design when the boundary layer effects are ignored.
The optimization calculation starts from the same initial configuration when the viscothermal effects are considered.
The $J$ value at the initial iteration was $-0.127$, which again means a small $\alpha$ value.
The optimization calculation is stopped at the 50th iteration, at which the 10-iteration moving average of the relative error between the $J$ values for two consecutive iterations is $1.50\times 10^{-2}$.
The $J$ value at the optimized design is $-0.992$, indicating the realization of an almost perfect sound absorption.
Different from the result considering the boundary layers, the optimized design surface was smooth, and no bump was observed during the optimization history shown in Fig.~\ref{fig:Bench_optconfig}(c).}

\begin{figure}[H]
	\centering
	\includegraphics[scale=0.5]{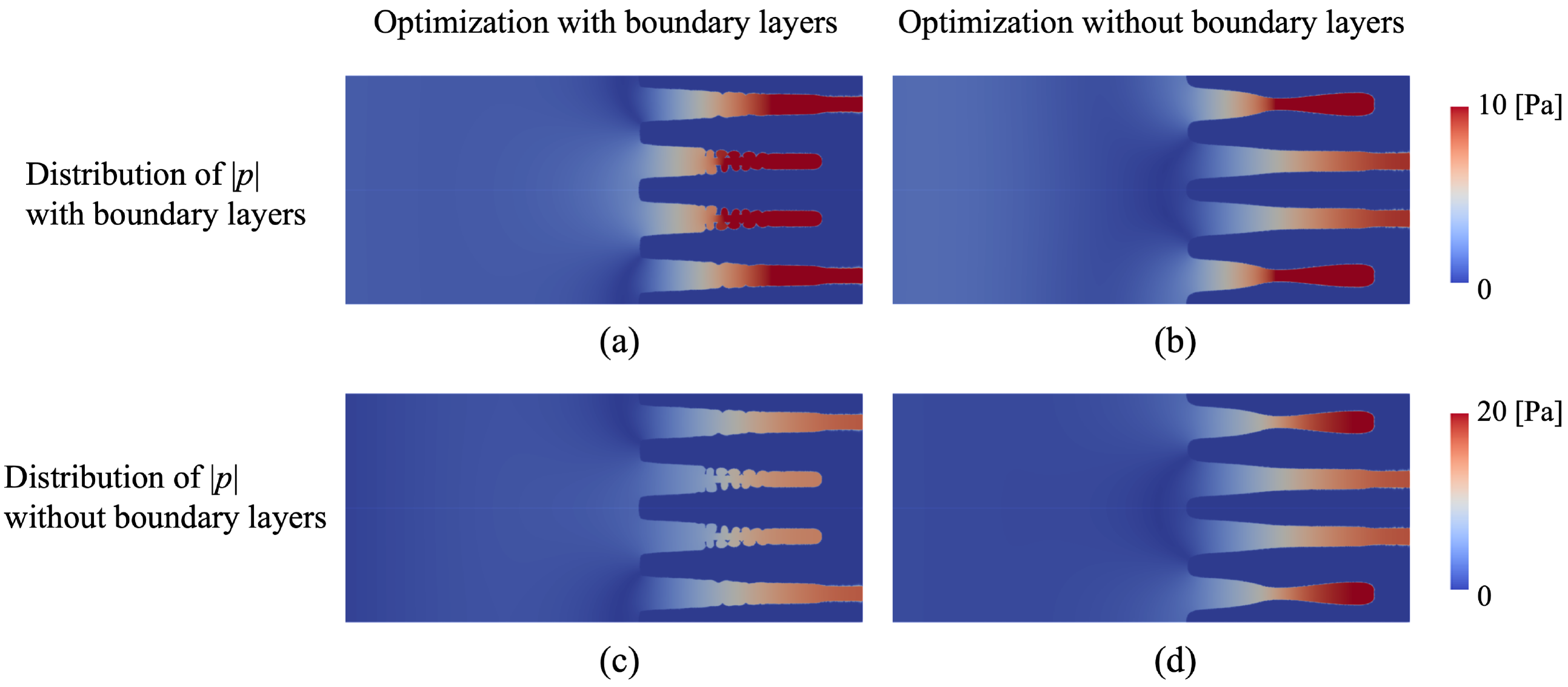}
	\caption{\textcolor {black}{
		Distributions of $|p|$ for the optimized results with and without considering the boundary layers at 2000 [Hz]:
		(a) distribution of $|p|$ obtained by the SLNS model for the optimization result considering the boundary layers;
		(b) distribution of $|p|$ obtained by the SLNS model for the optimization result ignoring the boundary layers;
		(c) distribution of $|p|$ obtained by the standard Helmholtz equation with bulk loss for the optimization result considering the boundary layers; and
		(d) distribution of $|p|$ obtained by the standard Helmholtz equation with bulk loss for the optimization result ignoring the boundary layers.
		These $|p|$ distributions are mirrored by considering the symmetry boundary condition on $\Gamma_\mathrm{sym}$.
	}
	}
	\label{fig:Bench_absp}       
\end{figure}

\begin{figure}[H]
	\centering
	\includegraphics[scale=0.8]{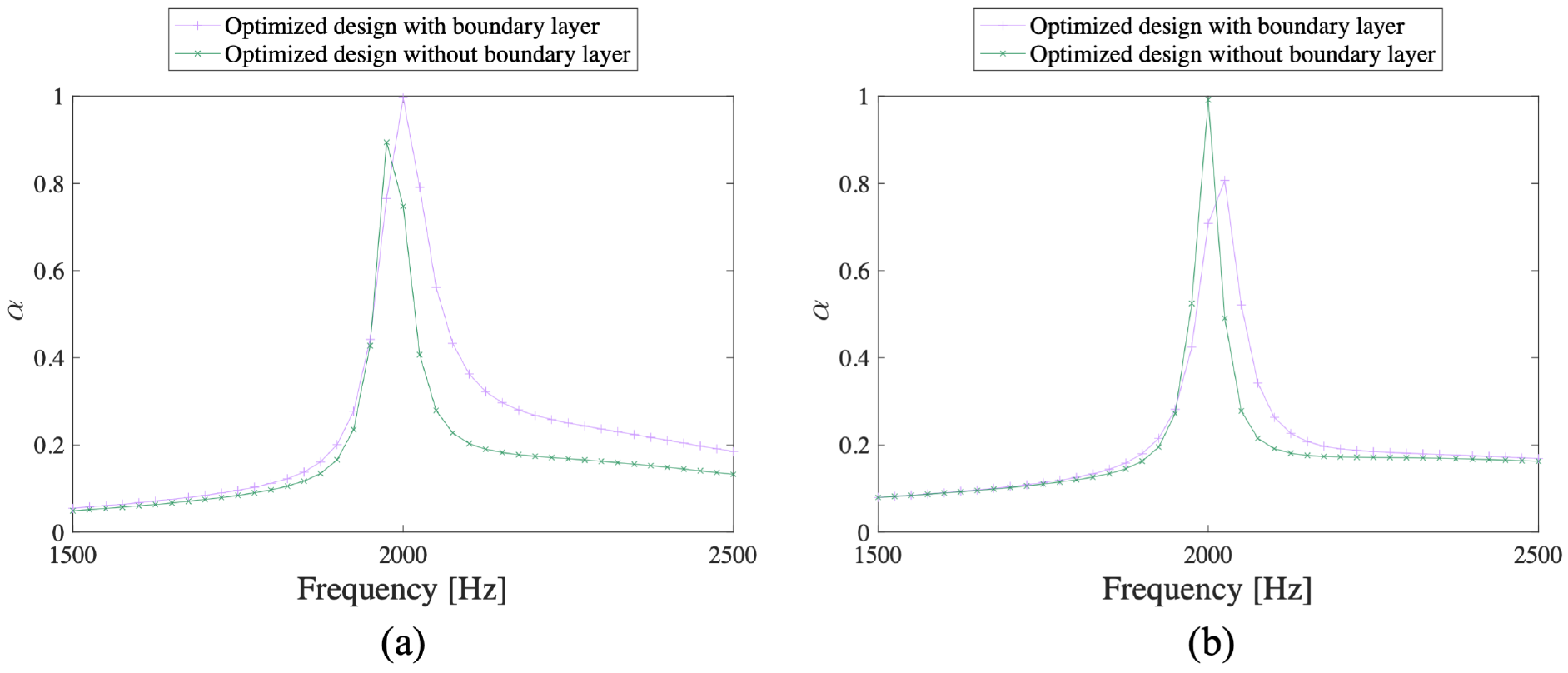}
	\caption{\textcolor {black}{Frequency responses of the sound absorption coefficient $\alpha$ with and without the boundary layers:
		(a) frequency responses of $\alpha$ evaluated by the SLNS model for the optimized designs shown in Fig.~\ref{fig:Bench_optconfig}(b) and (d), represented by the purple and green lines, respectively; and
		(b) frequency responses of $\alpha$ evaluated by the Helmholtz equation with bulk loss for the optimized designs shown in Fig.~\ref{fig:Bench_optconfig}(b) and (d), represented by the purple and green lines, respectively.}}
	\label{fig:Bench_freqs}       
\end{figure}

\textcolor {black}{Figure \ref{fig:Bench_absp} shows a comparison of the distributions of $|p|$ at 2000~[Hz] for the two optimized designs between the SLNS model-based analysis considering the boundary layers and the analysis ignoring them.
Figure \ref{fig:Bench_absp} (a) illustrates the distribution of $|p|$ with boundary layers for the optimized result considering the boundary layers, whereas (b) illustrates that for the optimized result ignoring the boundary layers.
Considering the symmetry condition on $\Gamma_\mathrm{sym}$, large $|p|$ values were confirmed at the four channels shown in (a); however, they are observed only at the external two channels in (b).
In contrast, Fig. \ref{fig:Bench_absp} (c) shows the distribution of $|p|$ without boundary layers for the optimized result considering the boundary layers, whereas (d) shows that for the optimized result ignoring the boundary layers.
As shown in (d), the large $|p|$ values can be seen in the external two channels, but not in (c).
In other words, the system with boundary layers has a different resonance frequency from that without boundary layers.
Figure~\ref{fig:Bench_freqs} shows the frequency responses of the sound absorption coefficient $\alpha$ for the optimized designs.
Figure~\ref{fig:Bench_freqs}(a) depicts the comparison of the frequency response of $\alpha$ estimated by the SLNS model for the optimized design considering the boundary layers and that ignoring the layers.
The optimized design considering the boundary layers has a peak value of $\alpha ~(\approx 1)$ at the target frequency of 2000~[Hz]; however, the peak of $\alpha$ for the design ignoring the boundary layers  is shifted from 2000~[Hz], and its value is smaller than 0.9.
The same trend is confirmed in Fig.~\ref{fig:Bench_freqs}(b), which shows a comparison of the frequency responses of $\alpha$ in both designs estimated by the Helmholtz equation with bulk loss.  
These behaviors conform to the distribution of $|p|$ shown in Fig.~\ref{fig:Bench_absp}.
The results reveal that the sound absorption property of the system with boundary layers is different from that of the system with bulk loss ignoring the boundary layers.
Therein, considering the boundary layer effects in topology optimization for sound absorption maximization is essential.
The same conclusion is obtained in \cite{andersen2019shape}, which verifies the validity of the proposed method.}

\section{Optimization of the metamaterial slab for the negative refraction}\label{sec: negative refraction}
\textcolor {black}{The losses induced by the boundary layers are maximized in the abovementioned numerical examples.
Here, we consider a different optimization setting, where the objective function is set to realize the negative refraction by optimizing the structure of the metamaterial slab.
Wave transmission through the metamaterial slab, which exhibits a negative refraction, is suppressed by the boundary layer effects \cite{christiansen2016experimental}.
Accordingly, we examine an optimization case with and without boundary layers to demonstrate that a negative refraction with a large value of the amplitude of the transmitted waves can be realized by using the proposed optimization method. By comparing these results, we discuss the importance of considering boundary layers in topology optimization.}
\begin{figure}[H]
	\centering
	\includegraphics[scale=0.5]{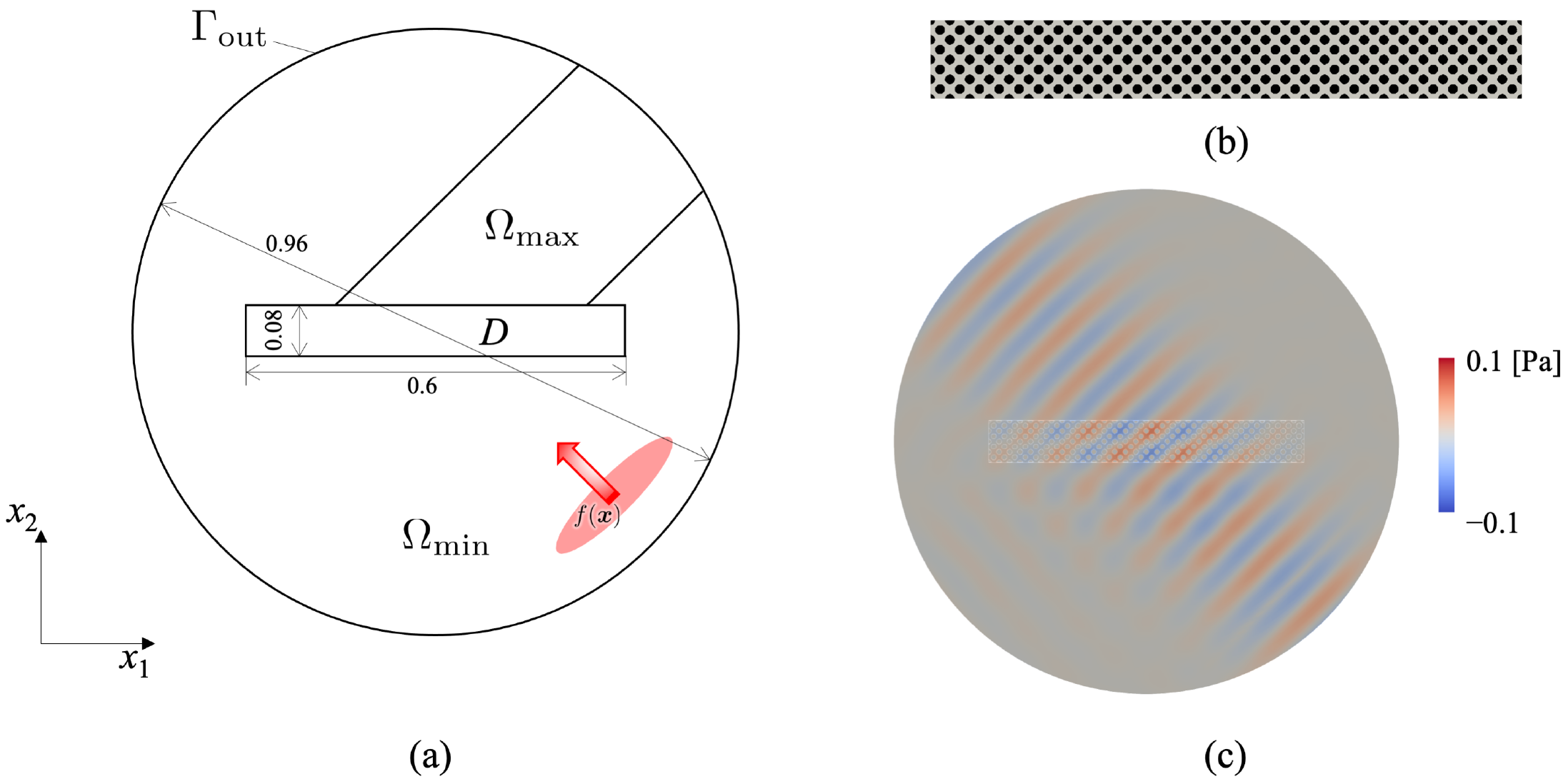}
	\caption{\textcolor {black}{(a) Geometrical settings and boundary conditions for the design of a metamaterial slab realizing a negative refraction. The units are in $\mathrm{[m]}$.
		(b) Initial configuration in $D$.
		(c) Distribution of $\mathrm{Re}(p)$ without boundary layers.}}
	\label{fig:Geom_negative}       
\end{figure}

\textcolor {black}{Figure~\ref{fig:Geom_negative}(a) shows the geometrical settings and the boundary conditions. We consider the acoustic waves propagating in the free space modeled by the circular-shaped computational domain with a non-reflecting boundary condition on $\Gamma_\mathrm{out}$. The incident wave is modeled by an additional term $f(\bm{x})$ in the right-hand side of Eq.~(\ref{eq: SLNS_p}). Similar to the previous research \cite{noguchi2021level}, $f(\bm{x})$ based on the Gaussian distribution is chosen, and an oblique incident wave is introduced. The design domain $D$ is set at the center of the computational domain. The rigid structure in $D$ is optimized such that the acoustic wave amplitude in $\Omega_\mathrm{max}$ is maximized, while that in $\Omega_\mathrm{min}$ is minimized. The objective function is formulated as follows:}

\textcolor {black}{
\begin{align}
J &= -w_{J_1} J_1 + w_{J_2} J_2,\nonumber\\
J_1 &=\frac{1}{|\Omega_\mathrm{max}|}\int_{\Omega_\mathrm{max}}|p|^2 d\Omega, \nonumber\\
J_2 &=\frac{1}{|\Omega_\mathrm{min}|}\int_{\Omega_\mathrm{min}}|p|^2 d\Omega, 
\end{align}
where, $w_{J_1}$ and $w_{J_2}$ are the weighting factors fixed to 0.5. 
This setting is similar to that in \cite{christiansen2016designing} and \cite{noguchi2021level}, where the negative refraction was successfully realized, but the boundary layer effects were not considered. 
The sensitivity analysis can be performed similar to the abovementioned examples, where the topological derivative formula is the same with Eq.~(\ref{eq: DtJ_mainpart}), and the source terms of the adjoint equation are defined in $\Omega_\mathrm{max}$ and $\Omega_\mathrm{min}$.
The optimization calculation starts from the structure shown in Fig.~\ref{fig:Geom_negative}(b), which is similar to \cite{christiansen2016designing}, and is based on a periodic array of unit cells.
The previous studies \cite{christiansen2016designing,noguchi2021level} conducted a topology optimization of metamaterials with a constraint forcing the periodicity in optimizing structures.
Herein, we do not impose such a constraint, and the overall structure is independently optimized.}

\textcolor {black}{As in the example in \ref{sec: Benchmark}, we focus on a single-frequency optimization targeting $\omega_0 = 2\pi \times 4000~\mathrm{[rad~s^{-1}]}$. 
The parameters representing air and rigid are similar to those in Sections \ref{sec: Case1} and \ref{sec: Case2}.
The regularization parameter is set to $\tau = 5\times 10^{-4}$, with the characteristic length $L_\phi = 2\times 10^{-2}$, which corresponds to the size of a unit cell in the initial configuration.
The boundary $\Gamma_{\phi n}$ is not used in this example, that is, the Dirichlet boundary condition in Eq.~(\ref{eq:reaction-diffusion equation with B.C}) was applied to $\partial D$. 
Time-directional sensitivity filtering is not applied to this example, and we set $\alpha_t = 1$.
The initial design contains multiple rigid bodies and will remain in the optimization procedure; thus, the remeshing procedure for the boundary layers as described in Section \ref{sec:Remesh and rigid domain}, bringing a large number of finite elements that would require heavy computational time.
To avoid this issue, coarse meshes for the boundary layers are used in this example.
We do not use the approximated solution of the Eikonal equation. We only use the level set function $\phi$, resulting in rigid body-fitted meshes with a zero-level isosurface of $\phi$.
The analysis results with finer meshes obtained by the procedure in Section \ref{sec:Remesh and rigid domain} will be shown for the optimized design to check its performance.}

\textcolor {black}{First, the optimization example without the boundary layers will be exhibited.
Different from the case in \ref{sec: Benchmark}, bulk loss is not added in this example. 
With the initial configuration shown in Fig.~\ref{fig:Geom_negative}(b), the acoustic pressure distribution without the boundary layer effects was obtained as shown in Fig.~\ref{fig:Geom_negative}(c), where a usual sound transmission was observed, and no wave concentration in $\Omega_\mathrm{max}$ was found. 
The $J$ value was $5.14\times 10^{-5}$, with $J_1 = 1.61\times 10^{-4}$ and $J_2 = 2.64\times 10^{-4}$.
}
\begin{figure}[H]
	\centering
	\includegraphics[scale=0.6]{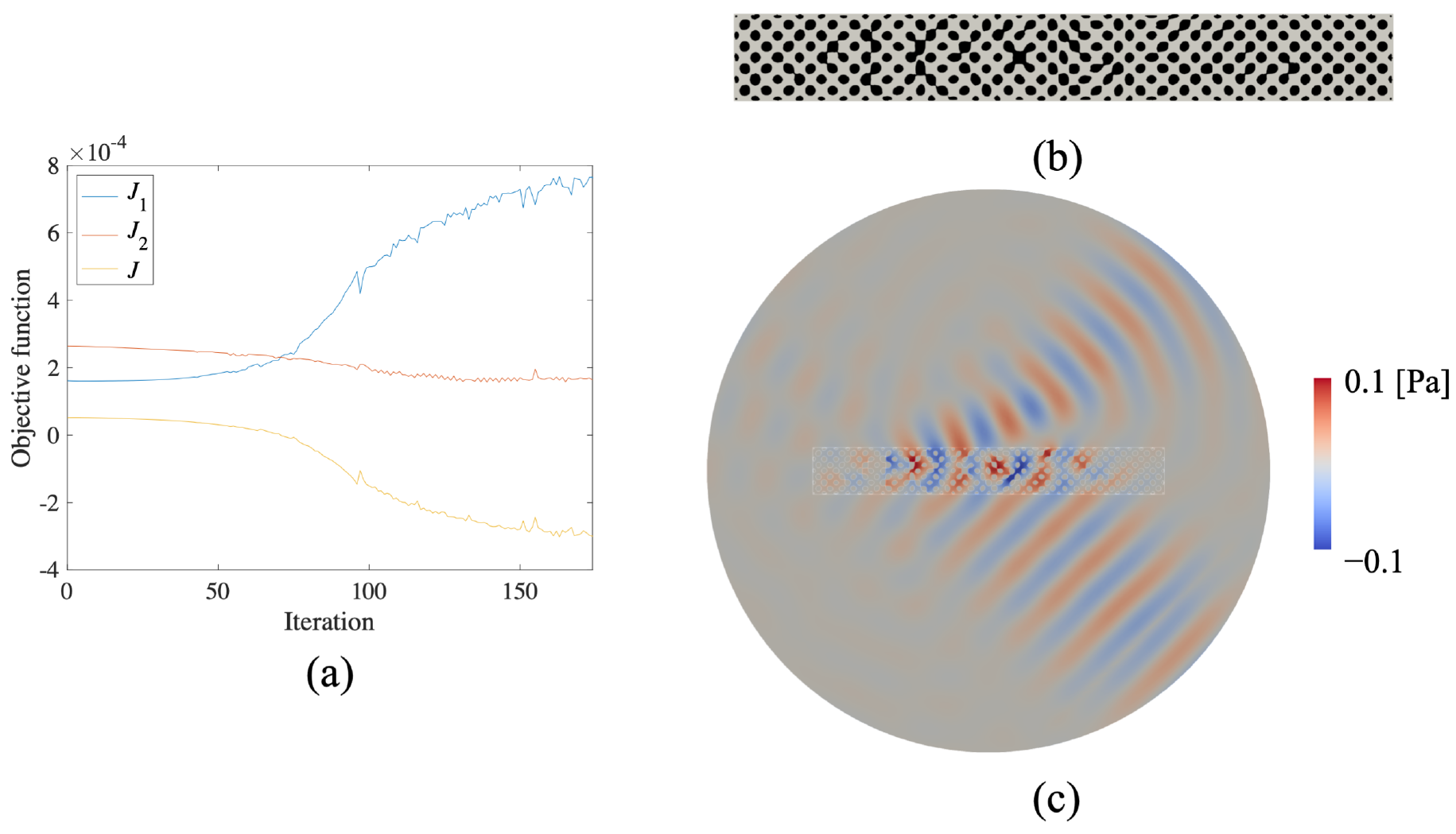}
	\caption{\textcolor {black}{
	Optimization results without boundary layers: (a) history of the objective function; (b) optimized design; and (c) distribution of $\mathrm{Re}(p)$ without boundary layers.}
	}
	\label{fig:history_woLayer}       
\end{figure}

\textcolor {black}{
Figure~\ref{fig:history_woLayer}(a) shows the history of the objective function. 
During the optimization, the $J_1$ values were maximized, while those of $J_2$ were minimized, as expected. 
We stopped the optimization calculation at the 174th iteration, where the 10-iteration moving average of the relative error between the $J$ values for two consecutive iterations was $2.70\times 10^{-2}$. 
Figure~\ref{fig:history_woLayer}(b) represents the optimized structure in the design domain $D$, while Fig.~\ref{fig:history_woLayer}(c) represents acoustic pressure distribution. 
The acoustic wave concentration in $\Omega_\mathrm{max}$ was observed, and negative refraction was realized. The $J$ value was $-3.01\times 10^{-4}$, with $J_1 = 7.65\times 10^{-4}$ and $J_2 = 1.64\times 10^{-4}$, which stands for the concentration.
}

\begin{figure}[H]
	\centering
	\includegraphics[scale=0.5]{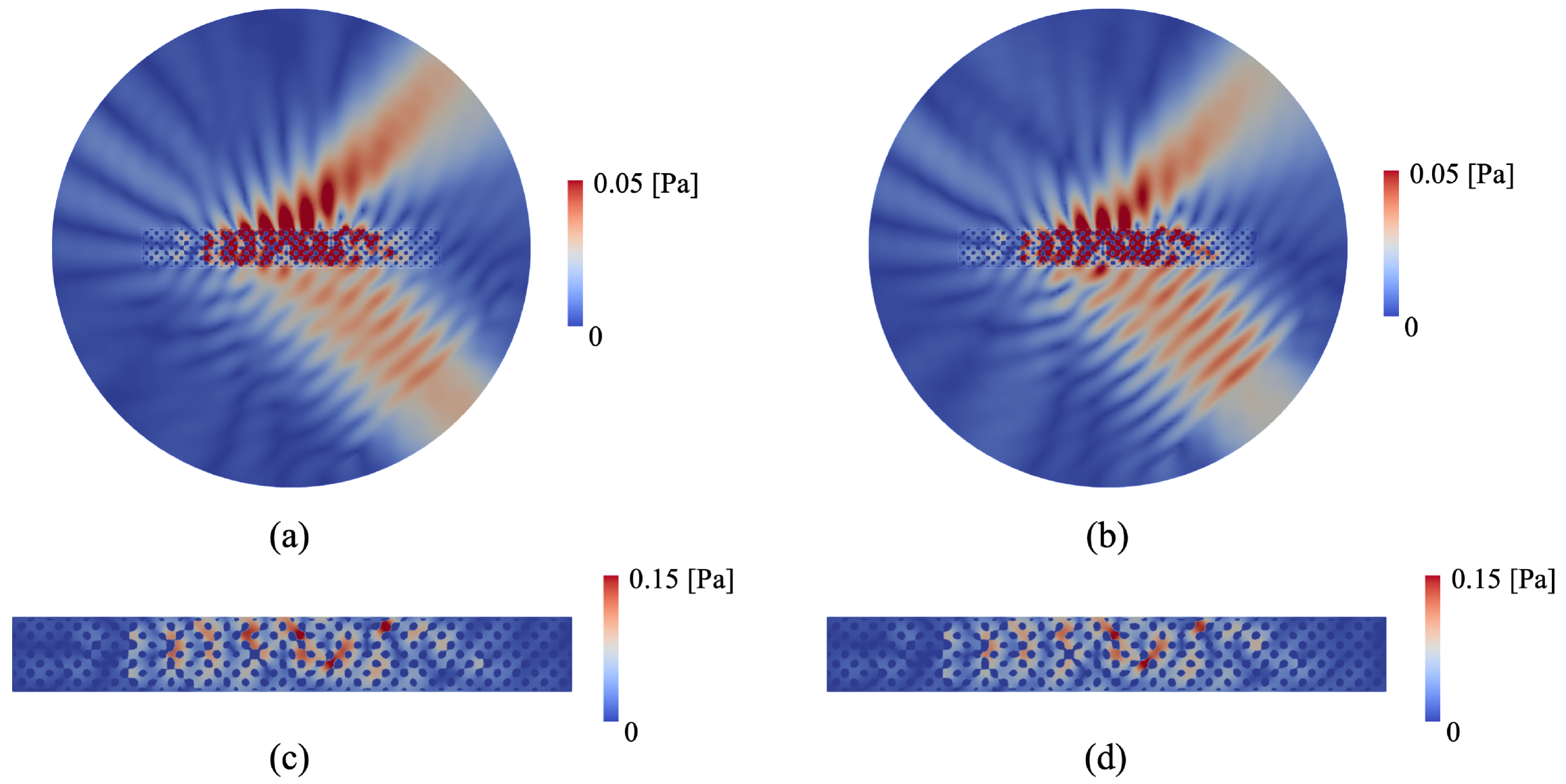}
	\caption{\textcolor {black}{
		Comparison of the distribution of $|p|$ in the optimized design shown in Fig.~\ref{fig:history_woLayer}(b) with and without the boundary layer effects: 
		(a) distribution of $|p|$ in the entire domain without boundary layers; 
		(b) distribution of $|p|$ in the entire domain with boundary layers; 
		(c) distribution of $|p|$ in $D$ without boundary layers;
		 and (d) distribution of $|p|$ in $D$ with boundary layers.	}
	}
	\label{fig:absp_compare}       
\end{figure}

\textcolor {black}{
	The results shown in Fig.~\ref{fig:history_woLayer} do not include the boundary layers. 
	The performance of the optimized slab is compared with that considering the boundary layers. 
	Figure~\ref{fig:absp_compare} shows a comparison of the distributions of the absolute value of acoustic pressure $|p|$ in the optimized design. 
	Figures~\ref{fig:absp_compare}(a) and (c) depict the distribution of $|p|$ without the effects of the boundary layers, while Figs.~\ref{fig:absp_compare}(b) and (d) illustrate that with boundary layers. 
	The overall distributions were similar between them; however, $|p|$ in $\Omega_\mathrm{max}$ in Fig.~\ref{fig:absp_compare}(b) was smaller than that in Fig.~\ref{fig:absp_compare}(a).
	Focusing on the design domain shown in Figs.~\ref{fig:absp_compare}(c) and (d), high values of $|p|$ can be confirmed between the optimized rigid bodies in both figures. 
	However, their peak value in (d) was smaller than that in (c). 
	The trends when considering the boundary layers worsened the performance of the optimized design. 
	The $J$ value considering the layer's effects was $-2.26\times 10^{-4}$, with $J_1 = 6.19\times 10^{-4}$ and $J_2 = 1.67\times 10^{-4}$, whereas the lossless analysis evaluated them as $J=-3.01\times 10^{-4}$ with $J_1 = 7.65\times 10^{-4}$ and $J_2 = 1.64\times 10^{-4}$.
	 In other words, the concentration in $\Omega_\mathrm{max}$ was weakened by the boundary layers.
 }

\begin{figure}[H]
	\centering
	\includegraphics[scale=0.6]{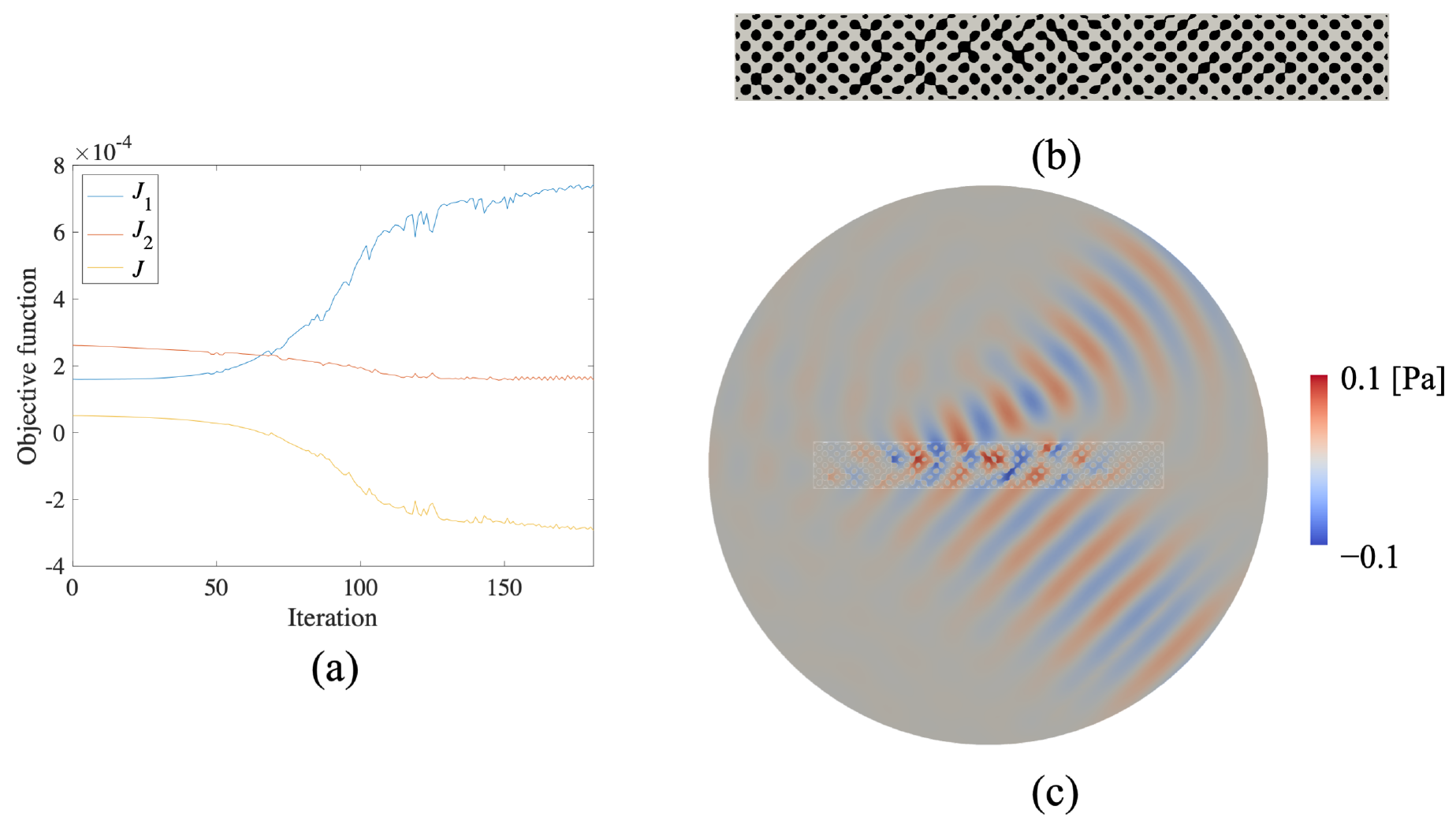}
	\caption{
		\textcolor {black}{
			Optimization results with boundary layers: (a) history of the objective function; (b) optimized design; and (c) distribution of $\mathrm{Re}(p)$ with boundary layers.
			}
	}
	\label{fig:history_wLayer}       
\end{figure}
\textcolor {black}{
	Next, the optimization case with boundary layers is exhibited. 
	Figure~\ref{fig:history_wLayer}(a) shows the history of the objective function. 
	Similar to the previous case, the $J_1$ values were maximized, while those of $J_2$ were minimized during the optimization. 
	The optimization calculation was halted at the 181st iteration, where the 10-iteration moving average of the relative error between the values of $J$ for two consecutive iterations was $2.06\times 10^{-2}$. Figure~\ref{fig:history_wLayer}(b) represents the optimized structure in the design domain $D$. 
	Figure~\ref{fig:history_wLayer}(c) represents the acoustic pressure distribution. 
	The negative refraction was also realized in this case. The $J$ value was $-2.92\times 10^{-4}$, with $J_1 = 7.42\times 10^{-4}$ and $J_2 = 1.58\times 10^{-4}$.
	}

\begin{figure}[H]
	\centering
	\includegraphics[scale=0.6]{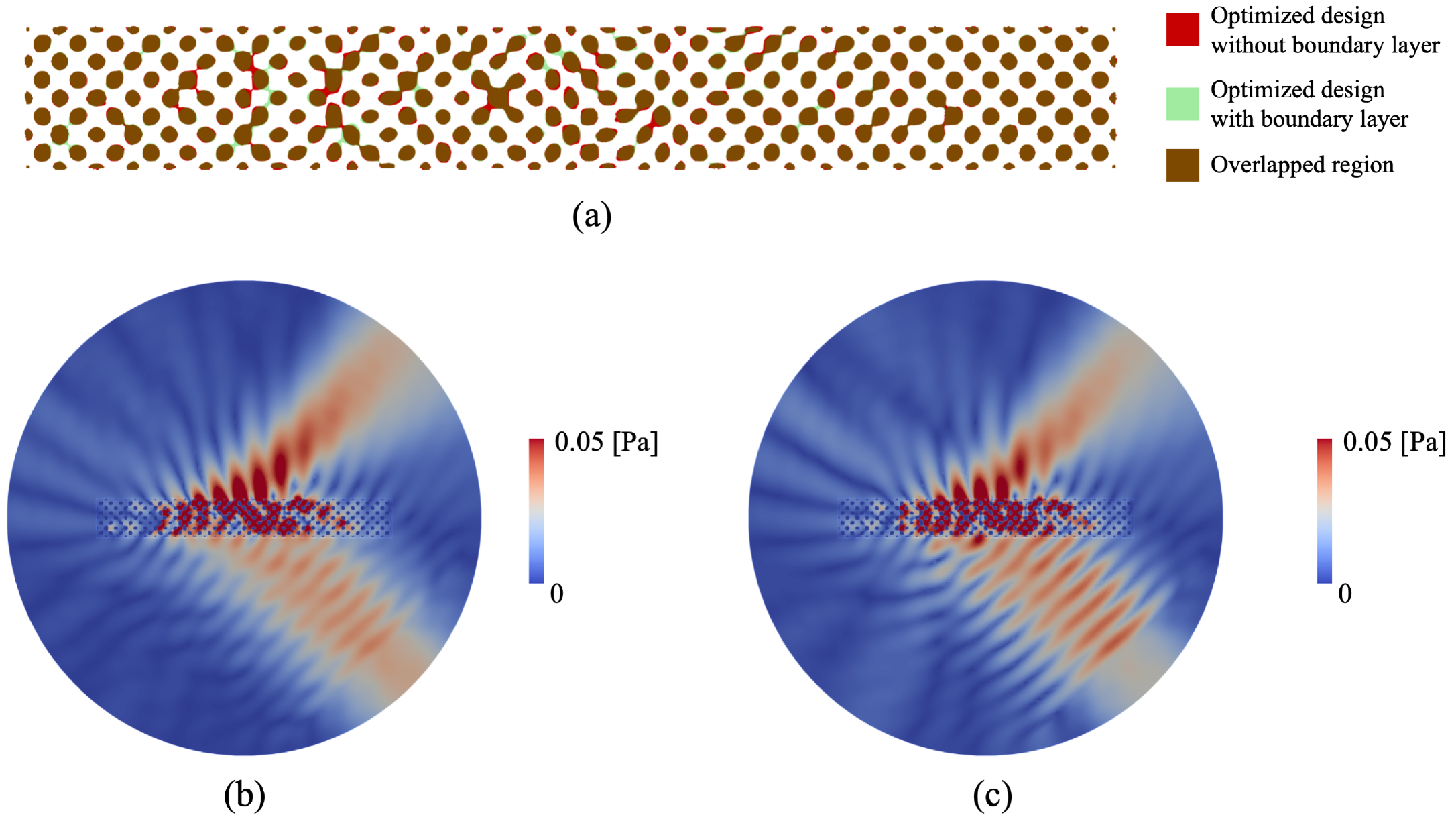}
	\caption{ \textcolor {black}{
			Comparison of the optimized results with and without the boundary layers: 
			(a) difference in the shape of the optimized designs; 
			(b) distribution of $|p|$, including the boundary layer effects for the optimized design in Fig.~\ref{fig:history_wLayer}(b); 
			and (c) distribution of $|p|$, including the boundary layer effects for the optimized design in Fig.~\ref{fig:history_woLayer}(b).
			}
	}
	\label{fig:optcompare}       
\end{figure}

\textcolor {black}{
	The obtained design is basically similar to that in the previous case without boundary layers, but with a slight difference. 
	Figure~\ref{fig:optcompare}(a) shows both of the optimized designs. 
	The optimized design without boundary layer is depicted by the red color, whereas that with the boundary layer is represented by green. 
	Their overlapped region is represented by brown. 
	Both optimized designs are mainly composed of rigid domains with a round shape; however, their area in the optimized design without boundary layers is larger than that in the optimized design with boundary layers. 
	Including the overlapped region, the ratio of the area of the rigid domain in the optimized design without boundary layers to $|D|$ is 35.2\%, whereas that in the optimized design containing boundary layers is 34.0\%. 
	In other words, the optimized design considering the boundary layers has wider air-filled channels than the lossless design. 
	This feature suppresses the boundary layer effect to absorb the sound energy through the designed metamaterial slab. 
	Figures~\ref{fig:optcompare}(b) and (c) show the distribution of $|p|$ with the boundary layer effects for both designs. 
	Apparently, the amplitude of the transmitted waves in the design considering the boundary layer, as shown in (b), is larger than that in the design ignoring the boundary layer, as shown in (c).
}

\textcolor {black}{
	This trend did not change with the FEM analysis using finer meshes for the boundary layers, where the optimized design was remeshed based on the remeshing parameter similar to that in Section \ref{sec: numerical examples}. 
	In this analysis, the objective function for the optimized design considering the boundary layers was obtained as $J=-2.65\times 10^{-4}$ with $J_1=6.96\times 10^{-4}$ and $J_2=1.65\times 10^{-4}$, whereas that for the design ignoring the boundary layers was $J=-2.07\times 10^{-4}$ with $J_1=5.69\times 10^{-4}$ and $J_2=1.54\times 10^{-4}$.
}

\textcolor {black}{In summary, the negative refraction with high transmission was achieved by the proposed method, even in the presence of boundary layers throughout the design domain.}

\section{Details on the derivation of the SLNS model from the FLNS model}\label{sec: Detail SLNS}
\textcolor {black}{Here, we present more details on the derivation of the SLNS model from the FLNS model. First,} the following linear acoustic assumption is introduced to linearize the system:
\begin{align}
	&\bm{v_0} = 0, \label{eq: linear assumption 1}\\
	&\left|\frac{\psi}{\psi_0} \right| \ll 1~~~\mathrm{for~}\psi = ({\rho}, {p},{T},{H},{s}), \label{eq: linear assumption 2}\\
	&\left|\frac{\bm{v}}{c_0} \right| \ll 1, \label{eq: linear assumption 3}
\end{align}
which indicates that zero mean flow is assumed \textcolor {black}{(Eq.~(\ref{eq: linear assumption 1})). The} perturbations from the quiescent values for $(\check{\rho}, \check{p},\check{T},\check{H},\check{s})$ are assumed to be small (Eq.~(\ref{eq: linear assumption 2})), and that for $\check{\bm{v}}$ is assumed to be much smaller than the sound speed, $c_0$ (Eq.~(\ref{eq: linear assumption 3})).

\textcolor {black}{The} linearized balance continuity equation, momentum equation, entropy equation, and the Gibbs relation are given as
\begin{align}
	&i \omega \rho = -\rho_0 \nabla \cdot \bm{v}~~~\mathrm{in~}\Omega_a,\nonumber\\
	&i \omega \rho_0 \bm{v} = \nabla \cdot \sigma + \bm{f}~~~\mathrm{in~}\Omega_a,\nonumber\\
	&i \omega \rho_0 T_0 s = -\nabla \cdot \bm{q} + Q~~~\mathrm{in~}\Omega_a,\nonumber\\
	&\rho_0 T_0 s = \rho_0 H -p~~~\mathrm{in~}\Omega_a,\label{eq: linearized navier stokes}
\end{align}
Applying the constitutive laws Eq.~(\ref{eq: constitutive laws}) to Eq.~(\ref{eq: linearized navier stokes}) \textcolor {black}{and} choosing $(\bm{v},T,p)$ as unknowns, \textcolor {black}{the set of equations for the FLNS model is obtained as Eqs.~(\ref{eq: FLNS 1})--(\ref{eq: FLNS 3}).}

The set of equations used in the SLNS model are derived by ignoring small terms considering the viscous and thermal \textcolor {black}{layers' property (Eq.~(\ref{eq: property of layers})).} 
To do this, the normalized equations of the FLNS model are introduced as
\begin{align}
	&\dbar{\bm{v}} + \dbar{k_v^{-2}}\dbar{\xi}\dbar{\nabla}(\dbar{\nabla}\cdot \dbar{\bm{v}})+\dbar{k_v^{-2}}\dbar{\nabla^2}\dbar{\bm{v}}=-\frac{1}{i\gamma}\dbar{\nabla}\dbar{p},\label{eq: nondim FLNS v}\\
	&\dbar{T} + \dbar{k_h^{-2}}\dbar{\nabla^2}\dbar{T} = \frac{\gamma -1}{\gamma}\dbar{p},\label{eq: nondim FLNS T}\\
	&\dbar{\nabla}\cdot \dbar{\bm{v}} - i\dbar{T} + i\dbar{p} = 0,\label{eq: nondim FLNS p}
\end{align}
where $\gamma$ is the ratio of specific heats, and the other nondimensional quantities are defined as
\begin{align}
	&\dbar{\bm{v}}=\frac{\bm{v}}{c_0},~~~\dbar{T}=\frac{T}{T_0},~~~\dbar{p}=\frac{p}{p_0},\nonumber\\
	&\dbar{\xi}=1+\frac{\lambda}{\mu},~~~\dbar{\nabla}=\frac{1}{k_0}\nabla.
\end{align}
In \cite{kampinga2010viscothermal}, 
the order of magnitude analysis revealed that the effect of the second term in the left-hand side of Eq.~(\ref{eq: nondim FLNS v}) is small compared to those of the other terms. Therefore, the following equations can be considered:
\begin{align}
	&\dbar{\bm{v}} +\dbar{k_v^{-2}}\dbar{\nabla^2}\dbar{\bm{v}}=-\frac{1}{i\gamma}\dbar{\nabla}\dbar{p}~~~\mathrm{in~}\Omega_a,\label{eq: nondim2 FLNS v}\\
	&\dbar{T} + \dbar{k_h^{-2}}\dbar{\nabla^2}\dbar{T} = \frac{\gamma -1}{\gamma}\dbar{p}~~~\mathrm{in~}\Omega_a,\label{eq: nondim2 FLNS T}\\
	&\dbar{\nabla}\cdot \dbar{\bm{v}} - i\dbar{T} + i\dbar{p} = 0~~~\mathrm{in~}\Omega_a,\label{eq: nondim2 FLNS p}
\end{align}
By re-applying the results of the order of magnitude analysis to \textcolor {black}{Eqs.~(\ref{eq: nondim2 FLNS v}) and (\ref{eq: nondim2 FLNS T})}, the normalized temperature and velocity with the nondimensional acoustic pressure and its gradient are expressed as
\begin{align}
	\dbar{\bm{v}}&=-\frac{u_v \dbar{\nabla}\dbar{p}}{i\gamma},\nonumber\\
	\dbar{T}&=u_h \frac{\gamma-1}{\gamma}\dbar{p},\label{eq: formula of v and T in SLNS}
\end{align}
\textcolor {black}{where,} $u_v$ and $u_h$ satisfy the nonhomogeneous dimensional Helmholtz equation, expressed in Eq.~(\ref{eq: SLNS_uphi}).
Substituting Eq.~(\ref{eq: formula of v and T in SLNS}) into Eq.~(\ref{eq: nondim2 FLNS p}) yields the dimensional continuity equation for $p$, as expressed in Eq.~(\ref{eq: SLNS_p}).

\section{Derivation of the topological derivative} \label{sec: Appendix: Derivation of DtJ}
This section explains the details of the derivation of the topological derivative. 
As explained, the topological derivative measures the change rate of the objective function when an infinitesimal circular void or inclusion domain $\Omega_\varepsilon$ with the radius $\varepsilon$ is inserted into the target system. 
Let $J$ denote the objective function before the domain $\Omega_\varepsilon$ appears, and 
$(J+\delta J)$ denotes that after $\Omega_\varepsilon$ appears. 
Then, the topological derivative $D_T J$ is defined as follows:
\begin{align}
	(J + \delta J) - J = g(\varepsilon)D_T J + o(g(\varepsilon)), \label{eq: Def of DtJ Appendix}
\end{align}
where $g(\varepsilon)$ is a scalar function that depends on the radius of $\Omega_\varepsilon$. 
In the following, we first estimate the left-hand side of Eq.~$(\ref{eq: Def of DtJ Appendix})$ in \ref{sec: estimation of dJ}. 
Then, in \ref{sec: numerical dJ}, we numerically confirm its validity and determine the function $g(\varepsilon)$ and obtain the explicit formula of the topological derivative, $D_T J$.

\subsection{Estimation of variations in the objective function}\label{sec: estimation of dJ}
To derive the topological derivative, we focus on the design settings shown in Figure~\ref{fig:Design settings}. 
First, it is assumed that there is no rigid domain that appeared in the design domain $D$. 
Corresponding to the example of Case 2, we consider the following objective function:
\begin{align}
	J = \int_{\Gamma_\mathrm{in}} f_1(p,p^\ast) d\Gamma + \int_{\Gamma_\mathrm{out}} f_2(p,p^\ast) d\Gamma,
\end{align}
where $f_1$ and $f_2$ are real-valued functions defined by the acoustic pressure. 
Although the objective function defined in Eq.~(\ref{eq: optimization problem}) is based on the multi-frequency setting, here we focus on the single-frequency setting of the objective function for simplicity. 
Note that the topological derivative for Case 1 in which there is no outlet $\Gamma_\mathrm{out}$ can also be derived by ignoring $f_2$.

We modify the objective function $J$ by adding terms equal to 0, as follows:
\begin{align}
	J &=\int_{\Gamma_\mathrm{in}} f_1(p,p^\ast) d\Gamma + \int_{\Gamma_\mathrm{out}} f_2(p,p^\ast) d\Gamma\nonumber\\
	  &+2\mathrm{Re}\left[\sum_{\varphi = v,h}\left\{\int_{\Omega}\left(k_\varphi^{-2}\nabla^2 u_\varphi + u_\varphi -1
	  \right)v_\varphi d\Omega
	  \right\} \right.\nonumber\\
	  &\left. +\int_{\Omega} \left\{ \nabla \cdot \left( \frac{u_v}{\rho_0}\nabla p \right) + \frac{\omega^2}{K_0}\{\gamma - (\gamma - 1)u_h\}p
	  \right\}q d\Omega \right]\nonumber\\
	  &=\int_{\Gamma_\mathrm{in}} f_1(p,p^\ast) d\Gamma + \int_{\Gamma_\mathrm{out}} f_2(p,p^\ast) d\Gamma\nonumber\\
	  &+2\mathrm{Re}\left[
	  \sum_{\varphi = v,h}\left\{ \int_{\Gamma_\mathrm{ext}}  k_\varphi^{-2} (\bm{n}\cdot \nabla u_\varphi)v_\varphi  d\Gamma 
	  -\int_{\Omega}k_\varphi^{-2}\nabla u_\varphi \cdot \nabla v_\varphi d\Omega\nonumber \right. \right.\\
	  &\left. +\int_{\Omega} (u_\varphi -1)v_\varphi d\Omega \right\} \nonumber\\
	  &\left.+\int_{\Gamma_\mathrm{ext}}\bm{n}\cdot \left( \frac{u_v}{\rho_0}\nabla p  \right)q d\Gamma
	  -\int_{\Omega}\frac{u_v}{\rho_0}\nabla p \cdot \nabla q d\Omega + \int_{\Omega} \frac{\omega^2}{K_0}\left\{
	  \gamma - (\gamma-1)u_h
	  \right\}pq d\Omega \right]
\end{align}
where $\Omega$ represents the entire domain, that is, $\Omega = \Omega_\mathrm{NDD} \cup D$.
$v_\varphi$ with $\varphi =v,h$ and $q$ are arbitrary functions belonging to $V_u = \left\{ \tilde{u_\varphi}\in H^1(\Omega) |~ \tilde{u_\varphi}=0~\mathrm{on~}\Gamma_\mathrm{wall}\right\}$ and $H^1(\Omega)$, respectively.
$\Gamma_\mathrm{ext}=\Gamma_\mathrm{in} \cup \Gamma_\mathrm{out} \cup \Gamma_\mathrm{sym} \cup \Gamma_\mathrm{wall}$ represents the external boundaries of $\Omega$ on which the following boundary conditions are satisfied:
\begin{align}
&u_\varphi = 0~~~\mathrm{on~}\Gamma_\mathrm{wall},\nonumber \\
&\bm{n}\cdot \nabla u_\varphi = 0~~~\mathrm{on~}\Gamma_\mathrm{in}\cup \Gamma_\mathrm{out}\cup \Gamma_\mathrm{sym},\nonumber\\
&\bm{n}\cdot \left(\frac{1}{\rho_0} \nabla p \right)=0 ~~~\mathrm{on~}\Gamma_\mathrm{wall}\cup \Gamma_\mathrm{sym},\nonumber\\
&\bm{n}\cdot \left(\frac{1}{\rho_0} \nabla p \right) + \frac{ik_0}{\rho_0} p =\frac{2ik_0}{\rho_0}P_\mathrm{in}~~~\mathrm{on~}\Gamma_\mathrm{in} ,\nonumber\\
&\bm{n}\cdot \left(\frac{1}{\rho_0} \nabla p \right) + \frac{ik_0}{\rho_0} p =0 ~~~\mathrm{on~}\Gamma_\mathrm{out}.
\label{eq: BC_u}
\end{align}
Next, we take the variation of $J$ when an infinitesimal rigid region $\Omega_\varepsilon$ with the radius $\varepsilon$ is inserted in $\Omega$. 
Let $(u_\varphi + \delta u_\varphi, p + \delta p)$ denote the state variables when $\Omega_\varepsilon$ appears, which correspond to $(u_\varphi, p)$.
These variables satisfy the governing equations defined in $\Omega\setminus\overline{\Omega_{\varepsilon}}$ and the following boundary conditions:
\begin{align}
	&u_\varphi + \delta u_\varphi = 0~~~\mathrm{on~}\Gamma_\mathrm{wall}\cup \Gamma_\varepsilon,\nonumber\\
	&\bm{n}\cdot \nabla (u_\varphi + \delta u_\varphi) = 0~~~\mathrm{on~}\Gamma_\mathrm{in}\cup \Gamma_\mathrm{out}\cup \Gamma_\mathrm{sym},\nonumber\\
	&\bm{n}\cdot \left\{\frac{1}{\rho_0} \nabla (p+\delta p) \right\}=0 ~~~\mathrm{on~}\Gamma_\mathrm{wall}\cup \Gamma_\mathrm{sym},\nonumber\\
	&\bm{n}\cdot \left\{\frac{1}{\rho_0} \nabla (p+\delta p) \right\} + \frac{ik_0}{\rho_0} (p+\delta p) =\frac{2ik_0}{\rho_0}P_\mathrm{in}~~~\mathrm{on~}\Gamma_\mathrm{in} ,\nonumber\\
	&\bm{n}\cdot \left\{\frac{1}{\rho_0} \nabla (p+\delta p) \right\} + \frac{ik_0}{\rho_0} (p+\delta p) =0 ~~~\mathrm{on~}\Gamma_\mathrm{out}.
	\label{eq: BC_udu}
\end{align}
Based on this, the objective function $J$ is changed to $J+\delta J$, as follows:
\begin{align}
	J+\delta J &=\int_{\Gamma_\mathrm{in}} f_1(p+\delta p, p^\ast +\delta p^\ast) d\Gamma + \int_{\Gamma_\mathrm{out}} f_2(p+\delta p,p^\ast + \delta p^\ast) d\Gamma\nonumber\\
	&+2\mathrm{Re}\left[
	\sum_{\varphi = v,h}\left\{ \int_{\Gamma_\mathrm{ext}}  k_\varphi^{-2} (\bm{n}\cdot \nabla (u_\varphi + \delta u_\varphi  ))v_\varphi  d\Gamma 
	-\int_{\Omega\setminus\overline{\Omega_\varepsilon}}k_\varphi^{-2}\nabla (u_\varphi + \delta u_\varphi) \cdot \nabla v_\varphi d\Omega\nonumber \right. \right.\\
	&\left. +\int_{\Omega\setminus\overline{\Omega_\varepsilon}} (u_\varphi + \delta u_\varphi -1)v_\varphi d\Omega
	+\int_{\Gamma_\varepsilon}  k_\varphi^{-2} (\bm{n}\cdot \nabla (u_\varphi + \delta u_\varphi  ))v_\varphi  d\Gamma
	 \right\} \nonumber\\
	&+\int_{\Gamma_\mathrm{ext}}\bm{n}\cdot \left( \frac{(u_v + \delta u_v) }{\rho_0}\nabla (p + \delta p)  \right)q d\Gamma
	-\int_{\Omega\setminus\overline{\Omega_\varepsilon}}\frac{(u_v + \delta u_v)}{\rho_0}\nabla (p + \delta p) \cdot \nabla q d\Omega \nonumber\\
	& + \int_{\Omega\setminus\overline{\Omega_\varepsilon}} \frac{\omega^2}{K_0}\left\{
	\gamma - (\gamma-1)(u_h + \delta u_h)
	\right\}(p + \delta p)q d\Omega \nonumber\\
	&\left.+\int_{\Gamma_\varepsilon}\bm{n}\cdot \left( \frac{(u_v + \delta u_v) }{\rho_0}\nabla (p + \delta p)  \right)q d\Gamma
	\right].
\end{align}
Then, the variation of the objective function $\delta J$ can be estimated by,
\begin{align}
	\delta J &=2\mathrm{Re}\left[
	\sum_{\varphi = v,h}\left\{ \int_{\Gamma_\mathrm{ext}}  k_\varphi^{-2} (\bm{n}\cdot \nabla \delta u_\varphi  )v_\varphi  d\Gamma 
	-\int_{\Omega \setminus \overline{\Omega_\varepsilon}}k_\varphi^{-2}\nabla \delta u_\varphi \cdot \nabla v_\varphi d\Omega\nonumber \right. \right.\\
	&+\int_{\Omega_\varepsilon}k_\varphi^{-2}\nabla u_\varphi \cdot \nabla v_\varphi d\Omega
	+\int_{\Omega\setminus \overline{\Omega_\varepsilon}} \delta u_\varphi v_\varphi d\Omega
	-\int_{\Omega_\varepsilon}(u_\varphi -1)v_\varphi d\Omega
	\nonumber\\
	&\left.
	+\int_{\Gamma_\varepsilon}  k_\varphi^{-2} (\bm{n}\cdot \nabla (u_\varphi + \delta u_\varphi  ))v_\varphi  d\Gamma
	\right\} \nonumber\\
	&+\int_{\Gamma_\mathrm{in}}\frac{\partial f_1}{\partial p} \delta p d\Gamma
	+\int_{\Gamma_\mathrm{out}}\frac{\partial f_2}{\partial p} \delta p d\Gamma\nonumber\\
	&+\int_{\Gamma_\mathrm{ext}}\bm{n}\cdot \left( \frac{(u_v + \delta u_v) }{\rho_0}\nabla (p + \delta p)
	-u_v \nabla p  \right)q d\Gamma\nonumber\\
	&-\int_{\Omega\setminus  \overline{\Omega_\varepsilon}}\frac{1}{\rho_0}\left\{(u_v + \delta u_v)\nabla (p + \delta p)-u_v \nabla p \right\}\cdot \nabla q d\Omega \nonumber
	+\int_{\Omega_\varepsilon}\frac{u_v}{\rho_0}\nabla p \cdot \nabla q d\Omega
	\\
	& + \int_{\Omega\setminus \overline{\Omega_\varepsilon}} \frac{\omega^2}{K_0}\left[\left\{
	\gamma - (\gamma-1)(u_h + \delta u_h)
	\right\}(p + \delta p)   
	-\left\{
	\gamma - (\gamma-1)u_h\right\}p 
	   \right]q d\Omega \nonumber\\
	& - \int_{\Omega_\varepsilon}\frac{\omega^2}{K_0}\left\{
	\gamma - (\gamma-1)u_h \right\}pq d\Omega \nonumber\\
	&\left.+\int_{\Gamma_\varepsilon}\bm{n}\cdot \left( \frac{(u_v + \delta u_v) }{\rho_0}\nabla (p + \delta p)  \right)q d\Gamma
	\right].
\end{align}
By ignoring terms containing the variations of the second order and integrating by parts, $\delta J$ can be modified as
$\delta J$ can be modified as,
\begin{align}
	\delta J &=2\mathrm{Re}\left[
	\int_{\Gamma_\mathrm{in} }\left\{ \left(-\frac{ik_0}{\rho_0}p + \frac{2ik_0}{\rho_0}P_\mathrm{in}\right)q - k_v^{-2}(\bm{n} \cdot \nabla v_v)\right\}\delta u_v d\Gamma \right.\nonumber\\
	&+\int_{\Gamma_\mathrm{out} }\left\{ \left(-\frac{ik_0}{\rho_0}p  \right)q - k_v^{-2}(\bm{n} \cdot \nabla v_v)\right\}\delta u_v d\Gamma 
	+\int_{\Gamma_\mathrm{sym} }\left\{  - k_v^{-2}(\bm{n} \cdot \nabla v_v)\right\}\delta u_v d\Gamma \nonumber\\
	&-\int_{\Gamma_\mathrm{in}\cup \Gamma_\mathrm{out} \cup \Gamma_\mathrm{sym}}k_h^{-2}(\bm{n}\cdot \nabla v_v) \delta u_hd\Gamma\nonumber\\
	&+\int_{\Omega\setminus\overline{\Omega_\varepsilon}}\left\{
	k_v^{-2}\nabla^2 v_v + v_v - \frac{1}{\rho_0}(\nabla p \cdot \nabla q)
	\right\}\delta u_v d\Omega\nonumber\\
	&+\int_{\Omega\setminus\overline{\Omega_\varepsilon}}\left\{
	k_h^{-2}\nabla^2 v_h + v_h - \frac{\omega^2}{K_0}(\gamma - 1)pq
	\right\}\delta u_h d\Omega\nonumber\\
	&+\sum_{\varphi = v,h}\left\{\int_{\Gamma_\mathrm{wall}}k_\varphi^{-2}(\bm{n}\cdot \nabla \delta u_\varphi )v_\varphi d\Gamma
	-\int_{\Gamma_\varepsilon}k_\varphi^{-2}(\bm{n}\cdot \nabla v_\varphi)\delta u_\varphi d\Gamma \right.\nonumber\\
	&+\left. \int_{\Omega_\varepsilon}\left(k_\varphi^{-2}\nabla u_\varphi \cdot \nabla v_\varphi - (u_\varphi -1)v_\varphi\right)d\Omega
	+\int_{\Gamma_\varepsilon}k_\varphi^{-2}\bm{n}\cdot \nabla(u_\varphi + \delta u_\varphi)v_\varphi d\Gamma
		\right\}\nonumber\\
	&-\int_{\Gamma_\mathrm{in}}\left\{\frac{ik_0}{\rho_0}u_vq + \bm{n}\cdot \left(\frac{u_v}{\rho_0}\nabla q\right) -\frac{\partial f_1}{\partial p}   \right\} \delta p d\Gamma\nonumber\\
	&-\int_{\Gamma_\mathrm{out}}\left\{\frac{ik_0}{\rho_0}u_vq + \bm{n}\cdot \left(\frac{u_v}{\rho_0}\nabla q\right) -\frac{\partial f_2}{\partial p}   \right\} \delta p d\Gamma
	-\int_{\Gamma_\mathrm{sym}\cup \Gamma_\mathrm{wall}}\bm{n}\cdot \left( \frac{1}{\rho_0} \nabla q  \right)u_v \delta p d\Gamma\nonumber\\
	&+\int_{ \Omega \setminus \overline{\Omega_\varepsilon} }\left\{
	\nabla \cdot \left(\frac{u_v}{\rho_0}  \nabla q \right) + \frac{\omega^2}{K_0}\left\{\gamma - (\gamma -1)u_h\right\}q
	\right\}\delta p d\Omega\nonumber\\
	&\left.-\int_{\Gamma_\varepsilon}\bm{n}\cdot \left(\frac{1}{\rho_0}\nabla q\right)u_v\delta p d\Gamma
	+\int_{\Omega_\varepsilon}\left\{
	\frac{u_v}{\rho_0}\nabla p \cdot \nabla q  - \frac{\omega^2}{K_0}\left\{\gamma - (\gamma -1)u_h\right\}pq
	\right\}d\Omega \right],
\end{align}
where we introduced the boundary conditions in Eq.~(\ref{eq: BC_u}) and Eq.~(\ref{eq: BC_udu}).

We assume that the variables $(v_\varphi, q)$ with $\varphi = v,h$ satisfy the adjoint equations defined as follows:
\begin{align}
&\nabla \cdot \left(\frac{u_v}{\rho_0}\nabla q\right) + \frac{\omega^2}{K_0}\left\{\gamma - (\gamma-1)u_h
\right\}q =0~~~\mathrm{in~}\Omega,\\
&\bm{n}\cdot \left(\frac{u_v}{\rho_0}\nabla q\right) + \frac{ik_0}{\rho_0}u_v q = \frac{\partial f_1}{\partial p}~~~\mathrm{on~}\Gamma_\mathrm{in},\\
&\bm{n}\cdot \left(\frac{u_v}{\rho_0}\nabla q\right) + \frac{ik_0}{\rho_0}u_v q = \frac{\partial f_2}{\partial p}~~~\mathrm{on~}\Gamma_\mathrm{out},\\
&\bm{n}\cdot \left(\frac{1}{\rho_0}\nabla q\right) = 0~~~\mathrm{on~}\Gamma_\mathrm{wall} \cup \Gamma_\mathrm{sym},\\
&k_v^{-2}\nabla^2 v_v + v_v = \frac{1}{\rho_0}\nabla p \cdot \nabla q~~~\mathrm{in~}\Omega,\\
&k_h^{-2}\nabla^2 v_h + v_h = \frac{\omega^2}{K_0}(\gamma -1)pq~~~\mathrm{in~}\Omega,\\
&k_v^{-2}(\bm{n}\cdot \nabla v_v) = q\left( -\frac{ik_0}{\rho_0}p + \frac{2ik_0}{\rho_0}P_\mathrm{in}  \right)
~~~\mathrm{on~}\Gamma_\mathrm{in},\\
&k_v^{-2}(\bm{n}\cdot \nabla v_v) = q\left( -\frac{ik_0}{\rho_0}p   \right)
~~~\mathrm{on~}\Gamma_\mathrm{out},\\
&k_v^{-2}(\bm{n}\cdot \nabla v_v) = 0
~~~\mathrm{on~}\Gamma_\mathrm{sym},\\
&k_h^{-2}(\bm{n}\cdot \nabla v_h) = 0
~~~\mathrm{on~}\Gamma_\mathrm{in}\cup\Gamma_\mathrm{out}\cup\Gamma_\mathrm{sym},\\
&v_v =0~~~\mathrm{on~}\Gamma_\mathrm{wall},\\
&v_h =0~~~\mathrm{on~}\Gamma_\mathrm{wall}.
\end{align}
These adjoint equations are weakly coupled, but the way of coupling is different from the state equations. 
Once the adjoint acoustic pressure $q$ is obtained, the adjoint viscous and thermal fields, $v_v$ and $v_h$, can be solved, which is the inverse procedure for obtaining $(u_v,u_h,p)$.

Using these adjoint variables, $\delta J$ can be simplified as
\begin{align}
\delta J &=2\mathrm{Re}\left[
\sum_{\varphi = v,h}\left\{
-\int_{\Gamma_\varepsilon}k_\varphi^{-2}(\bm{n}\cdot \nabla v_\varphi)\delta u_\varphi d\Gamma \right.\right.\nonumber\\
&+\left. \int_{\Omega_\varepsilon}\left(k_\varphi^{-2}\nabla u_\varphi \cdot \nabla v_\varphi - (u_\varphi -1)v_\varphi\right)d\Omega
+\int_{\Gamma_\varepsilon}k_\varphi^{-2}\bm{n}\cdot \nabla(u_\varphi + \delta u_\varphi)v_\varphi d\Gamma
\right\}\nonumber\\
&\left.-\int_{\Gamma_\varepsilon}\bm{n}\cdot \left(\frac{1}{\rho_0}\nabla q\right)u_v\delta p d\Gamma
+\int_{\Omega_\varepsilon}\left(
\frac{u_v}{\rho_0}\nabla p \cdot \nabla q  - \frac{\omega^2}{K_0}\left\{\gamma - (\gamma -1)u_h\right\}pq
\right)d\Omega \right].
\label{eq: dJ}
\end{align} 
To obtain the topological derivative, the asymptotic behavior of variables $(u_\varphi, v_\varphi, p, q)$ and the variation $(\delta u_\varphi, \delta p)$ when $\varepsilon \to 0$ are required.
Since $(u_\varphi, v_\varphi, p, q)$ are smooth solutions around the center of $\Omega_\varepsilon$, the Taylor expansions are introduced:
\begin{align}
	u_\varphi &= u_\varphi(\bm{x_0}) + \nabla u_\varphi(\bm{x_0}) \cdot (\bm{x - x_0}) + O(|\bm{x-x_0} |^2),\nonumber\\
	v_\varphi &= v_\varphi(\bm{x_0}) + \nabla v_\varphi(\bm{x_0}) \cdot (\bm{x - x_0}) + O(|\bm{x-x_0} |^2),\nonumber\\
	p &=p(\bm{x_0}) + \nabla p(\bm{x_0}) \cdot (\bm{x - x_0}) + O(|\bm{x-x_0} |^2),\nonumber\\
	q &=q(\bm{x_0}) + \nabla q(\bm{x_0}) \cdot (\bm{x - x_0}) + O(|\bm{x-x_0} |^2),
\end{align}
where $\bm{x_0}$ represents the coordinate at the center of $\Omega_\varepsilon$.

The boundary value problems for $(\delta u_\varphi, \delta p)$ are introduced to estimate their asymptotic behaviors.
At first, $\delta u_\varphi$ satisfies the following boundary value problems:
\begin{align}
	&k_\varphi^{-2}\nabla^2 \delta u_\varphi + \delta u_\varphi = 0 ~~~\mathrm{in~}\Omega\setminus \overline{\Omega_\varepsilon},\\
	&\bm{n}\cdot \nabla \delta u_\varphi = -\bm{n}\cdot \nabla u_\varphi ~~~\mathrm{on~}\Gamma_\mathrm{in}\cup \Gamma_\mathrm{out}\cup\Gamma_\mathrm{sym},\\
	&\delta u_\varphi = -u_\varphi~~~\mathrm{on~}\Gamma_\mathrm{wall} \cup \Gamma_\varepsilon,
\end{align}
Since $\delta J$ contains integrals over $\Omega_\varepsilon$ and $\Gamma_\varepsilon$, it is enough to obtain the behavior of $\delta u_\varphi$ around $\bm{x_0}$. 
Therefore, the above boundary value problem for $\delta u_\varphi$ is approximated as follows:
\begin{align}
	&\nabla^2 \delta u_\varphi = 0~~~\mathrm{in~}\mathbb{R}^2\setminus\overline{\Omega_\varepsilon},\\
	&\delta u_\varphi = -\left\{ u_\varphi(\bm{x_0}) + \varepsilon \left(
	\frac{\partial u_\varphi}{\partial x_1}(\bm{x_0}) \cos\theta
	+\frac{\partial u_\varphi}{\partial x_2}(\bm{x_0}) \sin\theta
	\right)   \right\}~~~\mathrm{on~}\Gamma_\varepsilon,
\end{align}
where we introduced $(r,\theta)$ so that $\bm{x-x_0}=r \cos\theta \bm{e_1} + r \sin\theta \bm{e_2}$.
This problem is so-called the exterior problem, which extracts the main part of $\delta u_\varphi$ when $\varepsilon\to 0$.
Using the solution of the Laplace's equation, $\delta u_\varphi$ can be expressed as
\begin{align}
	\delta u_\varphi \approx -\frac{\log |\bm{x}- \bm{x_0}|}{\log \varepsilon}u_\varphi (\bm{x_0})  -\frac{\varepsilon^2}{|\bm{x-x_0}|^2} (\bm{x-x_0})\cdot \nabla u_\varphi(\bm{x_0}).
	\label{eq: formula of du}
\end{align}

Similarly, $\delta p$ satisfies the following boundary value problem:
\begin{align}
	&\nabla \cdot \left( \frac{u_v + \delta u_v}{\rho_0}\nabla \delta p\right) + \frac{\omega^2}{K_0}
	\left\{\gamma - (\gamma -1)(u_h + \delta u_h)\right\}\delta p\nonumber\\
	&=-\nabla \cdot \left( \frac{u_v + \delta u_v}{\rho_0}\nabla p\right) - \frac{\omega^2}{K_0}
	\left\{\gamma - (\gamma -1)(u_h + \delta u_h)\right\} p~~~\mathrm{in~}\Omega\setminus\overline{\Omega_\varepsilon},\\
	&\bm{n}\cdot \left( \frac{1}{\rho_0}  \nabla \delta p\right) + \frac{ik_0}{\rho_0}\delta p = 0~~~\mathrm{on~}\Gamma_\mathrm{in}\cup\Gamma_\mathrm{out},\\
	&\bm{n}\cdot \left( \frac{1}{\rho_0}  \nabla \delta p\right) = -\bm{n}\cdot \left(\frac{1}{\rho_0}\nabla p\right)~~~\mathrm{on~}\Gamma_\mathrm{sym}\cup\Gamma_\mathrm{wall}\cup \Gamma_\varepsilon.
\end{align}
Here, we assume that the variations in solutions $u_v$ and $u_h$ are small, that is, $\delta u_v\approx 0$ and $\delta u_h\approx 0$. 
Then, the behavior of $\delta p$ around $\bm{x_0}$ can be approximately expressed by the following boundary value problem:
\begin{align}
	&\nabla^2 \delta p = 0~~~\mathrm{in~}\mathbb{R}^2\setminus\overline{\Omega_\varepsilon},\\
	&\bm{n}\cdot \left(\frac{1}{\rho_0} \nabla \delta p\right) = \frac{1}{\rho_0}\left(\frac{\partial p}{\partial x_1}(\bm{x_0})\cos\theta +  \frac{\partial p}{\partial x_2}(\bm{x_0})\sin\theta   \right)~~~\mathrm{on~}\Gamma_\varepsilon.
\end{align} 
Then, $\delta p$ is explicitly expressed as
\begin{align}
	\delta p \approx \frac{\varepsilon^2}{|\bm{x-x_0}|^2}(\bm{x-x_0})\cdot \nabla p(\bm{x_0}).\label{eq: formula of dp}
\end{align}
Mathematically, the assumptions that $\delta u_v \approx 0$ and $\delta u_h \approx 0$ are not true as Eq.~(\ref{eq: formula of du}) indicates.
However, we numerically show the utility of the obtained topological derivative based on these assumptions.

Substituting Eq.~(\ref{eq: formula of du}) and Eq.~(\ref{eq: formula of dp}) into Eq.~(\ref{eq: dJ}), the variation of the objective function can be estimated as follows:
\begin{align}
	\delta J &= 2\mathrm{Re}\left[
	\frac{2\pi}{\log \varepsilon}\left\{k_v^{-2}u_v(\bm{x_0}) v_v(\bm{x_0}) + k_h^{-2}u_h(\bm{x_0}) v_h(\bm{x_0})\right\}\right.\nonumber\\
	&+\pi \varepsilon^2\left\{
	-2k_v^{-2}\nabla u_v(\bm{x_0}) \cdot \nabla v_v(\bm{x_0}) -2k_h^{-2}\nabla u_h(\bm{x_0}) \cdot \nabla v_h(\bm{x_0}) \right.\nonumber\\
	&-(u_v(\bm{x_0}) -1)v_v-(u_h(\bm{x_0}) -1)v_h\nonumber\\
	&\left.\left. + 2\frac{u_v(\bm{x_0})}{\rho_0}\nabla p(\bm{x_0}) \cdot \nabla q(\bm{x_0}) 
	- \frac{\omega^2}{K_0}\left\{\gamma - (\gamma-1)u_h(\bm{x_0})\right\}  p(\bm{x_0}) q(\bm{x_0})
	\right\}
	\right]+o(\varepsilon^2)
	\label{eq: formulat of dJ}
\end{align}

\subsection{Numerical estimation of $\delta J$ and the explicit formula of the topological derivative}\label{sec: numerical dJ}
We numerically demonstrate that the estimation of $\delta J$ in Eq.~(\ref{eq: formulat of dJ}) is practically valid.
To do this, the following numerical difference corresponding to $\delta J$ is introduced:
\begin{align}
	\delta J_\mathrm{Num} = J(\Omega\setminus \overline{\Omega_{\varepsilon_0}}) - J(\Omega),
\end{align}
where $\Omega_{\varepsilon_0}$ is a circular rigid domain with a finite radius $0<\varepsilon_0 \ll 1$.
$J(\Omega\setminus \overline{\Omega_{\varepsilon_0}})$ represents the objective function with $\Omega_{\varepsilon_0}$, whereas $J(\Omega)$ represents that without $\Omega_{\varepsilon_0}$.

\begin{figure}[H]
	\centering
	\includegraphics[scale=0.6]{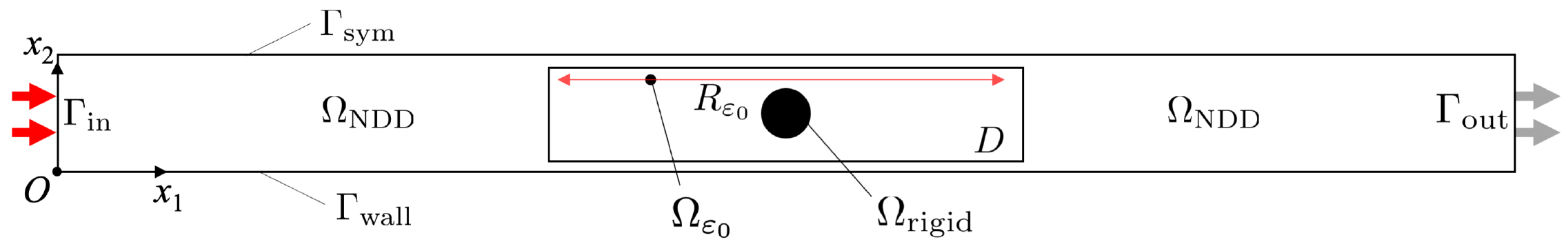}
	\caption{
		Computational domain setting to take the numerical difference, $\delta J_\mathrm{Num}$.}
	\label{fig:DtJ_comptmodel}       
\end{figure}
Figure~\ref{fig:DtJ_comptmodel} shows the computational model to take the numerical difference. 
The geometrical setting and boundary conditions are based on the example of Case 2. A rigid domain, whose radius is $0.1 D_\mathrm{ex}$ with $D_\mathrm{ex}=0.03$[m], is fixed to the center of the design domain. 
A small circular rigid domain
$\Omega_{\varepsilon_0}$ with radius $\varepsilon_0=\frac{D_\mathrm{ex}}{80}$[m] is placed in the range of $R_{\varepsilon_0}: 0.06375\mathrm{[m]}  \le x_1 \le 0.12225\mathrm{[m]}, x_2=0.012$[m]. 
The frequency of the incident wave is set to 2500 [Hz]. The rigid domains are expressed by eliminating their regions from the computational domain; thus, the method explained in Section~\ref{sec:Remesh and rigid domain} is not used here. Other settings are the same as those explained in Section~\ref{sec: numerical examples}.

For simplicity, we set the objective function $J$ as $J=|S_{11}|^2$ in this example.
We first solve the state and adjoint variables without the small rigid domain $\Omega_{\varepsilon_0}$ and 
estimate the formula of $\delta J$ by setting $\varepsilon=\varepsilon_0$, where the remainder $o(\varepsilon^2)$ is neglected.
Then, the rigid domain $\Omega_{\varepsilon_0}$ is set and numerical differences $\delta J_\mathrm{Num}$ are taken.
In addition to the comparison between $\delta J_\mathrm{Num}$ and $\delta J$,  
the following approximated variation of $J$ is introduced:
\begin{align}
	\delta J_p &= \pi \varepsilon^2 \times 2\mathrm{Re}\left[
	 \frac{2}{\rho_0}\nabla p(\bm{x_0}) \cdot \nabla q(\bm{x_0}) 
	- \frac{\omega^2}{K_0} p(\bm{x_0}) q(\bm{x_0})
	\right]
	\label{eq: formula of dJp}
\end{align}
This formula is obtained by ignoring terms with $O(\frac{1}{\log \varepsilon})$ and by setting $u_v(\bm{x_0})=1$ and $u_h(\bm{x_0})=1$. As a result, the viscous and thermal fields $(u_v,u_h)$ and corresponding adjoint fields $(v_v,v_h)$ are eliminated.
In fact, the terms with $O(\frac{1}{\log \varepsilon})$ in Eq.~(\ref{eq: formulat of dJ}) were derived from the terms corresponding to the weak forms for $u_h$ and $u_v$.
Since the objective function in this example and that used in the numerical examples explicitly depends on the acoustic pressure, the effects of these terms on the variation of the objective function are deduced to be small. 
Furthermore, $u_v$ and $u_h$ approach to 1 apart from the boundary layers because of their complex wavenumbers. 
Thus, if the position of interest $\bm{x_0}$ is apart from the surface of the rigid domain, setting them to 1 is justified.
\begin{figure}[H]
	\centering
	\includegraphics[scale=0.6]{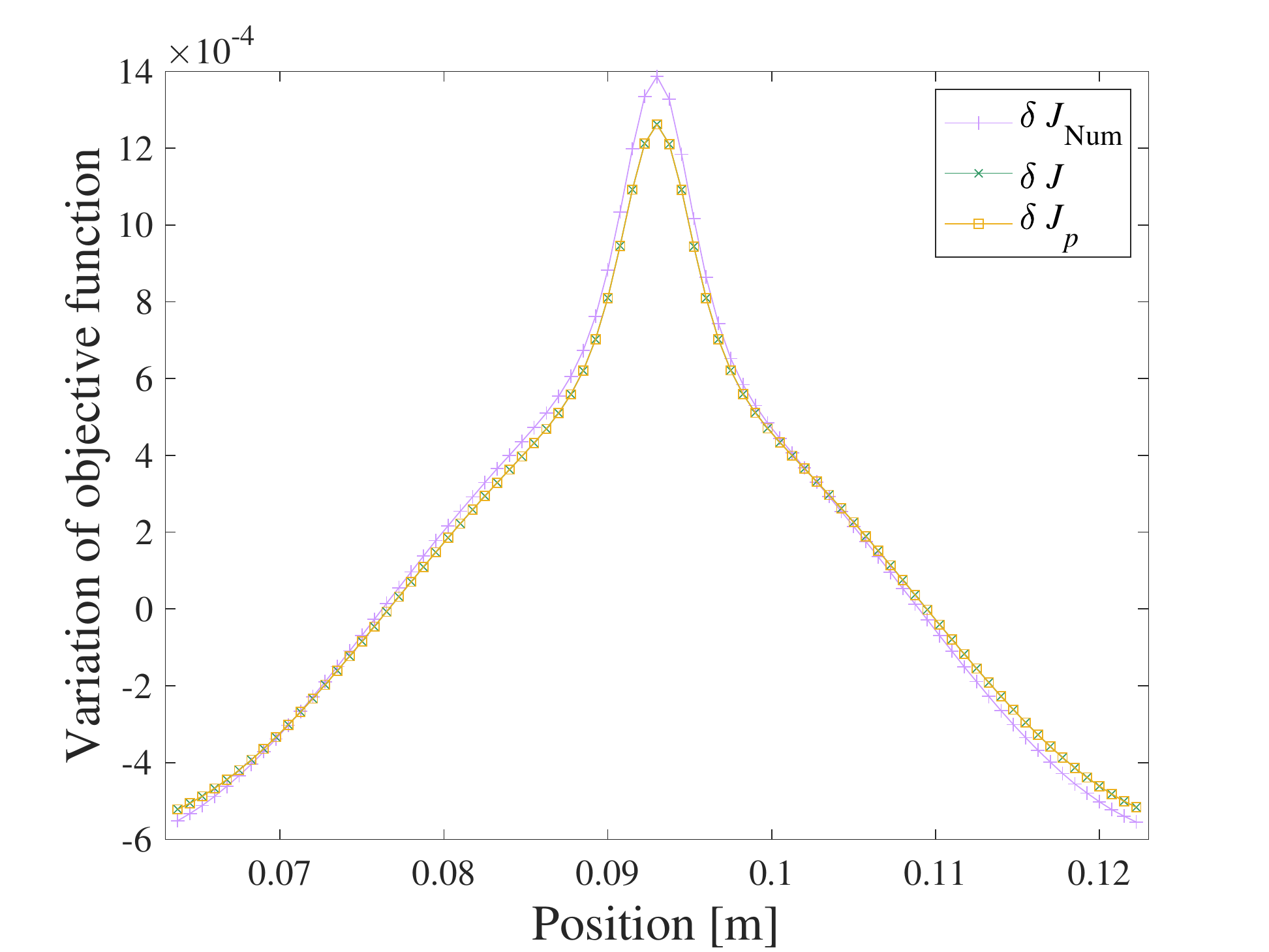}
	\caption{Comparison of the numerical differences $\delta J_\mathrm{Num}$, the derived variation $\delta J$, and the approximated variation $\delta J_p$.}
	\label{fig:DtJplots}       
\end{figure}
Figure~\ref{fig:DtJplots} presents the comparison results of $\delta J_\mathrm{Num}$, $\delta J$, and $\delta J_p$. 
Slight differences between the values of $\delta J_\mathrm{Num}$ and $\delta J$ can be seen, especially around their maximum and minimum values. 
These differences are likely due to the rough estimation of $\delta p$ in Eq.~(\ref{eq: formula of dp}); however, their trends in the magnitude relationship are the same throughout the examined range. 
Therefore, the derived formula $\delta J$ is practically usable for the estimation of the variation of $J$. 
Furthermore, both plots of $\delta J$ and $\delta J_p$ are almost identical. Therefore, we chose to use Eq.~(\ref{eq: formula of dJp}) as a variation of the objective function. By recalling the definition of the topological derivative, given by Eq.~(\ref{eq: Def of DtJ Appendix}), the topological derivative \textcolor {black}{$D_T J$} is finally obtained as
\begin{align}
	D_T J = 2\mathrm{Re}\left[
	\frac{2}{\rho_0}\nabla p(\bm{x_0}) \cdot \nabla q(\bm{x_0}) 
	- \frac{\omega^2}{K_0} p(\bm{x_0}) q(\bm{x_0})
	\right],
\end{align}
with $g(\varepsilon)=\pi \varepsilon^2$.

As we have noted, this topological derivative is effective when the objective function is formulated using the acoustic pressure at a certain frequency. In a multi-frequency setting, a summation must be performed for the given frequencies, as follows:
\begin{align}
	D_T J = \sum_{k=0}^{n}2\mathrm{Re}\left[
	\frac{2}{\rho_0}\nabla p(\omega_k)(\bm{x_0}) \cdot \nabla q(\omega_k)(\bm{x_0}) 
	- \frac{\omega^2}{K_0} p(\omega_k)(\bm{x_0}) q(\omega_k)(\bm{x_0})
	\right],
\end{align}
where $p(\omega_k)$ and $q(\omega_k)$ represent the pressure and adjoint pressure at $\omega=\omega_k$, respectively. 
By considering the arbitrary position $\bm{x}\in D$ instead of $\bm{x_0}$ as the central coordinate of $\Omega_{\varepsilon}$,
the topological derivative can be finally derived, as shown in Eq.~(\ref{eq: DtJ_mainpart}).
\bibliographystyle{elsarticle-num} 
\bibliography{reference}


%
%
%
\end{document}